\documentclass{emulateapj}

\usepackage{natbib}

\usepackage{graphicx}
\usepackage{latexsym}
\usepackage{amsfonts,amsmath,amssymb}
\usepackage{hyperref}
\usepackage[utf8]{inputenc}
\usepackage{longtable}

\newcommand{\jcap}{J. Cosm. Astropart. Phys.}

\newcommand{\be}{\begin{equation}}
\newcommand{\ee}{\end{equation}}
\newcommand{\bea}{\begin{eqnarray}}
\newcommand{\eea}{\end{eqnarray}}

\def\Sref#1{Section~\ref{#1}}
\def\Fref#1{Figure~\ref{#1}}
\def\Tref#1{Table~\ref{#1}}

\newcommand{\paperone}{Paper I}

\newcommand{\kms}{\ifmmode  \,\rm km\,s^{-1} \else $\,\rm km\,s^{-1}  $ \fi }
\newcommand{\kpc}{\ifmmode  {\rm kpc}  \else ${\rm  kpc}$ \fi  }
\newcommand{\pc}{\ifmmode  {\rm pc}  \else ${\rm pc}$ \fi  }
\newcommand{\Msun}{\ifmmode {\rm M_{\odot}} \else ${\rm M_{\odot}}$ \fi}
\newcommand{\Zsun}{\ifmmode {\rm Z_{\odot}} \else ${\rm Z_{\odot}}$ \fi}
\newcommand{\yr}{\ifmmode yr^{-1} \else $yr^{-1}$ \fi}
\newcommand{\hMsun}{\ifmmode h^{-1}\,\rm M_{\odot} \else $h^{-1}\,\rm M_{\odot}$ \fi}

\def\dt{\Delta t}
\def\ithree{i_3}

\def\SFinf{{\rm SF}_{\infty}}

\usepackage[usenames]{color}
\definecolor{purple}{RGB}{150,0,200}

\newcommand{\nteamsA}{Seven}
\newcommand{\nteamsaa}{seven}
\newcommand{\nmethods}{78}

\shorttitle{Strong Lens Time Delay Challenge: II. Results of TDC1}
\shortauthors{Liao et al.}

\begin{document}

\title{Strong Lens Time Delay Challenge: II. Results of TDC1}

\author{Kai~Liao\altaffilmark{1,2}$^*$}
\author{Tommaso~Treu\altaffilmark{2}$^{*}$}
\author{Phil~Marshall\altaffilmark{3}}
\author{Christopher~D.~Fassnacht\altaffilmark{4}}
\author{Nick Rumbaugh\altaffilmark{4}}
\author{Gregory~Dobler\altaffilmark{5,20}}
\author{Amir~Aghamousa\altaffilmark{9}}
\author{Vivien~Bonvin\altaffilmark{13}}
\author{Frederic~Courbin\altaffilmark{13}}
\author{Alireza~Hojjati\altaffilmark{6,7}}
\author{Neal~Jackson\altaffilmark{12}}
\author{Vinay Kashyap\altaffilmark{17}}
\author{S.~Rathna~Kumar\altaffilmark{14}}
\author{Eric~Linder\altaffilmark{8,18}}
\author{Kaisey Mandel\altaffilmark{17}}
\author{Xiao-Li Meng\altaffilmark{15}}
\author{Georges Meylan\altaffilmark{13}}
\author{Leonidas~A.Moustakas\altaffilmark{11}}
\author{Tushar~P.~Prabhu\altaffilmark{14}}
\author{Andrew~Romero-Wolf\altaffilmark{11}}
\author{Arman~Shafieloo\altaffilmark{9,10}}
\author{Aneta Siemiginowska\altaffilmark{17}}
\author{Chelliah~S.~Stalin\altaffilmark{14}}
\author{Hyungsuk Tak\altaffilmark{15}}
\author{Malte~Tewes\altaffilmark{19}}
\author{David van Dyk\altaffilmark{16}}

\altaffiltext{1}{Dept.\ of Astronomy, Beijing Normal University, Beijing 100875, China}
\altaffiltext{2}{Dept.\ of Physics, University of California, Santa Barbara, CA 93106, USA.}
\altaffiltext{3}{Kavli Institute for Particle Astrophysics and Cosmology, P.O. Box 20450, MS29, Stanford, CA 94309, USA.}
\altaffiltext{4}{Dept.\ of Physics, University of California, 1 Shields Ave., Davis, CA 95616, USA}.
\altaffiltext{5}{Kavli Institute for Theoretical Physics, University of California Santa Barbara, Santa Barbara, CA 93106, USA.}
\altaffiltext{6}{Dept.\ of Physics and Astronomy, University of British Columbia, 6224 Agricultural Road, Vancouver, B.C. V6T 1Z1, Canada}
\altaffiltext{7}{Dept.\ of Physics, Simon Fraser University, 8888 University Drive, Burnaby BC, Canada V5A1S6}
\altaffiltext{8}{Lawrence Berkeley National Laboratory and University of California, Berkeley, CA 94720}
\altaffiltext{9}{Asia Pacific Center for Theoretical Physics, Pohang, Gyeongbuk 790-784, Korea}
\altaffiltext{10}{Department of Physics, POSTECH, Pohang, Gyeongbuk 790-784, Korea}
\altaffiltext{11}{Jet Propulsion Laboratory, California Institute of Technology, M/S
169-506, 4800 Oak Grove Dr, Pasadena, CA 91109}
\altaffiltext{12}{University of Manchester, School of Physics \& Astronomy, Jodrell Bank Centre for Astrophysics, Manchester M13 9PL, UK}
\altaffiltext{13}{EPFL, Lausanne, Switzerland}
\altaffiltext{14}{Indian Institute of Astrophysics, II Block, Koramangala, Bangalore 560 034, India}
\altaffiltext{15}{Dept. of Statistics, Harvard University, 1 Oxford St., Cambridge, MA, 02138, USA}
\altaffiltext{16}{Department of Mathematics, Imperial College London, London SW7 2AZ UK}
\altaffiltext{17}{Harvard-Smithsonian Center for Astrophysics, 60 Garden St., Cambridge, MA 02138 USA}
\altaffiltext{18}{Korea Astronomy and Space Science Institute, Daejeon 305-248, South Korea }
\altaffiltext{19}{Argelander-Institut f\"ur Astronomie, Auf dem H\"ugel 71, D-53121 Bonn, Germany}

\altaffiltext{*}{Dept of Physics and Astronomy, University of California, Los Angeles CA 90095; tt@astro.ucla.edu}
\altaffiltext{20}{Center for Urban Science + Progress, New York University, Brooklyn, NY 11201, USA.}


\begin{abstract}
We present the results of the first strong lens time delay challenge.
The motivation, experimental design, and entry level challenge are
described in a companion paper. This paper presents the main
challenge, TDC1, which consisted of analyzing thousands of simulated
light curves blindly. The observational properties of the light curves
cover the range in quality obtained for current targeted efforts
(e.g.,~COSMOGRAIL) and expected from future synoptic surveys
(e.g.,~LSST), and include simulated systematic errors. \nteamsA\ teams
participated in TDC1, submitting results from \nmethods\ different
method variants. After a describing each method, we compute and
analyze basic statistics measuring accuracy (or bias) $A$, goodness of
fit $\chi^2$, precision $P$, and success rate $f$.  For some methods
we identify outliers as an important issue. Other methods show that
outliers can be controlled via visual inspection or conservative
quality control. Several methods are competitive, i.e., give
$|A|<0.03$, $P<0.03$, and $\chi^2<1.5$, with some of the methods
already reaching sub-percent accuracy.  The fraction of light curves
yielding a time delay measurement is typically in the range $f =
$20--40\%. It depends strongly on the quality of the data:
COSMOGRAIL-quality cadence and light curve lengths yield significantly
higher $f$ than does sparser sampling.  Taking the results of TDC1 at
face value, we estimate that LSST should provide around 400 robust
time-delay measurements, each with $P<0.03$ and $|A|<0.01$, comparable
to current lens modeling uncertainties. In terms of observing
strategies, we find that $A$ and $f$ depend mostly on season length,
while P depends mostly on cadence and campaign duration.
\end{abstract}



\section{Introduction}
\label{sec:intro}

The past decade has seen the emergence of a concordance cosmology,
$\Lambda$CDM, in which the contents of the universe are dominated by
dark matter and dark energy. Even though the basic
parameters appear to be robustly measured, more stringent measurements
are sought as a way to improve our understanding of the nature of
these mysterious components, as well as a way to test the model
against signatures of new physics \citep{SuyuEtal2012,2013PhR...530...87W}.

Achieving better cosmography means two things. On the one hand,
increasingly higher quality data are being obtained
\citep[e.g.][]{Planck2013} in order to improve the precision of each
method. On the other hand, independent observational methods are being
exploited to break the degeneracies inherent to each method and to uncover
unknown systematic uncertainties, thus improving accuracy. With
precision and accuracy rigorously under control, potential
inconsistencies might reveal new physics, such as the presence of
additional families of neutrinos or deviations from general relativity.

In the past few years, strong lens time delays \citep{Refsdal1964,Koc02} have
made something of a comeback, becoming an  increasingly popular probe of
cosmography
\citep{Oguritimedelay2007,timedelaycosmology2009,DobkeEtal2009,Hubble2010,
Tre++13,timedelaycosmology2014}. The configuration most suitable
for this work consists of a quasar with variable luminosity, being
lensed by a foreground elliptical galaxy that creates multiple images
of the quasar \citep[e.g.,][for a recent review]{Tre10}. Differences in optical paths
and gravitational potentials give rise to time delays between the
images.  In turn, the observable time delays, combined with a model of
the mass distribution in the main deflector and along the line of sight, provide
information on the so-called time-delay distance, which is a combination
of angular diameter distances. The time delay distance is primarily
sensitive to the Hubble constant \citep{SuyuEtal2013}, but can also
constrain other cosmological parameters, especially with large numbers
of time delay systems and in combination with other methods
\citep{Paraficz2009,Linder2011}.

At the time of writing, only a fraction of the hundred or so known
gravitationally lensed quasars has well-measured time delays, owing to
the considerable observational challenge associated with this
measurement. Accurate time delays in the optical require long and
well-sampled light curves as well as sophisticated algorithms that
account for data irregularities and astrophysical effects such as
microlensing \citep[e.g.,][]{TewesEtal2013a}. Radio wavelength light
curves have been used to determine time delays with great accuracy
\citep[e.g.,][]{FassnachtEtal2002}, but unfortunately are restricted to
the radio-loud subset of systems. In all cases, the success rate is
limited by the intrinsic variability of the sources.

The number of systems with known time delays is about to increase
dramatically. In the immediate future, as more lensed quasars are
discovered (e.g. via the STRIDES program\footnote{\url{strides.physics.ucsb.edu}}),
there will be more opportunities to identify highly variable systems in
cosmologically favorable configurations for targeted follow-up. The
state-of-the-art project COSMOGRAIL\footnote{\url{http://www.cosmograil.org}} with its newly developed
methods \citep{TewesEtal2013a} has shown the potential power of
extracting time delay data from sparsely sampled photometric data
\citep{TewesEtal2013b}. In the near future, the upcoming cadenced
optical imaging surveys will provide light curves for large samples of
lensed quasars.  For example, the Large Synoptic Survey Telescope
\citep[LSST;][]{LSSTSciBook,LSSTpaper} will repeatedly observe approximately
18000 deg$^2$ of sky for ten years, and is predicted to find and monitor
several thousand time delay lens systems \citep{OM10,DESCwhitepaper}.

In preparation for this wealth of light curves, it is crucial to carry
out a systematic study of the current algorithms for time delay
determination. Such an investigation has two main goals. The first is to
determine whether current methods have sufficient precision and accuracy
to exploit the kind of data anticipated in the next decade. Identifying
limitations and failure modes of current methods is a necessary step to
develop the next generation of measurement algorithms. In parallel, the
second goal is to test the impact of different observational strategies.
For example, what kind of cadence, duration, and sensitivity is required
to obtain precise and accurate time delays? Is the LSST baseline
strategy sufficient to meet the goals of time delay cosmography or can
we identify changes that would improve the outcome?

With these two goals in mind, a Time Delay Challenge (TDC) was
initiated in October 2013. The challenge ``Evil'' Team (GD, CDF, KL,
PJM, NR, TT) simulated large numbers of time delay light curves,
including all anticipated physical and experimental effects. The wider
community was then invited to extract time delay signals from these
mock light curves, blindly, using their own algorithms as ``Good
Teams.''\footnote{We note here that the tongue-in-cheek names ``evil''
and ``good'' teams do not denote any despicable intention or moral
judgment, but were chosen to capture the desire of the challenge
designers to produce significantly realistic (and difficult) light
curves as well as an incentive for the outside teams to participate.}
This invitation was made by the posting of an initial version of
\paperone\ of this series \citep{PaperI} on the arxiv.org preprint
server, and on the TDC website (\url{http://timedelaychallenge.org/}).

The two first ladders of this challenge are TDC0 and TDC1. TDC0
consisted of a small set of simulated data, which was used mostly as a
debugging and validation tool.  TDC0 is discussed in detail in \paperone.
Four statistics were used to evaluate the performance of every method's
submitted time delays $\tilde{\Delta t}_i$ and uncertainties $\delta_i$,
in light of the the true time delay value (defined as positive in the input),
$\Delta t_i$.
These four metrics are: the success fraction
\begin{equation}
f \equiv \frac{N_{\rm submitted}}{N},
\end{equation}
where $N$ is the total number of light curves available for analysis
in the ladder, the $\chi^2$ value:
\begin{equation}
\chi^2 = \frac{1}{fN}\sum_i
\left(\frac{\tilde{\Delta t}_i - \Delta t_i}{\delta_i}\right)^2,
\end{equation}
the ``precision''
\begin{equation}
P=\frac{1}{fN}\sum_i \left(\frac{\delta_i}{\Delta t_i}\right),
\end{equation}
and the ``accuracy'' or ``bias''
\begin{equation}
A=\frac{1}{fN} \sum_i \frac{\tilde{\Delta t}_i - \Delta
t_i}{\Delta t_i}.
\end{equation}


In addition
to the sample metrics we also define the analogous metrics for each
individual point $A_i$, $P_i$ and $\chi^2_i$.  Thus, \textbf{$A$, $P$ and $\chi^2$
defined above are the averages of the individual point values.}

Target thresholds in each of these sample metrics were set for the teams
entering TDC0. The \nteamsaa\ ``Good'' Teams whose methods passed these
thresholds were given access to the TDC1 dataset, which consisted of
several thousand light curves. This large number was motivated by the
goals of revealing the potential biases of each algorithm at the
sub-percent level and testing the ability of current pipelines to handle
large volumes of data.

To put this challenge in cosmological context, absolute distance
measurements with 1\% precision and accuracy are highly desirable for
the study of dark energy \citep{SuyuEtal2012,2013PhR...530...87W} and
other cosmological parameters. Therefore, in order for the time delay method to be
competitive it has to be demonstrated that the delays can be measured with
sub-percent accuracy {\it and} that the combination of precision for
each system and the available sample size is sufficient to bring the
statistical uncertainties to sub-percent level in the near future.
The total uncertainty on the time delay distance, and therefore on
the derived cosmology, depends on both the time delay and on the residual
uncertainties from modeling the lens potential and the structure along
the line of sight.  Thus, controlling the precision and accuracy of the
time delay measurement is a necessary, but not sufficient, condition. In
this first challenge we focus on just the time delay aspect of the
measurement. The assessment of residual systematic uncertainties in the
other components of time delay lens cosmography, and the distillation of
the time delay measurement biases and uncertainties into a single
cosmology metric is left for future work.

This paper focuses on TDC1, the analysis period of which closed on 1
July 2014, and it is structured as follows. \Sref{sec:tdc1} contains a
brief recap of the light curve generation process, and describes the
design of TDC1. In \Sref{sec:response} we describe the response of the
community to the challenge and give a brief summary of each method that
was applied, and then in \Sref{sec:analysis} we analyze the submissions.
We look at some of the apparent implications of the TDC1 results for
future survey strategies in \Sref{sec:strategy}, and  briefly discuss
our findings in \Sref{sec:discussion}. In \Sref{sec:summary} we
summarize our conclusions.


\section{Description of Time Delay Challenge TDC1}
\label{sec:tdc1}

In TDC1, the ``Evil'' Team simulated several thousand realistic mock
light curve pairs, using the methods outlined in \paperone.  In this
section, we first describe the general 5-rung design of TDC1, and then
describe the process of generating these light curves step by step,
revealing quantitative details of all the elements considered. {\bf We
emphasize that TDC1 was purely a light curve analysis challenge; no
additional information regarding the gravitational lensing
configuration, such as positions of the multiple images, or
redshifts of the source and deflector, was given. This choice was motivated by
the goal of performing the simplest possible test of time delay
algorithms. As discussed at the end of this paper, the inclusion of
additional lensing information could provide means to further improve
the performance of the methods.}


\subsection{The rungs of the challenge}

Each rung of TDC1 represents a possible wide-field survey that has
monitored sufficient sky area that we are in possession of light curves
for 1000 gravitationally-lensed AGN image pairs.  The number of lens
systems in this sample is somewhat less than 1000: quad systems are
presented as 2 pairs, flagged as coming from the same system but
enabling two independent time delay measurements.  The five rungs of
TDC1 span a selection of possible observing strategies, ranging from a
high cadence, long season dedicated survey (such as COSMOGRAIL might
evolve into), to the kind of ``universal cadence'' strategy that might
be adopted for an ``all-sky'' synoptic imaging survey (such as is being
designed for LSST). The challenge allows four control variables to be
investigated (within small plausible ranges): cadence, sampling
regularity, observing season length, and campaign duration.
\Tref{tab:obs} gives the values of these control variables for each
rung.

To make the mock data generation more efficient, and to better enable
comparison of results between the different rungs, we re-used the same
catalog of lenses for all the rungs. This trick was disguised from the
``Good'' Teams by randomly re-allocating the lightcurve identification
labels in each rung. In addition, the random noise was independently
generated in each rung. As a consequence, the submissions for different
rungs may be deemed independent, as if they had addressed 5000 lensed
image pairs.

\begin{table*}
\begin{center}
\begin{tabular}{cccccc} \hline\hline
  Rung &  Mean Cadence & Cadence Dispersion & Season   & Campaign & Length   \\
       &  (days)       & (days)             & (months) & (years)  & (epochs) \\ \hline
  0    &    3.0        &   1.0              &   8.0    &    5     & 400      \\
  1    &    3.0        &   1.0              &   4.0    &    10    & 400      \\
  2    &    3.0        &   0.0              &   4.0    &    5     & 200      \\
  3    &    3.0        &   1.0              &   4.0    &    5     & 200      \\
  4    &    6.0        &   1.0              &   4.0    &    10    & 200      \\
\hline\hline
\end{tabular}
\caption{The observing parameters for the five rungs of
TDC1.\label{tab:obs}}
\end{center}
\end{table*}


\subsection{Lens sample}
\label{sec:tdc1:sample}

\begin{figure}[!htbp]
\includegraphics[width=0.9\linewidth]{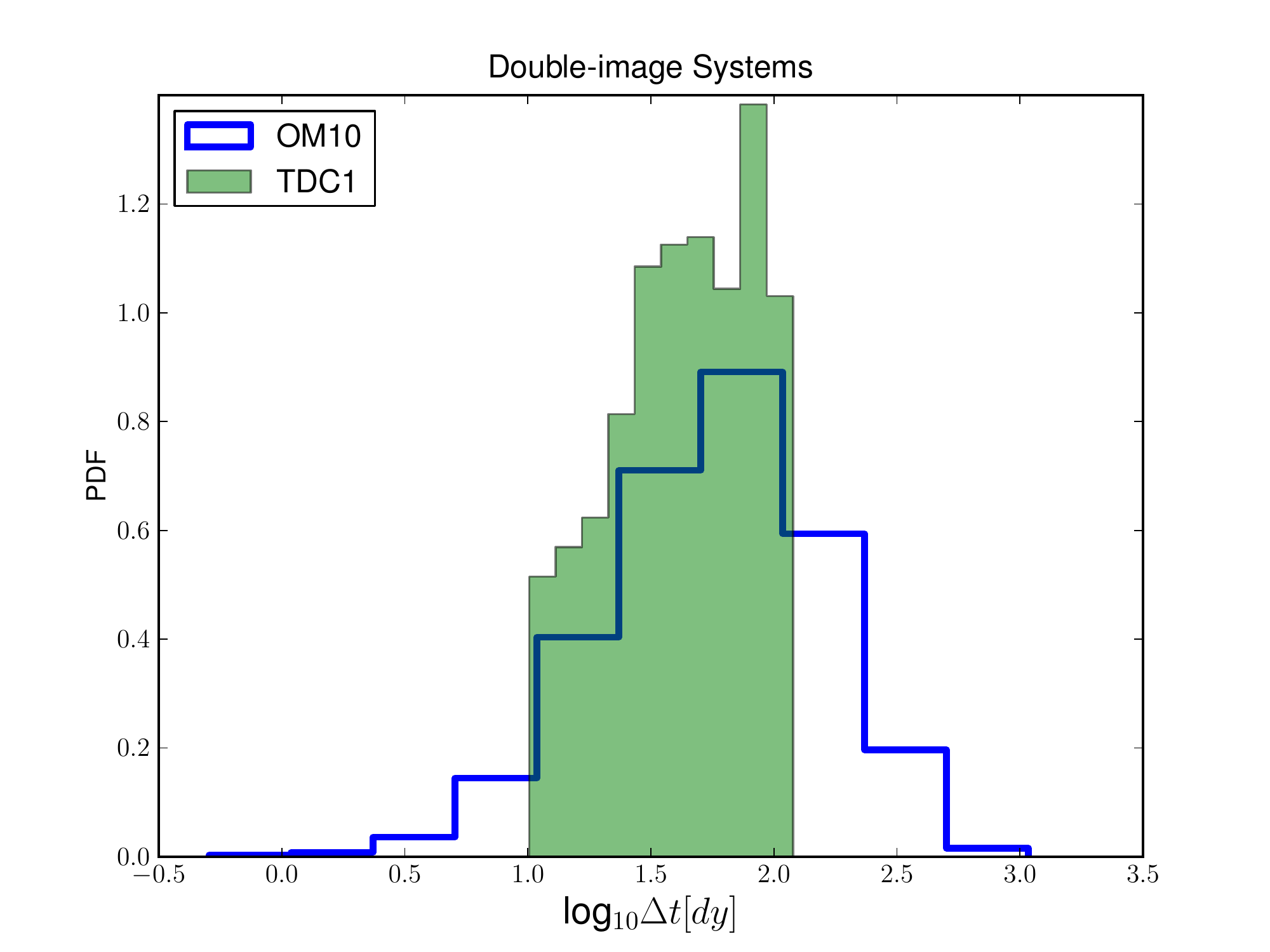}
\includegraphics[width=0.9\linewidth]{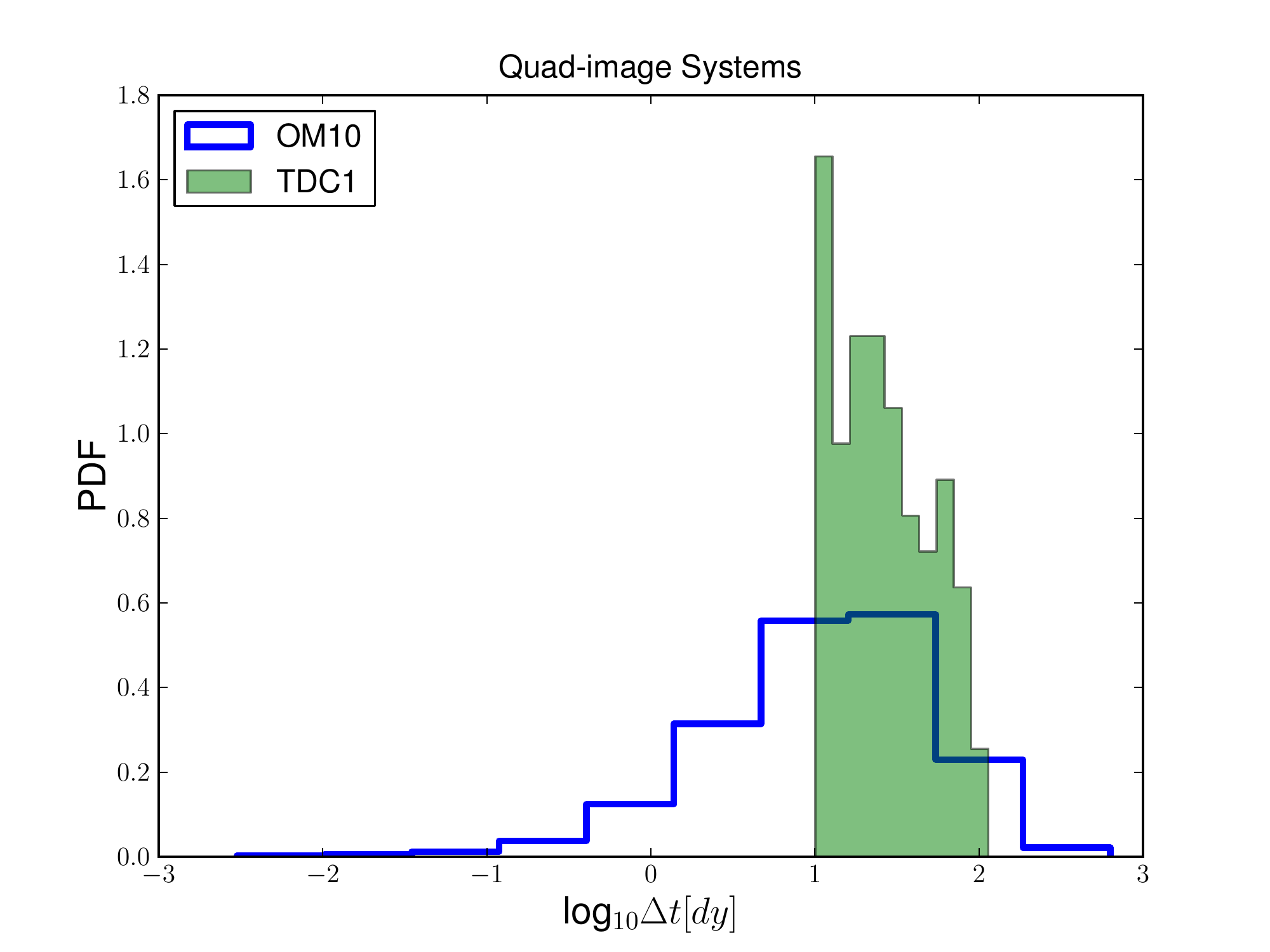}
\caption{Time delay distributions, from both the parent OM10 catalog
and the sample used in the TDC1 analysis, for the
double-image (top) and quad-image (bottom) systems.}
\label{fig:dt}
\end{figure}

The time delays between the light curves of gravitationally lensed
images  are determined primarily by the macro structure of the lens
galaxy. For the TDC1 sources and lenses we use the  mock LSST catalog of
lensed quasar systems prepared by \citet[][hereafter
OM10]{OM10}.\footnote{The OM10 catalog is available from
\url{https://github.com/drphilmarshall/OM10}} This sample was drawn from
plausible physical distributions for the various key  properties of
lensed quasar systems and very approximate observing conditions expected
with LSST, namely a characteristic angular resolution of 0.75 arcsec and
a 10-sigma limiting magnitude per monitoring epoch of 23.3 in the
$i$-band. Assuming a survey area of 18000 square degrees, these numbers
correspond to an OM10-predicted mock sample of some 2813 lenses. Given
these constraints, we randomly drew 720 doubly-imaged and 152
quadruply-imaged quasars from this catalog, to give a total of 1024
independent time delayed image pairs. As \Fref{fig:dt} shows, the mean
time delay in TDC1 is several tens of days. We rejected all time delays
outside the range 5 to 120 days as we drew the mock sample, since the
typical observing cadence and season length are expected to be a few
days and a few months respectively. The same time delay range constraint
reduced the parent  OM10 mock lens sample by 76\%, to 2124 lenses. When
analyzing the submissions, we found that very few accurate measurements
of time delays less than 10 days were possible, and so in the rest of
this paper we focus on the range $10 < \Delta t < 120$ days. Imposing
this narrower range on the OM10 mock LSST lens sample results in  1990
systems. While the image pairs with $5 < \Delta t < 10$ days were not
used in the analysis, they are still there in the TDC1 dataset for
potential future use.

\begin{figure}[!htbp]
\includegraphics[width=0.9\linewidth]{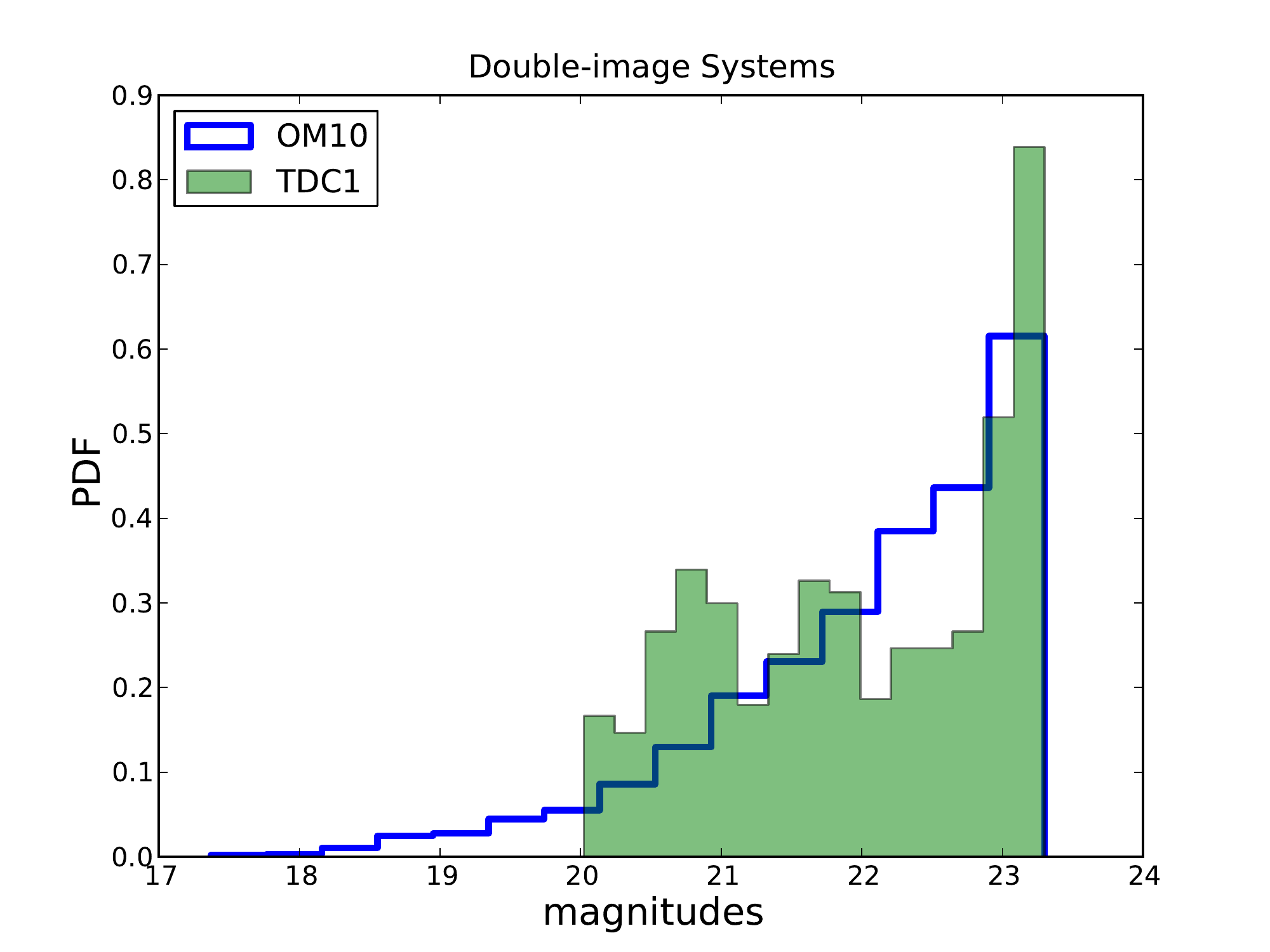}
\includegraphics[width=0.9\linewidth]{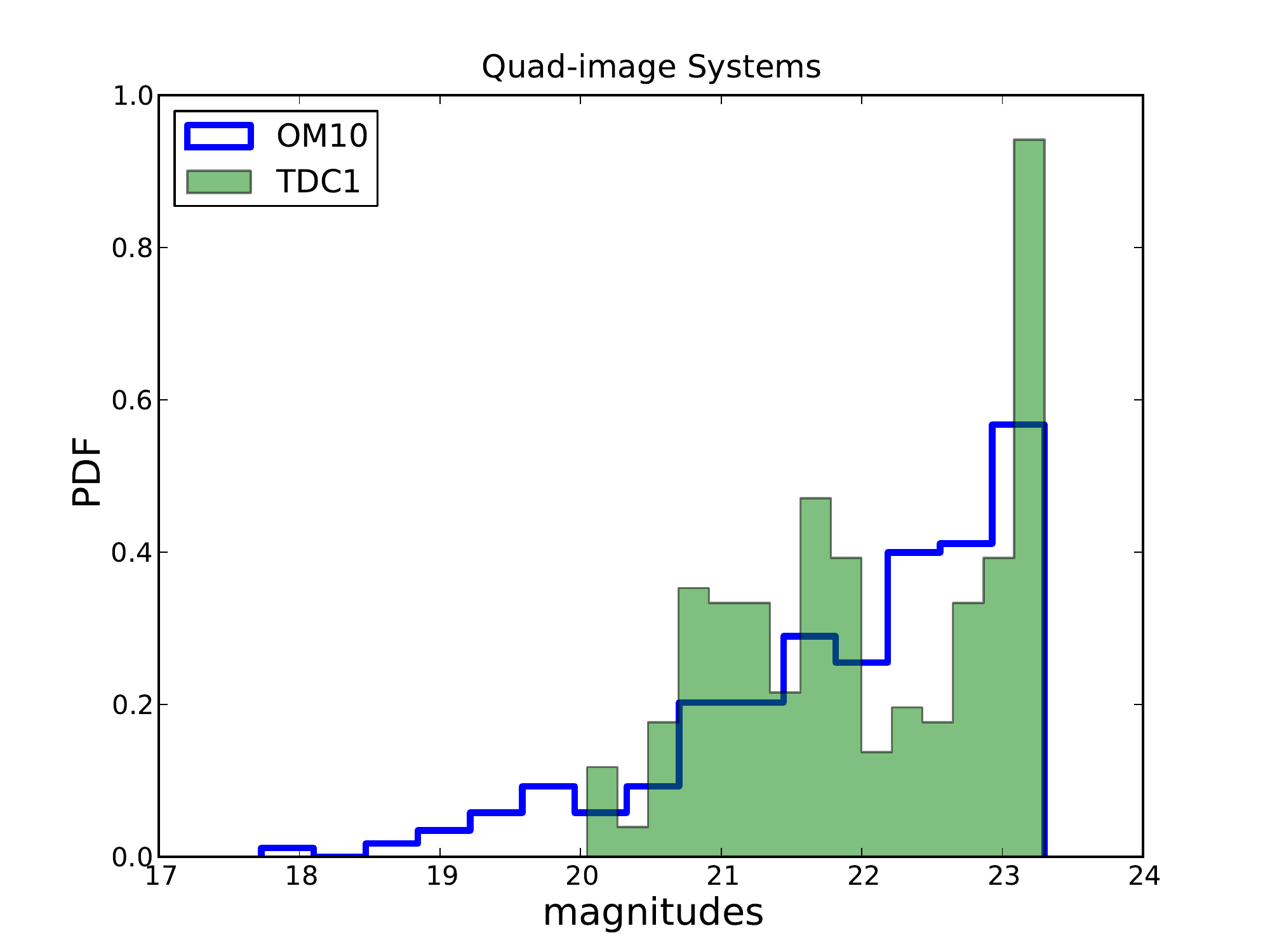}
\caption{Detection magnitude ``$\ithree$'' distributions for
the double (top) and quad (bottom) systems.  For doubles,
$\ithree$ is the magnitude of the fainter image, while for
quad systems it is the magnitude of the third-brightest image.
Distributions are shown both for the parent OM10 sample, and
the sample used for TDC1.}
\label{fig:i3}
\end{figure}

To give an overview of this sample, we show the distributions of
time delays $\dt$ between images in our 1024 image
pairs (in \Fref{fig:dt}), and detection magnitudes $\ithree$ in the 872
lens systems (in \Fref{fig:i3}). The $\ithree$ quantity is the $i$-band
magnitude of the third brightest image in a quad system or the
magnitude of the fainter image in a double-image system. (It is an
important parameter because it helped OM10 characterize the
detectability of lensed quasars: lenses are assumed to be measurable if
$\ithree$ is above the 10$\sigma$ limiting magnitude of a survey.) The
lens abundance rises fairly steeply with $\ithree$, so in order to probe
the relationship between it and the  time delay measurement accuracy, we
split the magnitude range 20-24 into four sub-ranges, and selected
approximately equal numbers of systems in each sub-range.

In summary, our sample is similar to OM10's, except that the brighter
lenses and intermediate time delays are somewhat over-represented. As we
will discuss later in this paper, this allows us to sample the range of
magnitudes more evenly, while introducing negligible bias in the
inferred performance of the methods.

\subsection{Generation of intrinsic light curves}

The mechanism for generating intrinsic light curves is described in
\paperone. In TDC1, we needed to simulate many more datasets; the most
time-consuming part was generating the damped random walk (DRW)
stochastic process with which we modeled the intrinsic AGN light curves. The
interval between discrete epochs had to be 0.01 days in order to enable
the counter-image light curve to be simulated with a time
delay precision sufficient to not affect the ensemble metrics. Each of these
intrinsic light curves took approximately 1-2 CPU hours to make, so for
efficiency we created just 500 intrinsic light curves, each of 10 years
length, and re-cycled them between several mock datasets, with different
starting epochs chosen relative to the season gaps, so that all the
release data could be considered to be independent.

The DRW light curves represent light curve fluctuations, and have zero
mean magnitude. They are determined by only two parameters: the
characteristic timescale $\tau$ and the characteristic amplitude of the
fluctuations $\sigma$. These were drawn from distributions designed to
match that observed for the spectroscopicly \textbf{confirmed} ($i < 19.1$ magnitude) quasars
in \citet{MacLeodEtal2010}. Their $\log\tau$ and
$\log\SFinf$ (asymptotic rms variability on long time scales) parameters
were drawn uniformly from the ranges $[1.5:3.0]$ and $[-1.1:-0.3]$
respectively. The endpoints of these ranges  correspond to 30 and 1000
days, and 0.08 and 0.5 magnitudes. The rms fluctuation level was derived
for each light curve via $\sigma = \SFinf / \sqrt{\tau}$.


\subsection{Modeling microlensing}

Microlensing is an important source of systematic error because it makes
the multiply-imaged light curves differ by more than the time delay and
the macrolens magnification ratio. In galaxy-scale lenses, the
variability of the microlensing typically has time scale significantly
larger than that of the quasar intrinsic variability (although
occasional caustic crossing events can provide some transient rapid
variability). We expect the most successful light curve measurement
algorithms to model an additional microlensing light curve component
individually at each image.

Given an OM10 catalog convergence $\kappa$, shear $\gamma$ and surface
density in stars $F_*$ at each image position, we generated a static
stellar field with a mean mass per star of $0.3\Msun$
\citep{Massfunction}. We then calculated its source plane magnification
map and convolved this with a Gaussian kernel to represent the extended
accretion disk of the source quasar; we drew source sizes $s$ (Gaussian
radii) uniformly from the range $[10^{14}$-$10^{16}]$ cm. When
calculating the microlensing light curves, we assumed Gaussian
distributions for the components of the relative velocity~$v$ between the
source and the stars in the lens, with standard deviation of 500
km$\cdot$s$^{-1}$ in each direction.\footnote{The microlensing code used
in this work, {\sc MULES} is freely available at
\url{https://github.com/gdobler/mules}.}
In the appendix we show how the scatter in microlensing variability
amplitude depends on $F_*$, $\kappa$, and source size. {\bf
Finally, we note that there are several characteristic timescales in
microlensing light curves, ranging from the crossing time of the mean
stellar mass Einstein Radius \citep{Paraficz2006} to the source
caustic crossing time, to the density of caustics in the network, and
those can give rise occasionally to quasi-periodic features.}


\subsection{Photometric and Systematic Errors}

Following \citet{TewesEtal2013a} we considered several sources of
observational error when generating the lightcurve fluxes. The main
source of statistical uncertainty is the sky brightness, which we assume
dominates the photometry. We used the approximate distribution of
5-sigma limiting point source magnitudes from one of the LSST project
operations simulator outputs (L.\ Jones, priv.\ comm.),  and converted
these to flux uncertainties. The mean and standard deviation of the
5-sigma $i$-band limiting flux was found to be 0.263 and 0.081 AB
nanomaggies\footnote{One ``AB maggy'' is the flux corresponding to an AB
magnitude of 0.0 \citep{maggies}. Thus, 0.263 nanomaggies is the flux
corresponding to an AB magnitude of 24.} respectively; to add
photometric noise to a lightcurve flux we  first drew an rms photometric
uncertainty from a Gaussian of mean 0.053 and width 0.016 nanomaggies
(dividing the above numbers by 5), and then drew a noise value from a
Gaussian of width equal to this rms. The minimum noise value was set to
be 0.001 nanomaggies.

Beyond this basic (though possibly epoch-dependent) Gaussian noise, we might expect
additional flux errors to be present as the  observing set-up changes
over a long monitoring campaign. To mimic such fluctuations, we added
the following three types of ``evilness'' to the light curves:
\begin{itemize}
\item Flux uncertainty under-estimation: for each pair of light curves
and for approximately 1 in every 10 epochs, we added noise that was 3
times larger than standard, but reported it as the normal one.
\item Calibration error: for each pair of light curves and for
approximately 1 every 10 epochs, we added correlated noise, i.e.\ both
points were higher or lower than in the normal case.
\item Episodic transparency loss: we took a subset of the data (a few
weeks every year), and offset the fluxes by 1\% or 3\%.
\end{itemize}

There could be more than one type of ``evilness'' present in any given
lightcurve: the  combinations applied to the TDC1 lightcurves were as
follows. 3\% of the light curves, selected randomly, were contaminated
with a single type of ``evilness.'' Another 1\% were contaminated
with two types, and 3\% were contaminated with all three. In total
then, 15\% of the light curves were contaminated with these simulated
bad observational conditions.


\subsection{Example TDC1 light curves}

\begin{figure*}[!htbp]
\begin{center}
\includegraphics[width=45mm]{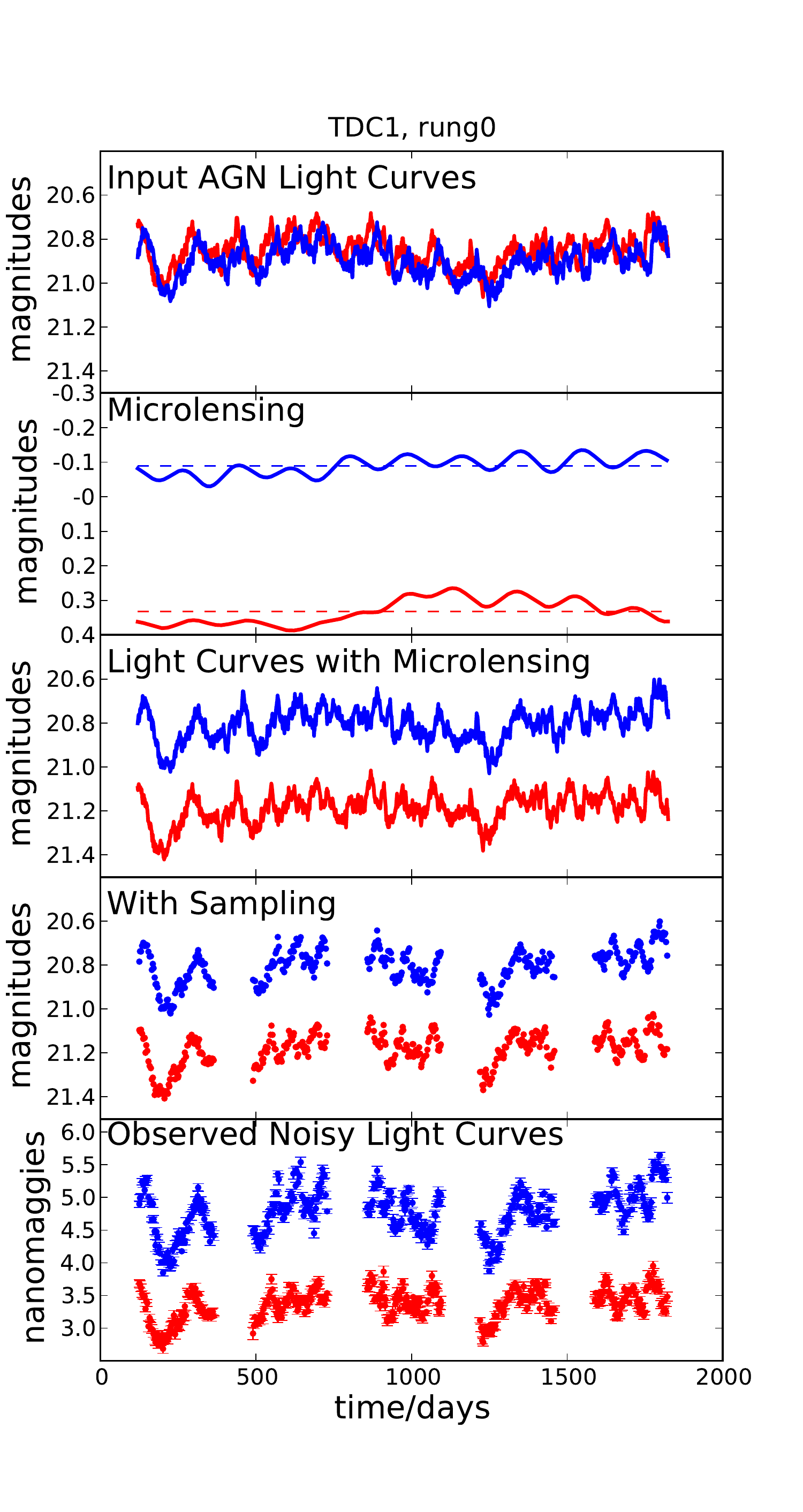}
\includegraphics[width=45mm]{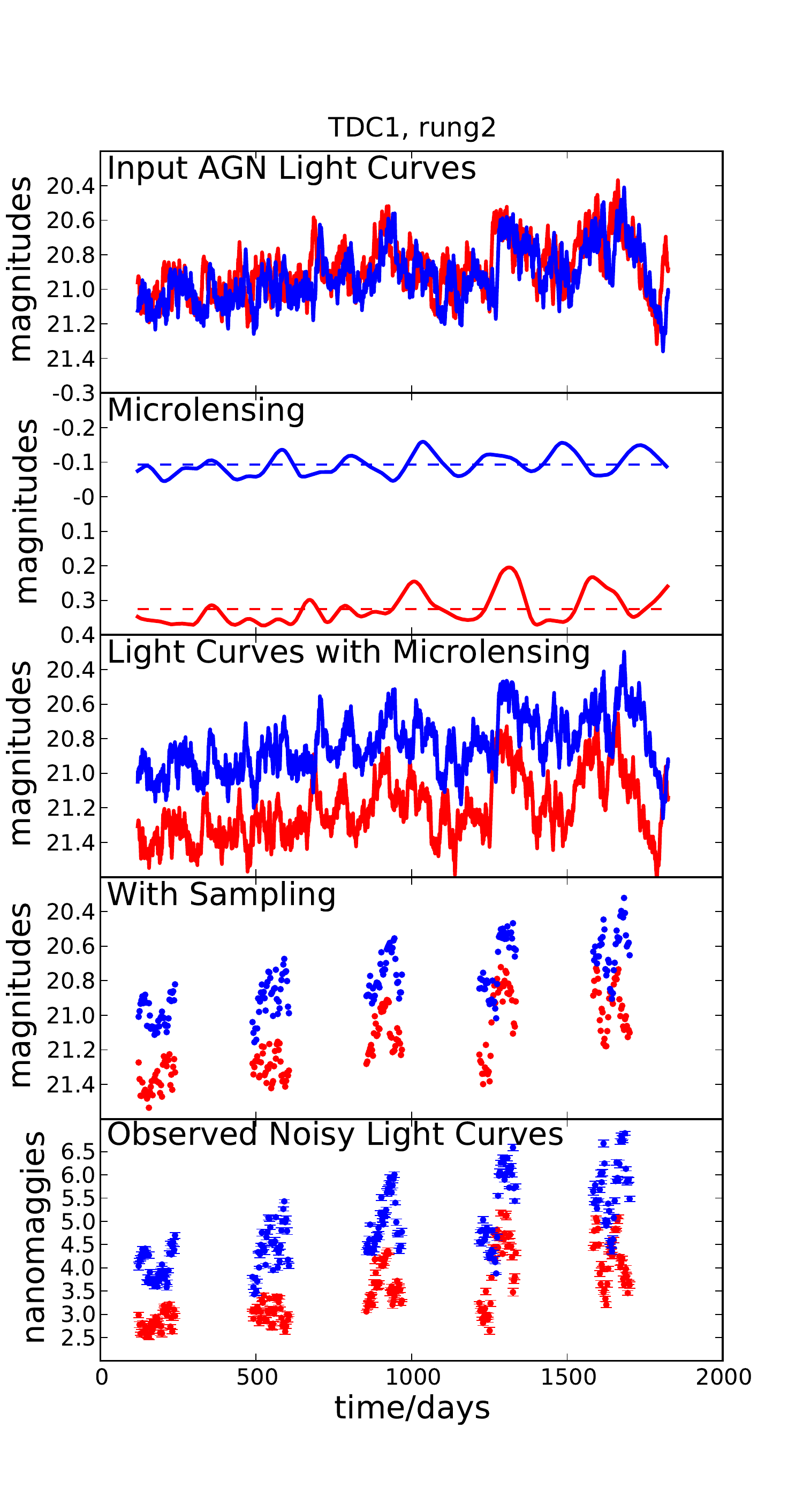}
\includegraphics[width=45mm]{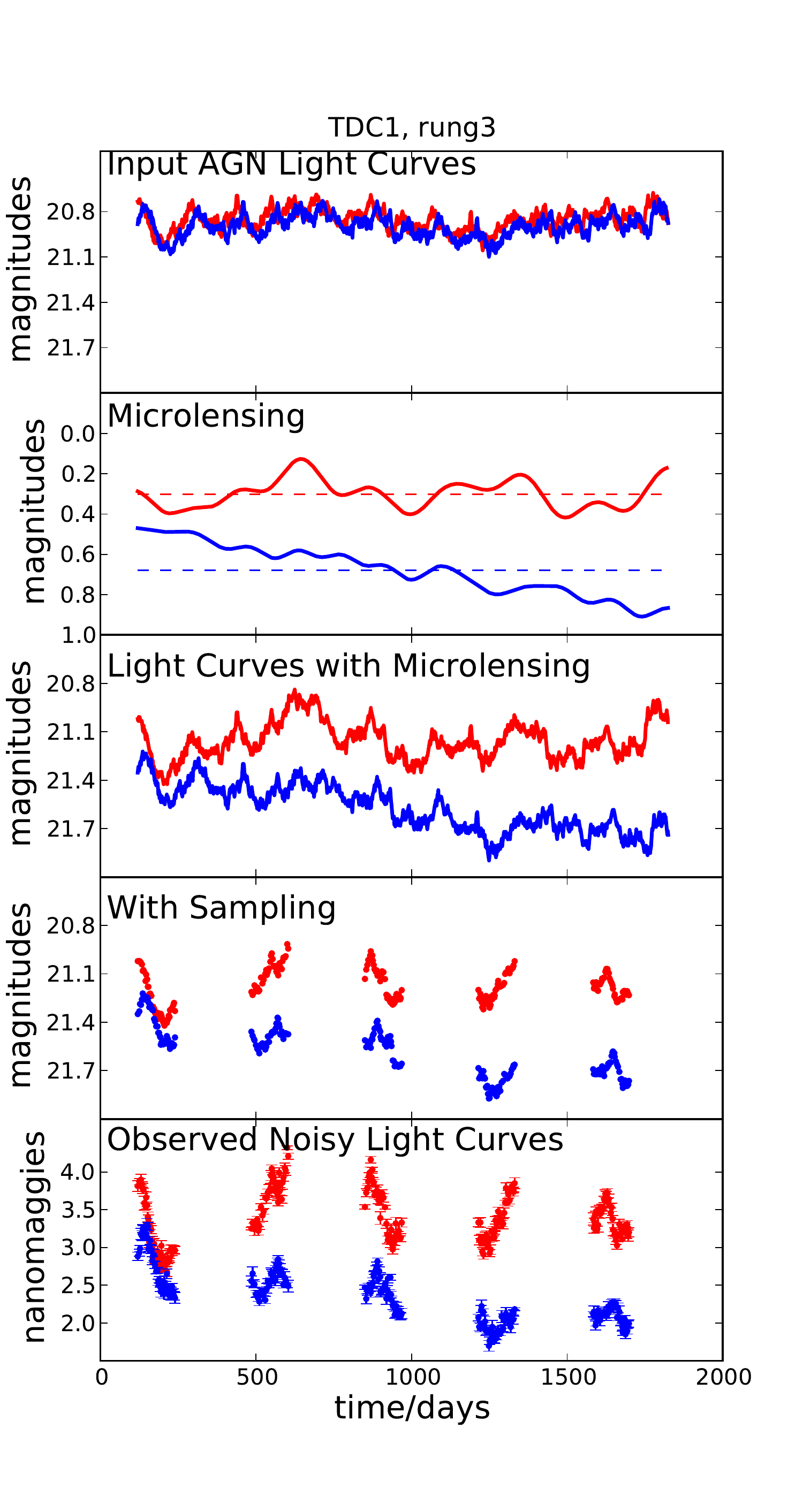}
\caption{Illustration of the process of generating time delay light
curves, with examples taken from the Rung 0 (left), Rung 2 (middle), and
Rung 3 (right) samples. The panels in each figure show, going from the
top to the bottom, (1) the input AGN light curves, (2) the microlensing
contributions in magnitudes, (3) the AGN light curves including the
microlensing contributions, (4) the result of down-sampling to the
required cadence and season length, and (5) the final sparsely sampled
noisy light curves. }
\label{fig:rung023}
\end{center}
\end{figure*}

\begin{figure*}[!htbp]
\begin{center}
\includegraphics[width=90mm]{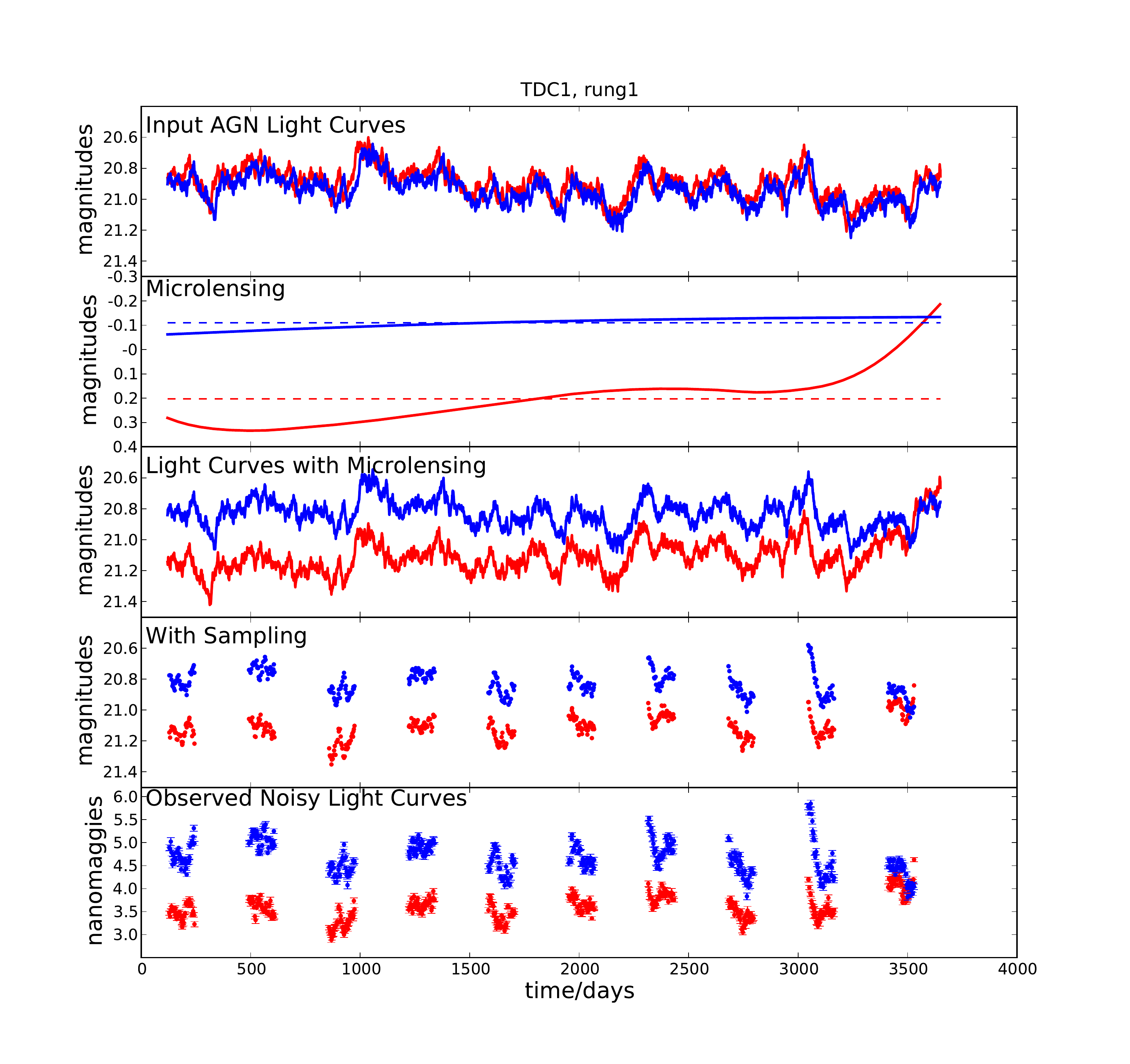}\hfill
\includegraphics[width=90mm]{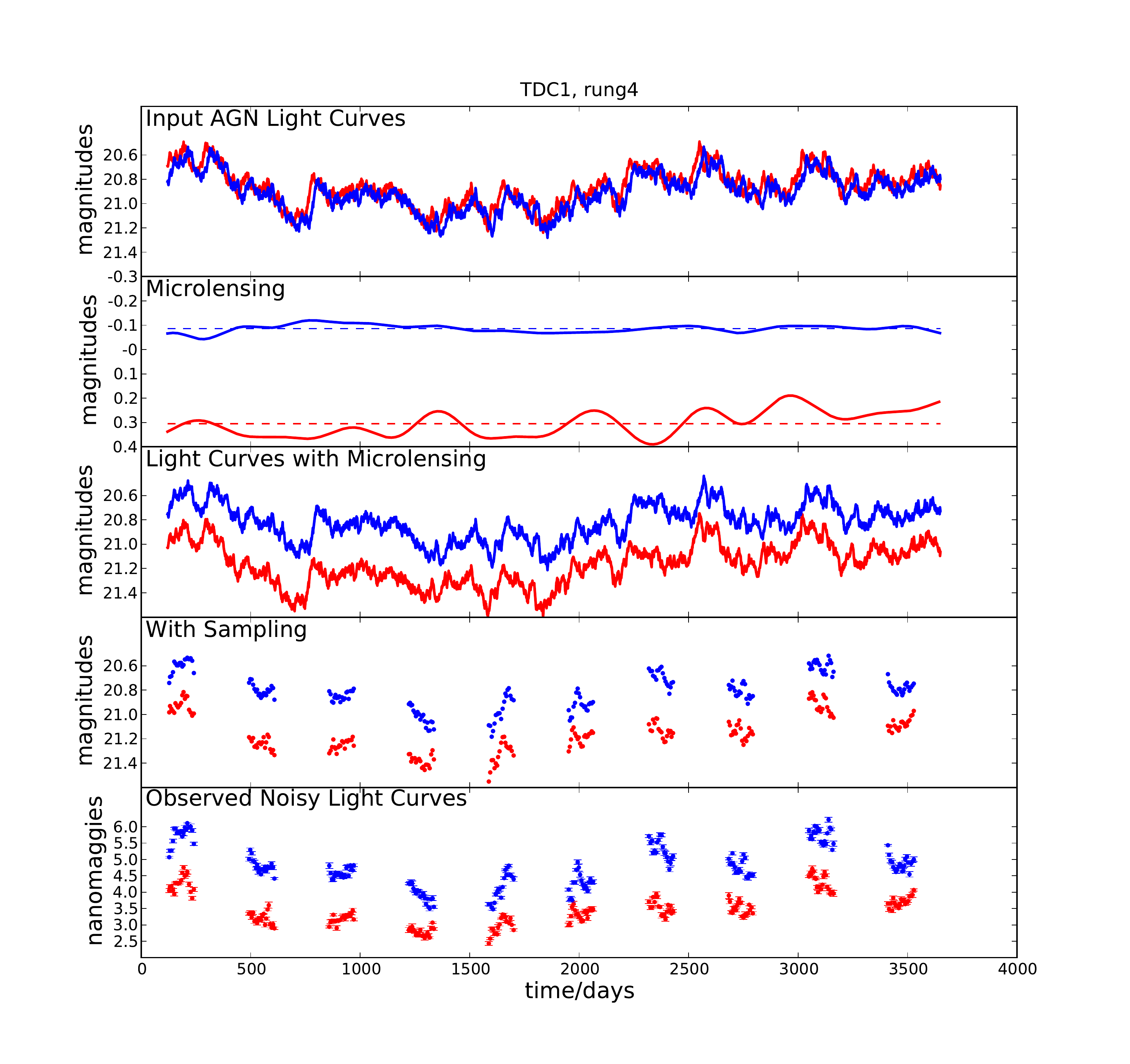}
\caption{Same as Figure~\ref{fig:rung023}, but for the longer
campaign-duration light curves of Rungs 1 and 4.}
\label{fig:rung14}
\end{center}
\end{figure*}

Figures \ref{fig:rung023} and \ref{fig:rung14} illustrate the process of
generating TDC1 data in each of the five rungs, using lightcurves
selected randomly from those datasets. The top panels show the AGN
intrinsic light curves in magnitudes. The panels beneath them show
the microlensing magnifications (also in magnitudes). The third panels
show the AGN light curves with microlensing effects, and the effect of
sampling is shown
in the fourth panels. Finally, the sparsely sampled noisy
mock lightcurves are shown on the bottom panels, in flux units.

Comparing panels 3 and 5, we can easily see how two similar curves
become difficult to associate by eye once the sparse sampling and the
addition of noise have been applied. Table~2 shows the values of the input parameters
$\tau$, $\sigma$, $v$, $s$, $F_*$, enabling some intuition to be
developed by comparing plots shown for the different rungs.

\begin{table*}
\begin{center}
\begin{tabular}{lllcclrr} \hline\hline
  Rung &  $\tau(day)$ &  $\sigma(mag/day$$^{-1/2}$) & $v(km/s)$   &  $s(10$$^{14}cm)$  &   $F_{*A}$   & $F_{*B}$  \\ \hline
  0    &  37.8        &    0.017                    &   731       &   3.87            &    0.037     &  0.062   \\
  1    &  83.0        &    0.017                    &   731       &   38.7            &    0.037     &  0.062   \\
  2    &  40.6        &    0.039                    &   1462      &   3.87            &    0.037     &  0.062   \\
  3    &  37.8        &    0.017                    &   731       &   3.87            &    0.019     &  0.031   \\
  4    &  178.0       &    0.017                    &   365       &   3.87            &    0.037     &  0.062   \\
\hline\hline
\end{tabular}
\caption{The parameters used to make the simulated data shown in
\Fref{fig:rung023} and \Fref{fig:rung14}, to enable study of their
effects on the light curves.}
\end{center}
\end{table*}


\section{Response to the challenge}
\label{sec:response}

As described in \Sref{sec:intro}, the Time Delay Challenge was
presented to the community as two ``ladders'', TDC0 and TDC1.  The
TDC0 data were used as a gateway to TDC1; in order to gain access to
the TDC1 data, each ``Good'' Team had to submit a set of time delays
inferred from TDC0 that met the targets described in \Sref{sec:intro},
and in more detail in \paperone. In total, 13 ``Good'' Teams
participated in TDC0, many of which submitted multiple sets of
solutions.  Seven teams passed TDC0 and, went on to participate in
TDC1.  One of the teams submitted results based on three different
algorithms: those were considered independent submissions. In
addition, the ``Evil'' Team did an in-house analysis of the TDC1 data,
using a relatively simple procedure, to serve as a baseline comparison
for the ``Good'' Team submissions.  All ten of these algorithms are
described below. It is worth noting that the teams continued to
develop their methods between TDC0 and TDC1 and beyond, and the
description given here is for the versions of the methods that were
applied to TDC1.


\subsection{Benchmark technique by Rumbaugh (``Evil'' Team)}

The baseline method used by the ``Evil'' Team was a $\chi^2$-based Markov
Chain Monte Carlo (MCMC) approach. While the member of the team that
wrote and executed this baseline method (NR) did not work directly on
simulating the light curves, this method should not be considered blind in the same
way as the ``Good'' Teams'.

In practice the method consists of comparing a shifted copy of one of
the light curves to the other light curve, and using a $\chi^2$
function to compute the posterior PDF for the time delay. Matching the
lightcurves requires some interpolation, which was carried out using a
boxcar kernel with a full width of ten days. This particular kernel was
chosen to save computational time; however, the choice of the kernel did
not have a significant effect on the accuracy or precision of the
method. In order to gain additional computational speed, the correlation
between temporally close data points introduced by the smoothing kernel
was neglected.  This approximation reduced the computation time by about
an order of magnitude, while providing only marginally worse accuracy.
The posterior was sampled using the {\it emcee}~\citep{emcee} software
package.  For each trial value of the time delay, only the overlapping
parts of the time-shifted lightcurves were used in the computation of
the change in $\chi^2$. To avoid calculations using small overlap
regions, a maximum time delay was imposed equal to 75\% of the shortest
season length of the dataset currently being analyzed. Time delay point
estimates were chosen to be the median of the output sample values, with
the uncertainties chosen to be half the width of the region containing
68.3\% of the chain surrounding the median.

Before applying the benchmark technique to TDC1 data, it was tested on
the TDC0 data, as well as on an additional set of simulated data
designed to be similar to TDC0. In this testing, the smoothing kernel
was varied, as well as several other aspects of the method as indicated
above (including whether or not the full covariance matrix was used).
The accuracy and precision of the inference were found to not depend
significantly on these choices.

Time delay estimates from three implementations of this method were
submitted, with the aim of producing answers of different degrees of
reliability. The three implementations were obtained by restricting the
submissions to those systems with estimated time delay uncertainty below
6, 10, and 20 days. The submissions resulting from these cuts are named
Gold, Silver, and Bronze, respectively.


\subsection{Gaussian Processes by Hojjati \& Linder}
\label{sec:response:HL}

This ``Good'' Team implemented Gaussian Process (GP) regression to estimate
the time delays \citep[see][for the basic approach]{Hojjati:2013jna}.
Gaussian Processes are widely used as a model-independent technique for
reconstructing an underlying function from noisy measurements. The GP is
specified by a mean function, and a covariance (kernel) function
characterized by a set of hyperparameters, describing the time delay,
relative magnitude shift, QSO variability and coherence length,
microlensing variability and coherence length, and measurement noise.
This approach is very flexible, not assuming a physical model for the
quasar or microlensing input, but allowing the data to decide how best
to describe the signal in terms of a GP.  The hyperparameters were
fitted to data using the GP likelihood through a Bayesian analysis. The
parallel and highly efficient fitting code employed two covariance
kernels, two optimization methods, and variation of priors to
cross-check the results for robustness. The team passed or rejected a
system, based on the consistency of fits and their likelihood weights,
and then assigned a final best fit, uncertainty, and confidence class to
the passed systems.

The overall philosophy emphasized complete automation and accuracy of
estimation, rather than precision (e.g., fitting down to five day
delays and placing no cut on precision) or numbers of fits. Within
this, the team fine-tuned samples based on their confidence in the
fit, and to a lesser extent the error estimation. Six samples were
submitted, with the basic three representing progressively more
inclusive fit confidence along the lines of, e.g., gold, silver,
bronze estimation. These correspond to the samples nicknamed
Lannister, Targaryen, and Baratheon, respectively. In addition, a more
conservative sample (nicknamed Tully) and one with tighter error
assignment (nicknamed Stark) were submitted. Catastrophic outliers
were identified by running selected samples (e.g., especially short or
long time delays) with controlled priors, and also an analysis of the
best-fit parameters for the selected systems. The sample nicknamed
``Freefolk'' was the result of such analysis.

A correction to the mean function treatment in the code significantly
increased the consistency of the fits. However, since this modification
was made after the TDC1 submission deadline, this is not reflected in
the results presented in this paper; see the updates and discussion by
\citet{HLinprep}. Furthermore, the method has benefited from, and was
improved after, a reanalysis of the fits and the investigation of the
hyperparameter behavior using the unblinded TDC1 data.


\subsection{FOT by Romero-Wolf \& Moustakas}

The Full of Time (FOT) team's Gaussian process (GP) inference
algorithm took a Bayesian approach to solve for the delay between a
pair of light curves. The probability of the light curve parameters
$\bar{M}$ (\textbf{mean magnitude}), $\sigma$ (\textbf{characteristic
amplitude of the fluctuations}), and $\tau$ (\textbf{characteristic
timescale}) given the data is proportional to the
product of the likelihood function for a CAR process
\citep{Kelly:2009,MacLeodEtal2010} and uniform priors. \textbf{Details about
the CAR
process can be also found in Paper I.} The {\it
emcee}~\citep{emcee} MCMC ensemble sampler provides an estimate of the
posterior probability distribution for the light curve parameters. To
reconstruct the delay, the pair of light curves were combined into a
single time series assuming a delay and magnitude offset. The
probability of the delay and magnitude offset, along with light curve
parameters, is given by the CAR process likelihood function of the
combined light curve and uniform priors. The light curve delay and its
uncertainty were then inferred from the marginalized posterior
distribution for the time delay given the light curves.  The algorithm
did not characterize or fit for microlensing, although it identifies
the datasets that are most likely to have microlensing variations. A
more thorough description of this method and internal tests are being
written up by Moustakas \& Romero-Wolf (2014, in preparation).

The procedure was tested by generating tens of thousands of ``blind''
time-delayed light curves through the CAR process, with varying
(irregular) observational patterns and campaigns, photometric
uncertainties, magnitude offsets, and time delays.  These were then
processed with the inference technique described above.  Both the
successful recovery rate and the precision of the (marginalized) time
delay and magnitude offset were then studied as a function of each
``observational'' parameter (i.e., the observational campaign factors
and the assumed photometric precision).

To avoid outliers, a set of consistency
requirements between the posterior distributions for the individual and
combined light curve parameters were required. A solution was rejected
if the mean of the posterior $\sigma$ distributions from each light
curve and their combinations differed by more than 2.6
root-sum-squared standard deviations. The means of the posterior
$\log_{10}\tau$ distributions for each light curve must also agree to
within one standard deviation, forcing a consistency in the physical
behavior of the reconstructed ``stitched'' data set compared to the
input data.  Additional quality cuts were included from inspection of
the reconstructed time delay and time delay uncertainty scatter
relation. These required that delays less than 100 days have
uncertainties smaller than 10 days.  The ratio of the delay
uncertainty to the delay was also required to be smaller than 2.


\subsection{Smoothing and Cross-Correlation by Aghamousa \& Shafieloo}

This ``Good'' Team combined various statistical methods of data analysis in
order to estimate the time delay between different light curves. At
different stages of their analysis they used iterative smoothing,
cross-correlation, simulations and error estimation, bias control and
significance testing to prepare their results. Given the limited
timeframe (they started the project in early May 2014), they had to make
some approximations in their error analysis.

In their approach to estimate the time delay between a pair of light
curves $A_1$ and $A_2$, they first smoothed over both light curves
using an iterative smoothing method
\citep{Shafieloo:2005nd,Shafieloo:2007cs,Shafieloo:2009hi,Shafieloo:2012yh},
producing the smoothed light curves $A^{\rm smooth}_1$ and $A^{\rm
smooth}_2$. During smoothing, they recorded the ranges with no data
available (which would have resulted in unreliable smoothing). The
algorithm was set to automatically detect such ranges. Then, they
calculated the cross-correlation between $A_1$ and $A^{\rm smooth}_2$
and also between $A_2$ and $A^{\rm smooth}_1$ for different time
delays, and found the maximum correlations. These two maximum
correlations should be for the same time delays (that is, the absolute
values of the time delays should be consistent with each other). The
difference between these two estimated time delays (with maximum
correlations) was part of the total uncertainty considered for each
pair (in the estimated time delay). To estimate the error on each
derived time delay, the team also simulated many realizations of the
data for each rung, and for various time delays. Knowing the fiducial
values, they derived the expected uncertainties in the estimated
values of the time delays.

This team also performed bias control, since long time delays have a
limited data overlap between the two light curves. In the case of the
quad sample, they used different combinations of the smoothed and raw
light curves to test the internal consistency of the results and
relative errors. These internal consistency relations can be used to
adjust the estimated error-bars for each pair (considering the
consistency of all light curves as a prior). The team selected for
cross-correlations between the two light curves with more than $50\%$ or
$60\%$ correlation coefficients. Pairs with potentially high bias were
cut as well. In this methodology the light curves are compared in multi-segments.
The effect of micro-lensing can be considered as a linear distortion
in these segments. While the correlation coefficient is unchanged
under linear transformation, there is no concern for micro-lensing in
this algorithm and the method is unaffected.
Additional details of this method will be described in a
separate paper \citet{Arman2014}.


\subsection{Supervised Pelt by Jackson}

All pairs of joint lightcurves were inspected by eye by this team, using
a Python tool developed for the purpose. An initial \citet{PeltEtal1994}
statistic was calculated for a large range of time delays, and its
minimum found, but this resulted in catastrophic errors in many cases
and was frequently over-ridden by visual inspection. Time delays were
regarded as believable if (1) at least three coincident points of
inflection were detected in the lightcurves, (2) if no discordant
features were seen (i.e., differences between the lightcurves which
could not be plausibly attributed to microlensing) and (3) if the plot
of the Pelt statistic against time delay showed a smooth and
well-defined minimum.

In the process of assessing the lightcurves by eye, the following
operations were available to find a time delay fitting the above
criteria: (1) smoothing of either lightcurve to match the scatter of the
other; (2) adjustment of the zero-point of each segment of the
lightcurve to match the zero-point of the segment of the other
lightcurve that it overlapped using the current time delay; (3) manual
adjustment of the current time delay; (4) deletion of one or more
segments of the lightcurve if they were judged to be severely affected
by microlensing. \textbf{In practice, this was the case if a simple rescaling
of a whole segment of data between the two lightcurves produced residuals
much larger than those of other rescaled segments. This will happen if the
microlensing produces a large change in flux over the period of one data
segment; the method therefore roughly corresponds to assuming that
microlensing produces variations on a timescale larger than those of
the intrinsic brightness variations of the quasar, and deleting regions
of data where this is not the case.} In most cases, the delay and its error bar were
calculated after this process using 100 instances of resampling of the
dataset using the observed flux errors and a small Gaussian error in
each time stamp.  This allowed the calculation of a set of delays, in
each case using the delay from the Pelt statistic minimum, from which
the mean and scatter was used for the delay and its error. In a few
cases, mostly those in which the Pelt statistic vs.\ time delay plot had
a local minimum around the optimum, an additional error, or in some
cases a minor adjustment to the value, was estimated by eye. The error
bar was also adjusted in cases where the optimization using smoothing
and adjustment of the zero point resulted in a significant reduction of
the error estimated by the resampling process. With practice, about 100
pairs of lightcurves per hour could be processed, so that thousands or
tens of thousands of lightcurve pairs could in principle be analyzed
using this method.

The same basic algorithm was used for all submissions, but different
submissions were made by separating the objects into three categories,
again by eye, according to confidence that the time delay was correct
within the stated error. Evaluations with less confidence corresponded
to violation of one or more of the believability conditions, and the
least certain category usually involved light-curves with only two
clearly detected points of inflection. (For each of the three
categories, subsidiary submissions were also made with a smaller number
of rungs). Three catastrophic errors in rung 0 of the original blind
submission were due to incorrect entry of a minus sign during the manual
adjustment process in three objects; these were corrected in a non-blind
submission which consisted of the original blind submission for all
rungs, and all three confidence levels with the three signs corrected.
The program was accordingly modified to question the user in the case of
large changes imposed by hand.


\subsection{PYCS by Bonvin, Tewes, Courbin \& Meylan}

The PyCS team made submissions using three time delay measurement
methods: {\tt d3cs}, {\tt spl}, and {\tt sdi}. The latter two build upon
initial estimations provided by the former. The following subsections
summarize each of these three methods.

\subsubsection{{\tt d3cs}: D3 curve shifting}

This first method is based on human inspection of the light curves, in
the spirit of citizen science projects. The PyCS team has developed a
dedicated browser-based visualization interface, using the D3.js
JavaScript library\footnote{Data-Driven Documents,
\url{http://www.d3js.org/}} by \citet{d3}. The tool is now publicly
available
online.\footnote{\url{http://www.astro.uni-bonn.de/~mtewes/d3cs/tdc1/}
-- see ``Read me first'' for help.}

The main motivation behind this time-consuming yet simple approach were
to obtain, for each light curve pair, (1) a rough initial estimate for
the time delay and its associated uncertainty, and (2) a robust
characterization of the confidence that this estimate is not a
catastrophic error.
The interface asks each user to pick a confidence category for the
proposed solution, among four choices:
\begin{enumerate}
\item ``doubtless'' if a catastrophic error can be virtually excluded,
\item ``plausible'' if the solution yields a good fit and no other
solutions are seen,
\item ``multimodal'' if the proposed solution is only one among two or
more possible solutions,
\item ``uninformative'' if the data does not reveal any delay.
\end{enumerate}

At least two human estimates were obtained for each pair of curves. The
database of {\tt d3cs} estimates was then carefully reduced to a single
estimate per pair, resolving any conflicts between estimates in a
conservative way. A key result of this step was a sample of 1628
``doubtless'' time-delay estimates, which the team hoped to be free from
any catastrophic outliers. Through this exercise, the team  demonstrated
that such an approach remains tractable for about 5000 light curves,
with typical human inspection times of a minute per light curve pair and
user.

\subsubsection{{\tt spl}: free-knot spline fit}

The {\tt spl} method is a simplified version of the ``free-knot spline
technique'' described by \citet{TewesEtal2013a} and implemented in the PyCS
software package. Using the {\tt d3cs} estimate as the starting point,
the method simultaneously fits a single spline representing the
intrinsic QSO variability, and a smoother ``extrinsic'' spline
representing the differential microlensing variability, to the light
curves. During this iterative process, the curves were shifted in time
so as to optimize the fit. The fit was repeated 20 times, starting from
different initial conditions, to test and improve the robustness of the
resulting delay against local minima of the $\chi^2$ hyper surface. Such
a model fit was then used to generate 40 simulated noisy light curves
with a range of true time delays around the best-fit solution. By
re-running the spline fit on these simulated curves, and comparing the
resulting delays with the true input time delays, the delay measurement
uncertainty was estimated.

The {\tt spl} method for TDC1 is simpler, faster, and significantly less
conservative in the uncertainty estimation than the free-knot spline
technique that was applied to the COSMOGRAIL
data\footnote{\url{http://www.cosmograil.org}} by \citet{TewesEtal2013b}
and \citet{Kumar++2013}. In particular, the temporal density of
spline knots was automatically determined from signal-to-noise ratios
measured on the two light curves, and only white noise was used in the
generative model.
With these simplifications, the team expects the resulting TDC1 error
estimates to be rather optimistic. The entire {\tt spl} analysis took
about 5 CPU-minutes for an average TDC1 pair.

\subsubsection{{\tt sdi}}

The third method, {\tt sdi} (for spline difference) was inspired by the
``regression difference technique'' of \citet{TewesEtal2013a}, replacing the
Gaussian process regressions by spline fits to speed up the analysis.
The method involves fitting a different spline to each of the two light
curves, and then minimizing the variability of the difference between
these two splines by shifting them in time with respect to each other.
The advantage of this approach is that it does not require an explicit
microlensing model. To estimate the uncertainty, this method uses the
simulated light curves provided by the {\tt spl} technique. As in the
{\tt spl} technique, the estimates from {\tt d3cs} were used as the
starting point to define the time delay intervals in which {\tt sdi}
optimizes its cost function.

\subsubsection{Identification of catastrophic failures}

To prevent catastrophic failures, this team relied solely on the {\tt
d3cs} ``doubtless'' sample. The {\tt spl} and {\tt sdi} methods do not
alter this confidence classification. Furthermore, a small number of
{\tt spl} and {\tt sdi} measurements that did not lie within 1.5$\sigma$
of the corresponding {\tt d3cs} estimates were rejected.

\subsubsection{Differences between submissions}

For all three methods, the submissions were named following the
scheme {\tt A-B-C-D.dt}, where:
\begin{description}
\item[A] gives the method, {\tt d3cs}, {\tt spl} or {\tt sdi}.
\item[B] gives the method parameters, with {\tt vanilla} denoting the
\emph{a priori} best or simplest.
\item[C] gives the confidence category, with {\tt dou} for doubtless and
{\tt doupla} for both doubtless and plausible light curve pairs. The
{\tt doupla} submissions are expected to be contaminated by some
catastrophic outliers, but feature more than twice the number of time
delays than the {\tt dou} sample.
\item[D] gives the filter that selects systems according to different
criteria across all rungs, mostly based on the blind relative precision
$\delta_i / |\widetilde{\Delta t_i}|$. The code {\tt full} corresponds
to no filter. {\tt XXXbestP} selects the XXX ``best'' systems in terms
of blind relative precision, {\tt P3percent} selects the largest number
of systems so that the average blind relative precision is
approximately 3\%, and {\tt 100largestabstd} is the selection of the 100
largest delays.
\end{description}

Submissions that share the same method and method parameters (A and B)
differ only in the selection of systems, and not in the numerical values
of the estimates. They can thus be seen as subsamples of the
A-B-dou/doupla-full submissions.

\begin{table}
\begin{center}
\begin{tabular}{lc} \hline\hline
Method 				&            Microlensing            \\
Rumbaugh  			&                    No              \\
Shafieloo 	 		&                    Yes             \\
PyCS-d3cs		  	&                    Yes             \\
PyCS-sdi		  	&                    Yes             \\
PyCS-spl		  	&                    Yes             \\
Jackson-manchester	  	&                    Yes             \\
Kumar  				&                    Yes             \\
JPL  				&                    No              \\
Hojjati  			&                    Yes             \\
DeltaTBayes  			&                    No              \\
\hline\hline
\end{tabular}
\caption{Summary of methods explicitly accounting for microlensing.}
\label{tab:microlensingsum}
\end{center}
\end{table}


\subsection{Difference-smoothing by Rathna Kumar, Stalin, \& Prabhu}

The difference-smoothing technique, originally introduced by
\citet{Kumar++2013}, is based on the principle of minimizing the
residuals of a high-pass filtered difference light curve between the
lensed quasar images. The method is a point estimator that determines
an optimal time delay between two given light curves, and an optimal
shift in flux to one of the light curves, besides allowing for smooth
extrinsic variability. To estimate the uncertainty of the measured time
delay in \citet{Kumar++2013}, this team made use of simulations
produced and adjusted according to \citet{TewesEtal2013a}. However,
for participation in the TDC, they made use of a modified version of the
difference-smoothing technique as presented by
\citet[][submitted]{2014arXiv1404.2920R}. In that paper, they describe an optimal
way to adjust the two free parameters in the technique according to the
peculiarities of the light curves under analysis and also introduce a
recipe for simulating light curves having true delays at discrete
intervals in a plausible range around the optimal time delay found.
These simulations were used to estimate the uncertainty of the measured
value of the time delay. Outliers were identified by noting when the team's
technique was found to return random time delays which were uncorrelated
with the true delays in their simulated light curves.

\textbf{The free parameters in the technique are decorrelation length and
smoothing time scale. For participation in the Time Delay Challenge, the
value of decorrelation length was set equal to the mean temporal
sampling of the light curves and the value of smoothing time scale was
set equal to the largest integer multiple of decorrelation length for
which the amplitude of residual extrinsic variability was less than the
3$\sigma$ level of noise for each of the light curves. In the absence of
significant extrinsic variability between the light curves, the value of
smoothing time scale was set equal to $\infty$.}

\subsection{$Delta t$-Bayes by Tak, Meng, van Dyk, Siemiginowska, Kashyap,
\& Mandel}

A fully Bayesian approach was developed by this team, based on the key assumption that
one of the unobserved underlying light curves is a shifted version of
the other. The horizontal shift is the time delay ($\Delta t$), and the
vertical shift is the magnitude offset ($c$). Both shifts are treated as
unknown parameters. Specifically, from the state-space modeling
perspective, it was observed that $\textbf{x(t)}\equiv\{x(t_1), x(t_2),
\ldots, x(t_n)\}$ and $\textbf{y(t)}$, transformed into the logarithm of
flux, around the irregularly sampled underlying light curves,
$\textbf{X(t)}$ and $\textbf{Y(t)}\equiv\textbf{X(t} -
\Delta t\textbf{)}+c$ each, with measurement errors in log scale. The
posterior distribution for $\Delta t$ is of primary interest.  Also, it
was assumed that the unobserved true process $\textbf{X(t)}$ follows an
Ornstein-Uhlenbeck process (also known as
CAR) as described by \citet{KellyEtal2009},  although a different parameterization was used for
more efficient model fitting. \citet{Harva2006} proposed a similar idea,
but they assumed a different model for the underlying process.

This Bayesian approach treats the unknown parameters as random variables
and this team uses specific prior distributions for the time delay and magnitude
offset: $p(\Delta t, c) \propto \delta_{\{\vert\Delta t\vert ~\in ~[0,~ (t_n
- t_1)]\}}$. A uniform prior on $c$ is a typical choice because this
$y$-shift is related to the mean of observed data or the underlying
process. The uniform prior on $\Delta t$ constrains its values to ensure
that the shifted light curves overlap in time. This naively-informative
hyperprior distribution on the parameters governing the underlying
process is $p(\bar{M}, \sigma, \tau)\propto\tau^{-2}e^{-1/\tau}$,
where $\bar{M}$, $\sigma$, $\tau$ are CAR parameters as defined above and in Paper I.
This puts a uniform prior on $\bar{M}$ and $\sigma$, and an inverse-$\Gamma(1, 1)$
prior on $\tau$.

The full posterior distribution was obtained by multiplying together (1)
the likelihood for the state-space representation, (2) the prior for the
underlying process, $\Delta t$, and $c$, and (3) the hyperpriors for $\bar{M},
\sigma$, and $\tau$. The team proposed a Gibbs sampler for this full
posterior distribution (algorithm 2) and its approximation (algorithm 1)
in TDC1.  Details of the two samplers were submitted to the ``Evil'' Team
and will appear in a separate paper, in preparation. In order to obtain
the time delay from its posterior distribution, three Markov chains were
combined with starting values chosen randomly around the most likely
values. Rigorous convergence checks of the Markov chains were conducted
using trace plots, autocorrelation plots, and the Gelman-Rubin
diagnostic statistic, applied to all of the model parameters.

\textbf{The model did not account for the microlensing. However, when it was
suspected it after a visual inspection, this team accounted for its polynomial long-term
effect (linear or quadratic) by the regression and ran the model on the residuals.
This worked well because the intrinsic variability of quasar data did not disappear
even after the long-term trend was removed.}



\section{Analysis of the submissions}
\label{sec:analysis}

\begin{table*}
\begin{center}
\begin{tabular}{lcccccccc} \hline\hline
Method &  Rung & $f$ & $\chi^2$ & $P$ & $A$ &  $\chi^2_{median}$ & $P_{median}$ & $A_{median}$\\
0    &                       0 & 0.36 &   195000$\pm$76000   &   0.078$\pm$0.004 &  -0.181$\pm$0.065   & $0.085_{0.078}^{189}$ & $0.055_{0.036}^{0.083}$  &$-0.004_{0.86}^{0.025}$\\
0    &                         1 & 0.36 &   390000$\pm$150000   &   0.08$\pm$0.005 &  -0.281$\pm$0.061   & $0.47_{0.46}^{2046}$ & $0.052_{0.039}^{0.088}$  &$-0.021_{0.98}^{0.04}$\\
0    &                        2 & 0.32 &   3996$\pm$1052   &   0.082$\pm$0.005 &  -0.28$\pm$0.042   & $0.42_{0.4}^{1199}$ & $0.059_{0.041}^{0.088}$  &$-0.02_{0.97}^{0.05}$\\
0    &                        3 & 0.33 &   920000$\pm$500000  &   0.08$\pm$0.005 &  -0.247$\pm$0.053   & $0.37_{0.36}^{2527}$ & $0.05_{0.036}^{0.098}$  &$-0.013_{0.97}^{0.034}$\\
0    &                         4 & 0.35 &   950000$\pm$240000   &   0.042$\pm$0.004 &  -0.712$\pm$0.03   & $16136_{16136}^{671657}$ & $0.008_{0.007}^{0.087}$  &$-1.0_{0.007}^{0.99}$\\
1    &                       0 & 0.53 &   0.579$\pm$0.047   &   0.038$\pm$0.001 &  -0.018$\pm$0.001   & $0.26_{0.22}^{0.77}$ & $0.034_{0.016}^{0.028}$  &$-0.015_{0.024}^{0.016}$\\
1    &                         1 & 0.37 &   0.543$\pm$0.049   &   0.045$\pm$0.001 &  -0.022$\pm$0.001   & $0.24_{0.22}^{0.69}$ & $0.04_{0.015}^{0.025}$  &$-0.02_{0.022}^{0.017}$\\
1    &                        2 & 0.35 &   0.89$\pm$0.19   &   0.053$\pm$0.001 &  -0.025$\pm$0.002   & $0.23_{0.21}^{0.92}$ & $0.047_{0.021}^{0.034}$  &$-0.02_{0.038}^{0.024}$\\
1    &                        3 & 0.34 &   0.524$\pm$0.077   &   0.059$\pm$0.002 &  -0.021$\pm$0.002   & $0.17_{0.15}^{0.67}$ & $0.051_{0.02}^{0.037}$  &$-0.018_{0.029}^{0.025}$\\
1    &                         4 & 0.35 &   0.608$\pm$0.072   &   0.056$\pm$0.002 &  -0.024$\pm$0.002   & $0.2_{0.18}^{0.84}$ & $0.051_{0.024}^{0.036}$  &$-0.019_{0.035}^{0.024}$\\
2    &                       0 & 0.53 &   0.125$\pm$0.011   &   0.205$\pm$0.007 &  -0.017$\pm$0.004   & $0.043_{0.039}^{0.178}$ & $0.151_{0.078}^{0.198}$  &$-0.008_{0.062}^{0.046}$\\
2    &                         1 & 0.27 &   0.138$\pm$0.016   &   0.233$\pm$0.01 &  -0.025$\pm$0.006   & $0.054_{0.05}^{0.216}$ & $0.19_{0.1}^{0.17}$  &$-0.008_{0.086}^{0.05}$\\
2    &                        2 & 0.21 &   0.043$\pm$0.004   &   0.242$\pm$0.01 &  -0.015$\pm$0.004   & $0.021_{0.019}^{0.058}$ & $0.201_{0.092}^{0.207}$  &$-0.009_{0.056}^{0.04}$\\
2    &                        3 & 0.3 &   0.099$\pm$0.013   &   0.247$\pm$0.011 &  -0.03$\pm$0.006   & $0.039_{0.035}^{0.121}$ & $0.17_{0.085}^{0.266}$  &$-0.013_{0.08}^{0.046}$\\
2    &                         4 & 0.21 &   0.178$\pm$0.018   &   0.363$\pm$0.015 &  -0.059$\pm$0.011   & $0.097_{0.084}^{0.252}$ & $0.32_{0.15}^{0.27}$  &$-0.04_{0.15}^{0.12}$\\
3    &                       0 & 0.53 &   1.068$\pm$0.069   &   0.043$\pm$0.003 &  -0.0$\pm$0.003   & $0.46_{0.4}^{1.67}$ & $0.022_{0.012}^{0.041}$  &$0.0_{0.025}^{0.025}$\\
3    &                         1 & 0.26 &   1.031$\pm$0.097   &   0.04$\pm$0.003 &  0.008$\pm$0.003   & $0.49_{0.46}^{1.47}$ & $0.027_{0.014}^{0.034}$  &$0.004_{0.026}^{0.033}$\\
3    &                        2 & 0.21 &   1.02$\pm$0.13   &   0.043$\pm$0.004 &  -0.002$\pm$0.004   & $0.38_{0.34}^{1.43}$ & $0.026_{0.013}^{0.037}$  &$0.003_{0.033}^{0.02}$\\
3    &                        3 & 0.3 &   0.813$\pm$0.074   &   0.068$\pm$0.006 &  -0.004$\pm$0.006   & $0.39_{0.37}^{1.04}$ & $0.034_{0.019}^{0.066}$  &$-0.002_{0.032}^{0.032}$\\
3    &                         4 & 0.21 &   1.07$\pm$0.23   &   0.098$\pm$0.014 &  0.0$\pm$0.008   & $0.24_{0.22}^{1.41}$ & $0.064_{0.034}^{0.06}$  &$0.003_{0.04}^{0.054}$\\
4    &                       0 & 0.53 &   0.497$\pm$0.047   &   0.033$\pm$0.002 &  -0.0$\pm$0.001   & $0.15_{0.14}^{0.75}$ & $0.018_{0.011}^{0.038}$  &$0.0_{0.012}^{0.012}$\\
4    &                         1 & 0.27 &   0.528$\pm$0.066   &   0.028$\pm$0.002 &  0.0$\pm$0.002   & $0.16_{0.15}^{0.78}$ & $0.02_{0.01}^{0.021}$  &$-0.001_{0.012}^{0.015}$\\
4    &                        2 & 0.21 &   0.464$\pm$0.069   &   0.028$\pm$0.002 &  -0.001$\pm$0.002   & $0.15_{0.13}^{0.54}$ & $0.02_{0.011}^{0.023}$  &$0.0_{0.01}^{0.013}$\\
4    &                        3 & 0.3 &   0.542$\pm$0.074   &   0.042$\pm$0.003 &  -0.003$\pm$0.003   & $0.16_{0.14}^{0.76}$ & $0.023_{0.013}^{0.038}$  &$-0.001_{0.015}^{0.017}$\\
4    &                         4 & 0.21 &   0.665$\pm$0.065   &   0.045$\pm$0.003 &  0.001$\pm$0.003   & $0.31_{0.29}^{0.94}$ & $0.032_{0.015}^{0.035}$  &$-0.001_{0.028}^{0.035}$\\
5    &                       0 & 0.68 &   0.91$\pm$0.092   &   0.032$\pm$0.001 &  0.003$\pm$0.002   & $0.24_{0.23}^{1.19}$ & $0.024_{0.014}^{0.034}$  &$0.001_{0.015}^{0.022}$\\
5    &                         1 & 0.27 &   1.76$\pm$0.42   &   0.037$\pm$0.002 &  -0.002$\pm$0.003   & $0.39_{0.36}^{1.86}$ & $0.03_{0.015}^{0.029}$  &$-0.001_{0.026}^{0.026}$\\
5    &                        2 & 0.32 &   1.57$\pm$0.21   &   0.043$\pm$0.001 &  -0.003$\pm$0.004   & $0.44_{0.41}^{1.93}$ & $0.036_{0.017}^{0.036}$  &$-0.001_{0.043}^{0.036}$\\
5    &                        3 & 0.35 &   1.89$\pm$0.31   &   0.036$\pm$0.001 &  0.002$\pm$0.003   & $0.42_{0.4}^{2.3}$ & $0.029_{0.015}^{0.03}$  &$0.001_{0.031}^{0.029}$\\
5    &                         4 & 0.18 &   7.2$\pm$2.7   &   0.05$\pm$0.003 &  -0.021$\pm$0.007   & $1.5_{1.4}^{4.3}$ & $0.043_{0.021}^{0.04}$  &$-0.016_{0.068}^{0.072}$\\
6    &                       0 & 0.04 &   0.32$\pm$0.071   &   0.077$\pm$0.017 &  0.005$\pm$0.011   & $0.11_{0.1}^{0.66}$ & $0.044_{0.027}^{0.06}$  &$0.0_{0.037}^{0.027}$\\
6    &                         1 & 0.02 &   66$\pm$64   &   0.175$\pm$0.055 &  2.3$\pm$2.2   & $0.36_{0.27}^{0.25}$ & $0.093_{0.037}^{0.13}$  &$0.042_{0.047}^{0.056}$\\
6    &                        2 & 0.03 &   0.71$\pm$0.21   &   0.142$\pm$0.021 &  0.027$\pm$0.032   & $0.37_{0.36}^{0.78}$ & $0.117_{0.064}^{0.098}$  &$0.029_{0.089}^{0.077}$\\
6    &                        3 & 0.02 &   1.7$\pm$1.2   &   0.168$\pm$0.031 &  0.14$\pm$0.1   & $0.33_{0.3}^{0.9}$ & $0.118_{0.068}^{0.079}$  &$0.02_{0.056}^{0.119}$\\
6    &                         4 & 0.01 &   0.19$\pm$0.1   &   0.55$\pm$0.12 &  0.169$\pm$0.058   & $0.066_{0.051}^{0.169}$ & $0.48_{0.25}^{0.2}$  &$0.16_{0.19}^{0.24}$\\
7    &                       0 & 0.33 &   65$\pm$51   &   0.04$\pm$0.003 &  -0.011$\pm$0.009   & $0.6_{0.55}^{3.44}$ & $0.021_{0.015}^{0.057}$  &$-0.0_{0.034}^{0.029}$\\
7    &                         1 & 0.24 &   2.71$\pm$0.5   &   0.036$\pm$0.003 &  0.002$\pm$0.006   & $0.67_{0.62}^{3.14}$ & $0.021_{0.015}^{0.045}$  &$0.001_{0.029}^{0.034}$\\
7    &                        2 & 0.37 &   3.21$\pm$0.55   &   0.04$\pm$0.003 &  0.008$\pm$0.006   & $0.74_{0.69}^{3.39}$ & $0.023_{0.015}^{0.051}$  &$-0.0_{0.029}^{0.036}$\\
7    &                        3 & 0.3 &   2.39$\pm$0.39   &   0.051$\pm$0.004 &  0.02$\pm$0.012   & $0.65_{0.6}^{2.49}$ & $0.025_{0.018}^{0.067}$  &$-0.0_{0.035}^{0.035}$\\
7    &                         4 & 0.22 &   185$\pm$119   &   0.062$\pm$0.005 &  -0.03$\pm$0.02   & $0.6_{0.53}^{3.47}$ & $0.035_{0.026}^{0.104}$  &$-0.001_{0.064}^{0.061}$\\
8    &                       0 & 0.44 &   109$\pm$58   &   0.047$\pm$0.004 &  -0.025$\pm$0.032   & $0.16_{0.15}^{1.21}$ & $0.025_{0.016}^{0.047}$  &$0.0_{0.021}^{0.019}$\\
8    &                         1 & 0.22 &   88$\pm$38   &   0.101$\pm$0.05 &  -0.02$\pm$0.019   & $0.17_{0.16}^{2.4}$ & $0.029_{0.016}^{0.066}$  &$0.0_{0.04}^{0.026}$\\
8    &                        2 & 0.18 &   91$\pm$72   &   0.07$\pm$0.006 &  -0.006$\pm$0.019   & $0.14_{0.14}^{0.81}$ & $0.046_{0.028}^{0.076}$  &$0.0_{0.032}^{0.032}$\\
8    &                        3 & 0.19 &   27$\pm$21   &   0.059$\pm$0.004 &  -0.008$\pm$0.013   & $0.24_{0.23}^{1.34}$ & $0.041_{0.025}^{0.064}$  &$0.001_{0.041}^{0.033}$\\
8    &                         4 & 0.16 &   2.6$\pm$1.1   &   0.068$\pm$0.004 &  -0.0$\pm$0.006   & $0.3_{0.28}^{1.29}$ & $0.055_{0.032}^{0.07}$  &$0.001_{0.049}^{0.045}$\\
9    &                         4 & 0.27 &   8.7$\pm$3.5   &   0.035$\pm$0.002 &  0.003$\pm$0.006   & $0.55_{0.47}^{2.79}$ & $0.024_{0.014}^{0.037}$  &$0.0_{0.042}^{0.031}$\\
\hline\hline
\end{tabular}
\caption{Mean and median statistics for the ``representative'' submissions.
Method 0:Rumbaugh-Gold, 1:Shafieloo-Arman7, 2:PyCS-d3cs-vanilla-dou-full,
3:PyCS-sdi-vanilla-dou-full, 4:PyCS-spl-vanilla-dou-full,
5:Jackson-manchester2\_0\_3\_4, 6:Kumar, 7:JPL, 8:Hojjati-Stark,
9:DeltaTBayes-DeltaTBayes1. }
\label{tab:unfiltered_table}
\end{center}
\end{table*}

\begin{table*}
\begin{center}
\begin{tabular}{lccccccccccc} \hline\hline
Method &  Rung & $f_{3.3\sigma}$ & $\chi^2_{3.3\sigma}$ & $P_{3.3\sigma}$ & $A_{3.3\sigma}$ & $X$ & $f_{A}$ & $\chi^2_{A}$ & $P_{A}$ & $A_{A}$ & $X_{A}$\\
0    &                        0 & 0.29 &   0.379$\pm$0.072   &   0.087$\pm$0.005 &  -0.003$\pm$0.004  & 0.8 & 0.28 & 0.299$\pm$0.056 & 0.08$\pm$0.004  &-0.0$\pm$0.002 &0.77\\
0    &                        1 & 0.23 &   0.577$\pm$0.095   &   0.096$\pm$0.006 &  -0.01$\pm$0.007  & 0.65 & 0.22 & 3.9$\pm$2.3 & 0.082$\pm$0.005  &-0.004$\pm$0.002 &0.62\\
0    &                        2 & 0.23 &   0.8$\pm$0.11   &   0.098$\pm$0.005 &  -0.007$\pm$0.006  & 0.73 & 0.21 & 0.74$\pm$0.23 & 0.09$\pm$0.005  &-0.002$\pm$0.003 &0.66\\
0    &                        3 & 0.22 &   0.59$\pm$0.1   &   0.097$\pm$0.006 &  0.0$\pm$0.007  & 0.66 & 0.21 & 1.26$\pm$0.4 & 0.087$\pm$0.006  &-0.002$\pm$0.002 &0.64\\
0    &                        4 & 0.11 &   0.37$\pm$0.11   &   0.119$\pm$0.009 &  -0.009$\pm$0.006  & 0.3 & 0.1 & 0.26$\pm$0.058 & 0.112$\pm$0.008  &-0.003$\pm$0.004 &0.28\\
1    &                        0 & 0.53 &   0.552$\pm$0.04   &   0.038$\pm$0.001 &  -0.017$\pm$0.001  & 1.0 & 0.52 & 0.53$\pm$0.038 & 0.038$\pm$0.001  &-0.017$\pm$0.001 &0.99\\
1    &                        1 & 0.37 &   0.543$\pm$0.049   &   0.045$\pm$0.001 &  -0.022$\pm$0.001  & 1.0 & 0.36 & 0.497$\pm$0.041 & 0.044$\pm$0.001  &-0.021$\pm$0.001 &0.99\\
1    &                        2 & 0.35 &   0.673$\pm$0.068   &   0.053$\pm$0.001 &  -0.025$\pm$0.002  & 0.99 & 0.33 & 0.73$\pm$0.19 & 0.052$\pm$0.001  &-0.021$\pm$0.002 &0.95\\
1    &                        3 & 0.34 &   0.458$\pm$0.039   &   0.059$\pm$0.002 &  -0.02$\pm$0.002  & 1.0 & 0.33 & 0.419$\pm$0.036 & 0.058$\pm$0.002  &-0.018$\pm$0.002 &0.97\\
1    &                        4 & 0.35 &   0.559$\pm$0.052   &   0.056$\pm$0.002 &  -0.024$\pm$0.002  & 1.0 & 0.33 & 0.535$\pm$0.069 & 0.055$\pm$0.002  &-0.021$\pm$0.002 &0.97\\
2    &                        0 & 0.53 &   0.125$\pm$0.011   &   0.205$\pm$0.007 &  -0.017$\pm$0.004  & 1.0 & 0.45 & 0.081$\pm$0.008 & 0.17$\pm$0.006  &-0.005$\pm$0.002 &0.83\\
2    &                        1 & 0.27 &   0.138$\pm$0.016   &   0.233$\pm$0.01 &  -0.025$\pm$0.006  & 1.0 & 0.21 & 0.078$\pm$0.01 & 0.191$\pm$0.008  &-0.006$\pm$0.003 &0.79\\
2    &                        2 & 0.21 &   0.043$\pm$0.004   &   0.242$\pm$0.01 &  -0.015$\pm$0.004  & 1.0 & 0.19 & 0.033$\pm$0.004 & 0.217$\pm$0.009  &-0.007$\pm$0.003 &0.9\\
2    &                        3 & 0.3 &   0.099$\pm$0.013   &   0.247$\pm$0.011 &  -0.03$\pm$0.006  & 1.0 & 0.25 & 0.056$\pm$0.005 & 0.201$\pm$0.01  &-0.007$\pm$0.003 &0.83\\
2    &                        4 & 0.21 &   0.178$\pm$0.018   &   0.363$\pm$0.015 &  -0.059$\pm$0.011  & 1.0 & 0.12 & 0.063$\pm$0.008 & 0.287$\pm$0.018  &-0.007$\pm$0.005 &0.55\\
3    &                        0 & 0.53 &   1.048$\pm$0.066   &   0.043$\pm$0.003 &  -0.0$\pm$0.003  & 1.0 & 0.5 & 0.956$\pm$0.068 & 0.037$\pm$0.003  &0.001$\pm$0.001 &0.94\\
3    &                        1 & 0.26 &   0.977$\pm$0.081   &   0.04$\pm$0.003 &  0.006$\pm$0.003  & 1.0 & 0.25 & 0.858$\pm$0.069 & 0.037$\pm$0.003  &0.004$\pm$0.002 &0.95\\
3    &                        2 & 0.21 &   0.94$\pm$0.1   &   0.043$\pm$0.004 &  -0.002$\pm$0.004  & 0.99 & 0.2 & 0.92$\pm$0.13 & 0.035$\pm$0.002  &-0.003$\pm$0.002 &0.93\\
3    &                        3 & 0.3 &   0.813$\pm$0.074   &   0.068$\pm$0.006 &  -0.004$\pm$0.006  & 1.0 & 0.27 & 0.747$\pm$0.073 & 0.05$\pm$0.004  &-0.003$\pm$0.002 &0.92\\
3    &                        4 & 0.21 &   0.804$\pm$0.096   &   0.098$\pm$0.015 &  0.005$\pm$0.006  & 0.99 & 0.19 & 0.64$\pm$0.11 & 0.069$\pm$0.004  &0.005$\pm$0.003 &0.86\\
4    &                        0 & 0.53 &   0.472$\pm$0.04   &   0.033$\pm$0.002 &  -0.0$\pm$0.001  & 1.0 & 0.52 & 0.483$\pm$0.048 & 0.029$\pm$0.001  &0.0$\pm$0.001 &0.98\\
4    &                        1 & 0.27 &   0.528$\pm$0.066   &   0.028$\pm$0.002 &  0.0$\pm$0.002  & 1.0 & 0.27 & 0.467$\pm$0.051 & 0.027$\pm$0.002  &-0.0$\pm$0.001 &0.99\\
4    &                        2 & 0.21 &   0.464$\pm$0.069   &   0.028$\pm$0.002 &  -0.001$\pm$0.002  & 1.0 & 0.21 & 0.431$\pm$0.064 & 0.028$\pm$0.002  &-0.001$\pm$0.001 &0.99\\
4    &                        3 & 0.3 &   0.494$\pm$0.057   &   0.042$\pm$0.003 &  -0.001$\pm$0.003  & 1.0 & 0.29 & 0.455$\pm$0.052 & 0.037$\pm$0.003  &-0.001$\pm$0.001 &0.97\\
4    &                        4 & 0.21 &   0.665$\pm$0.065   &   0.045$\pm$0.003 &  0.001$\pm$0.003  & 1.0 & 0.2 & 0.571$\pm$0.056 & 0.041$\pm$0.002  &0.0$\pm$0.002 &0.95\\
5    &                        0 & 0.68 &   0.741$\pm$0.053   &   0.032$\pm$0.001 &  0.004$\pm$0.002  & 0.99 & 0.65 & 0.659$\pm$0.054 & 0.03$\pm$0.001  &0.002$\pm$0.001 &0.95\\
5    &                        1 & 0.27 &   0.926$\pm$0.098   &   0.037$\pm$0.002 &  -0.003$\pm$0.003  & 0.97 & 0.26 & 1.42$\pm$0.42 & 0.034$\pm$0.002  &-0.001$\pm$0.002 &0.93\\
5    &                        2 & 0.31 &   1.083$\pm$0.096   &   0.043$\pm$0.001 &  -0.002$\pm$0.003  & 0.97 & 0.29 & 1.08$\pm$0.13 & 0.04$\pm$0.001  &-0.001$\pm$0.002 &0.92\\
5    &                        3 & 0.34 &   1.165$\pm$0.099   &   0.036$\pm$0.001 &  0.002$\pm$0.003  & 0.98 & 0.32 & 1.23$\pm$0.17 & 0.032$\pm$0.001  &0.0$\pm$0.002 &0.91\\
5    &                        4 & 0.16 &   2.12$\pm$0.2   &   0.052$\pm$0.003 &  -0.015$\pm$0.007  & 0.92 & 0.15 & 5.4$\pm$3.1 & 0.044$\pm$0.002  &-0.011$\pm$0.004 &0.82\\
6    &                        0 & 0.04 &   0.32$\pm$0.071   &   0.077$\pm$0.017 &  0.005$\pm$0.011  & 1.0 & 0.04 & 0.32$\pm$0.073 & 0.063$\pm$0.01  &-0.004$\pm$0.006 &0.97\\
6    &                        1 & 0.02 &   0.334$\pm$0.051   &   0.121$\pm$0.016 &  0.04$\pm$0.014  & 0.95 & 0.02 & 0.31$\pm$0.053 & 0.111$\pm$0.016  &0.027$\pm$0.012 &0.86\\
6    &                        2 & 0.03 &   0.71$\pm$0.21   &   0.142$\pm$0.021 &  0.027$\pm$0.032  & 1.0 & 0.02 & 0.333$\pm$0.087 & 0.111$\pm$0.012  &0.019$\pm$0.011 &0.75\\
6    &                        3 & 0.02 &   0.51$\pm$0.15   &   0.155$\pm$0.03 &  0.037$\pm$0.02  & 0.95 & 0.02 & 0.278$\pm$0.095 & 0.13$\pm$0.034  &-0.003$\pm$0.011 &0.64\\
6    &                        4 & 0.01 &   0.19$\pm$0.1   &   0.55$\pm$0.12 &  0.169$\pm$0.058  & 1.0 & 0.0 & 0.024$\pm$0.011 & 0.358$\pm$0.075  &-0.005$\pm$0.026 &0.33\\
7    &                        0 & 0.31 &   1.42$\pm$0.12   &   0.041$\pm$0.003 &  -0.001$\pm$0.004  & 0.95 & 0.3 & 1.82$\pm$0.28 & 0.033$\pm$0.003  &-0.001$\pm$0.002 &0.89\\
7    &                        1 & 0.23 &   1.39$\pm$0.13   &   0.037$\pm$0.003 &  -0.0$\pm$0.006  & 0.95 & 0.22 & 2.25$\pm$0.47 & 0.028$\pm$0.002  &0.002$\pm$0.002 &0.91\\
7    &                        2 & 0.35 &   1.41$\pm$0.1   &   0.04$\pm$0.003 &  0.006$\pm$0.004  & 0.94 & 0.33 & 2.06$\pm$0.34 & 0.032$\pm$0.002  &-0.001$\pm$0.002 &0.89\\
7    &                        3 & 0.28 &   1.28$\pm$0.11   &   0.051$\pm$0.004 &  0.007$\pm$0.007  & 0.95 & 0.26 & 1.82$\pm$0.32 & 0.033$\pm$0.002  &-0.003$\pm$0.002 &0.87\\
7    &                        4 & 0.21 &   1.33$\pm$0.14   &   0.063$\pm$0.005 &  0.003$\pm$0.007  & 0.93 & 0.18 & 1.93$\pm$0.44 & 0.043$\pm$0.004  &0.002$\pm$0.003 &0.79\\
8    &                        0 & 0.42 &   0.531$\pm$0.054   &   0.047$\pm$0.004 &  -0.0$\pm$0.002  & 0.95 & 0.41 & 0.81$\pm$0.14 & 0.041$\pm$0.003  &-0.001$\pm$0.001 &0.93\\
8    &                        1 & 0.2 &   0.596$\pm$0.087   &   0.105$\pm$0.056 &  -0.004$\pm$0.004  & 0.9 & 0.2 & 0.76$\pm$0.14 & 0.101$\pm$0.057  &-0.001$\pm$0.002 &0.89\\
8    &                        2 & 0.17 &   0.62$\pm$0.11   &   0.07$\pm$0.006 &  0.003$\pm$0.004  & 0.96 & 0.16 & 0.354$\pm$0.064 & 0.064$\pm$0.005  &-0.001$\pm$0.003 &0.88\\
8    &                        3 & 0.18 &   0.78$\pm$0.12   &   0.06$\pm$0.004 &  -0.003$\pm$0.005  & 0.96 & 0.17 & 1.03$\pm$0.34 & 0.053$\pm$0.004  &0.0$\pm$0.003 &0.89\\
8    &                        4 & 0.16 &   0.89$\pm$0.14   &   0.07$\pm$0.004 &  0.002$\pm$0.005  & 0.98 & 0.15 & 1.59$\pm$0.69 & 0.063$\pm$0.004  &0.002$\pm$0.003 &0.9\\
9    &                        4 & 0.25 &   1.2$\pm$0.1   &   0.036$\pm$0.003 &  -0.006$\pm$0.004  & 0.94 & 0.25 & 3.7$\pm$1.4 & 0.03$\pm$0.002  &-0.002$\pm$0.002 &0.91\\
\hline\hline
\end{tabular}
\caption{Filtered statistics for the ``representative'' submissions. Method
0:Rumbaugh-Gold, 1:Shafieloo-Arman7, 2:PyCS-d3cs-vanilla-dou-full,
3:PyCS-sdi-vanilla-dou-full, 4:PyCS-spl-vanilla-dou-full,
5:Jackson-manchester2\_0\_3\_4, 6:Kumar, 7:JPL, 8:Hojjati-Stark,
9:DeltaTBayes-DeltaTBayes1. }
\label{tab:filtered_table}
\end{center}
\end{table*}

\begin{figure*}[!htbp]
\begin{center}
\includegraphics[width=0.9\linewidth]{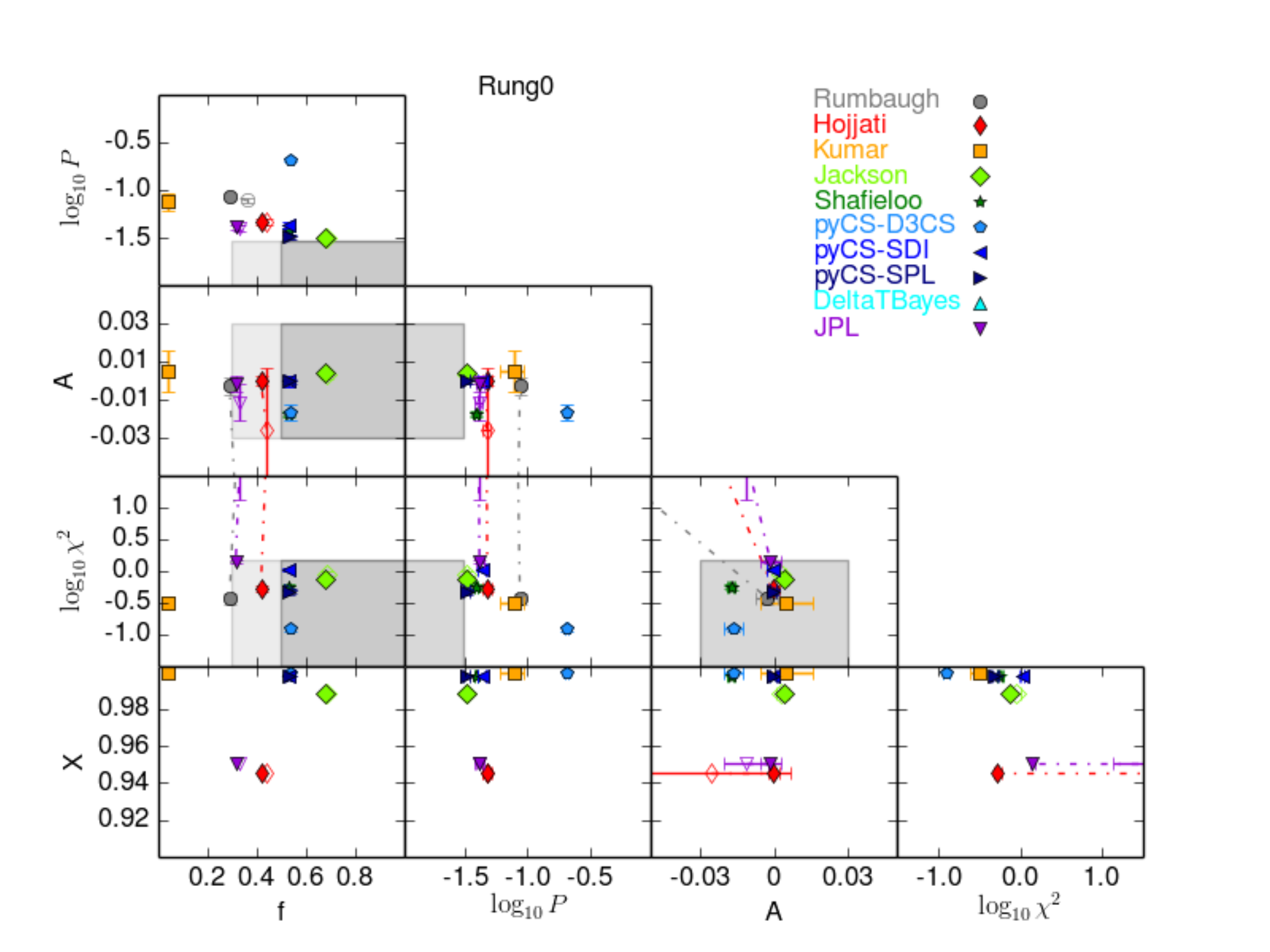}
\caption{Results for TDC1 Rung~0, showing metrics for the ``representative''
submission for each of the 10 algorithms.  This includes the baseline
submission by the ``Evil'' Team (``Rumbaugh'').  The $f$, $P$, $A$, and
$\chi^2$~metrics are defined in \Sref{sec:intro}, while $X$~is defined in
\Sref{sec:analysis:lessons}.  The shaded regions of each plot represent the
soft targets for TDC1, as presented in the TDC0 paper. Both unfiltered results
(open symbols) and results filtered by $\chi_i^2<10$ (solid symbols) are
presented, and they are connected by dashed lines to show the improvements.
Rung~0 simulates 3-day cadence and 8-month seasons over a 5 year campaign with
400 observations in total (\Tref{tab:obs}).}
\label{fig:TDC1-rung0}
\end{center}
\end{figure*}

\begin{figure*}[!htbp]
\begin{center}
\includegraphics[width=0.9\linewidth]{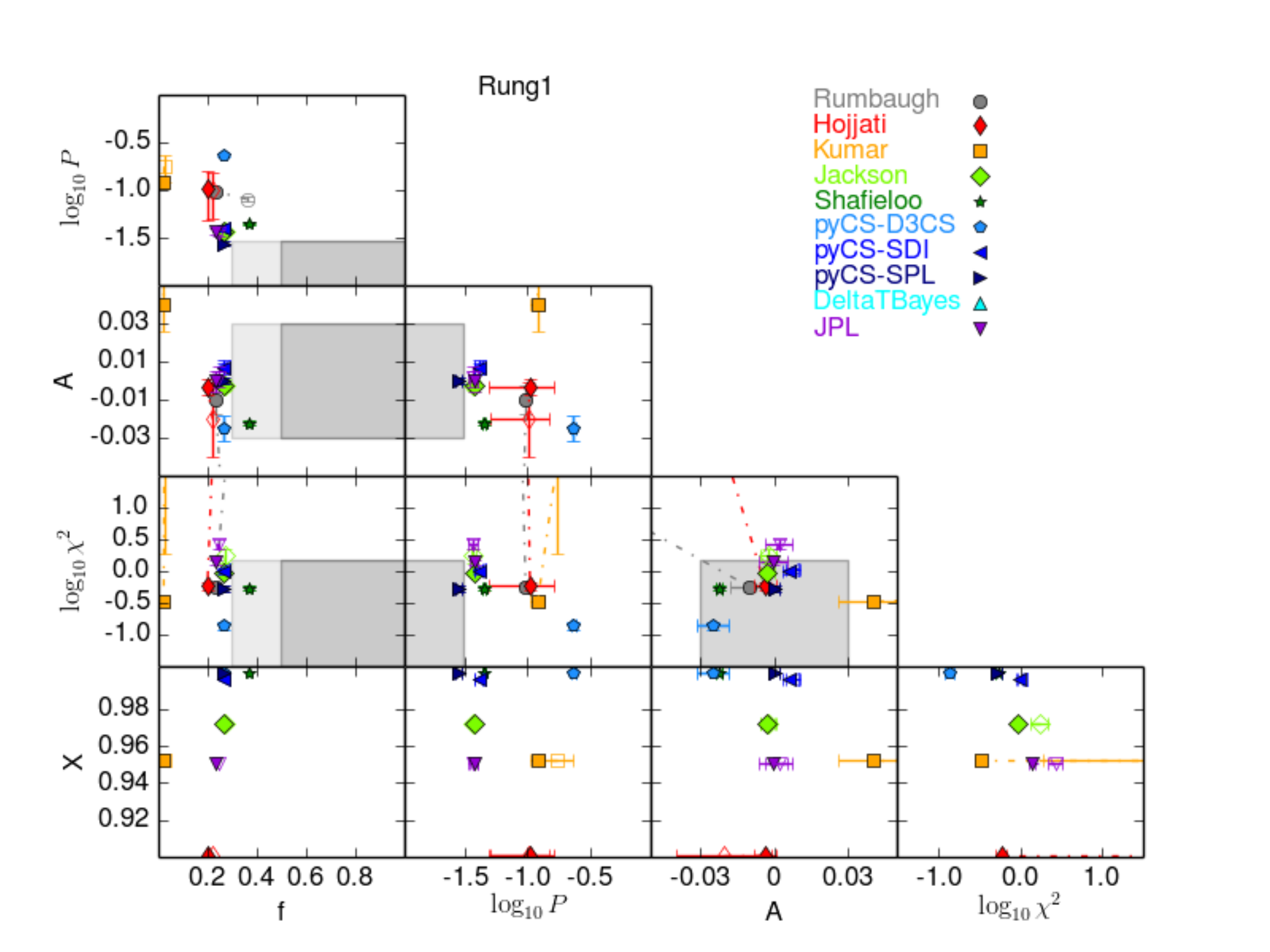}
\caption{Same as Figure~\ref{fig:TDC1-rung0}, but showing the results of
TDC1 Rung 1, which simulates 3-day cadence and 4-month seasons over a 10
year campaign with 400 observations in total (\Tref{tab:obs}).}
\label{fig:TDC1-rung1}
\end{center}
\end{figure*}

\begin{figure*}[!htbp]
\begin{center}
\includegraphics[width=0.9\linewidth]{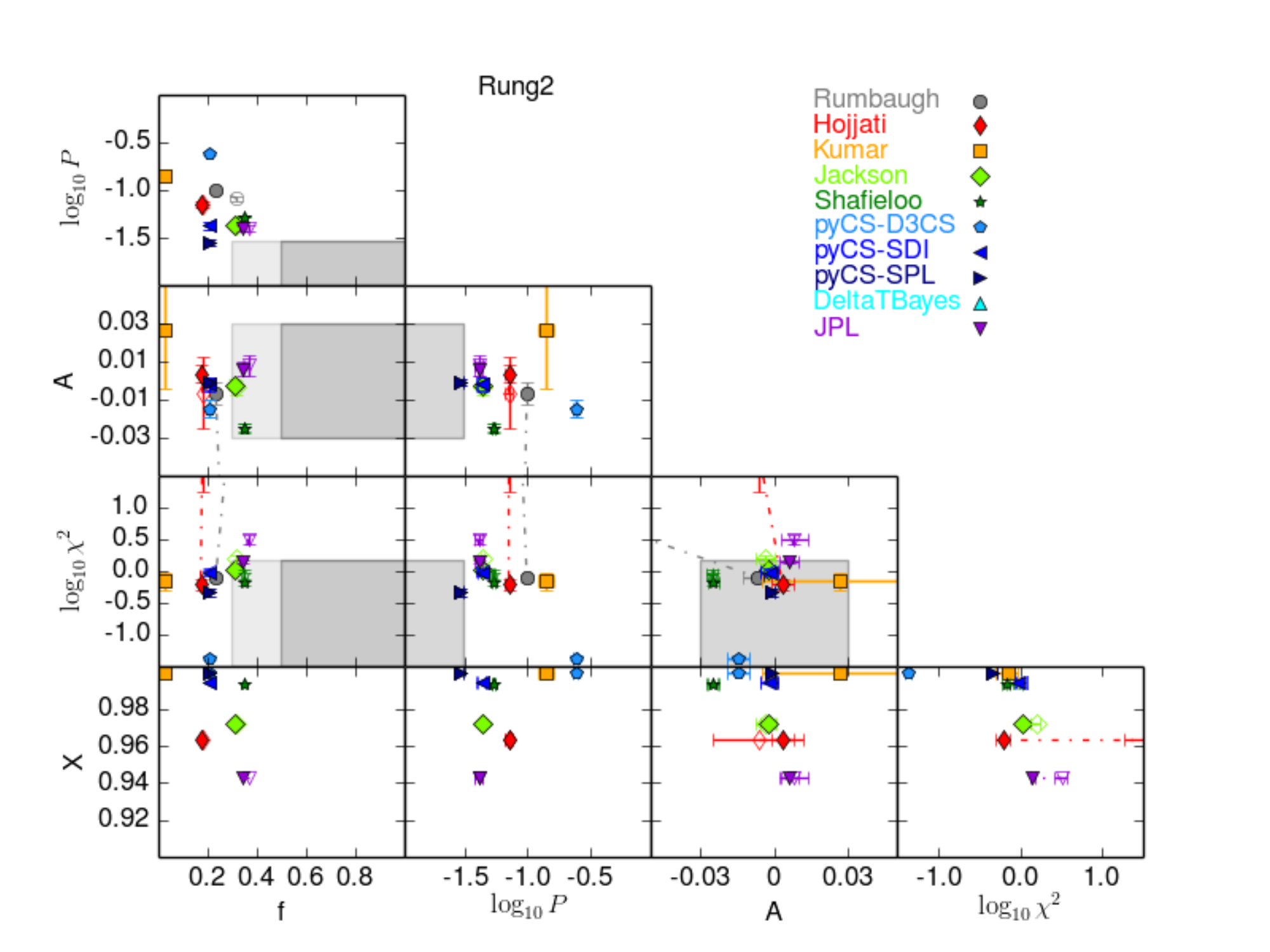}
\caption{Same as Figure~\ref{fig:TDC1-rung0}, but showing the results of
TDC1 Rung~2, which simulates 3-day cadence and 4-month seasons over a 5
year campaign with 200 observations in total (\Tref{tab:obs}), and
exactly regular time sampling.}
\label{fig:TDC1-rung2}
\end{center}
\end{figure*}

\begin{figure*}[!htbp]
\begin{center}
\includegraphics[width=0.9\linewidth]{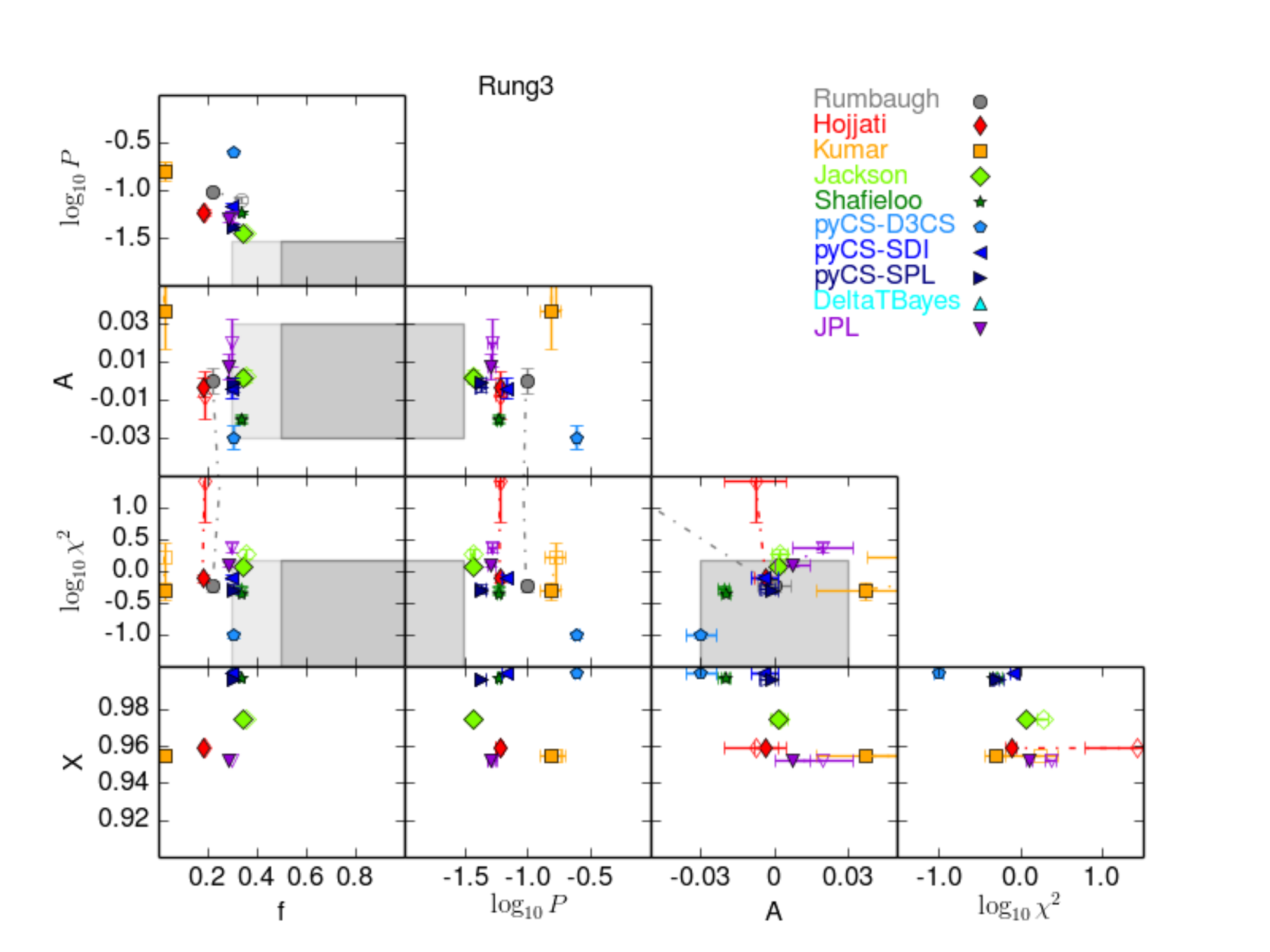}
\caption{Same as Figure~\ref{fig:TDC1-rung0}, but showing the results of
TDC1, Rung 3, which simulates 3-day cadence and 4-month seasons over a 5
year campaign with 200 observations in total (\Tref{tab:obs}), and with
1-day scatter in the separations between observations.}
\label{fig:TDC1-rung3}
\end{center}
\end{figure*}

\begin{figure*}[!htbp]
\begin{center}
\includegraphics[width=0.9\linewidth]{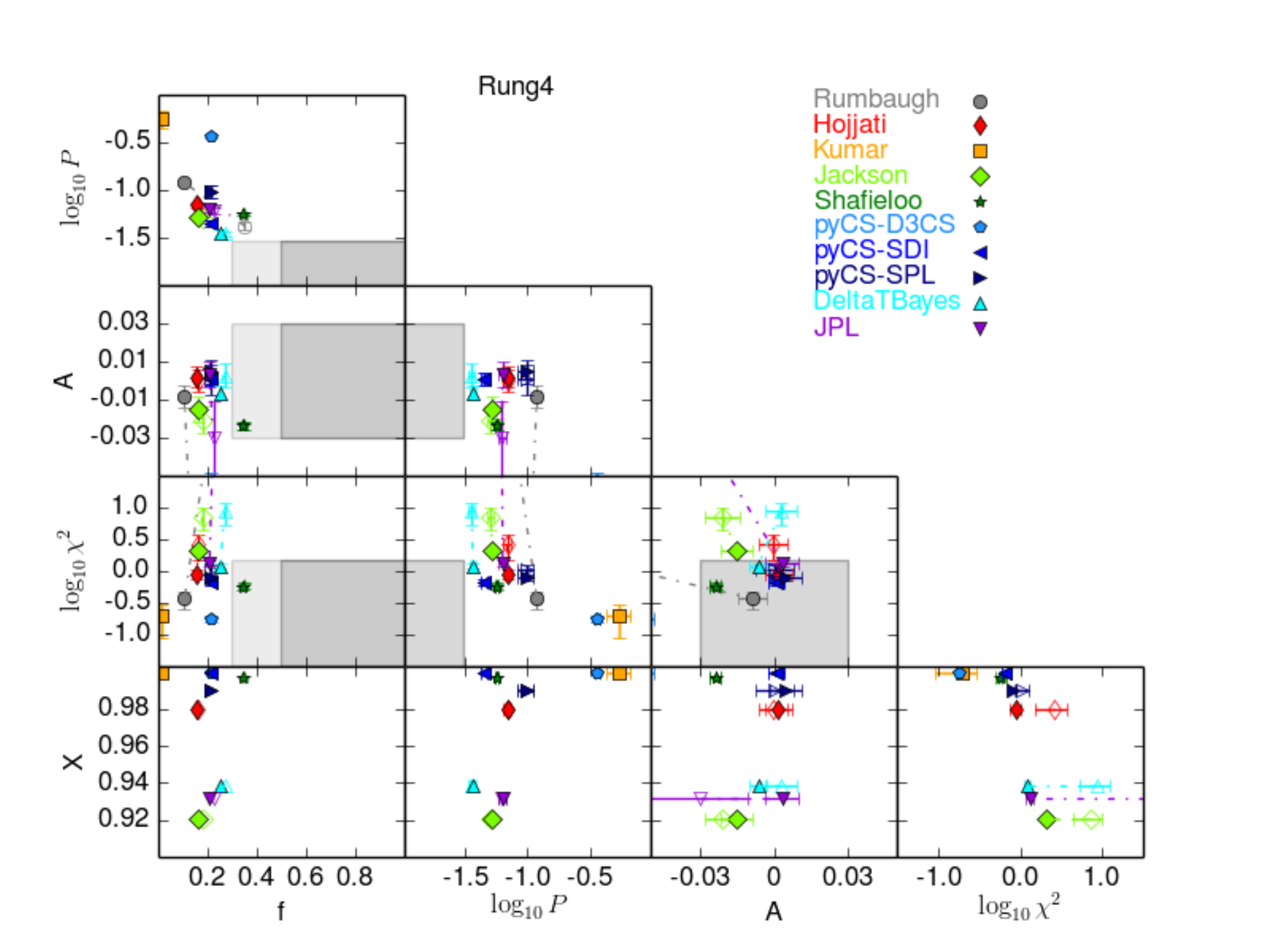}
\caption{Same as Figure~\ref{fig:TDC1-rung0}, but showing the results of
TDC1, Rung~4, which simulates 6-day cadence and 4-month seasons over a
10 year campaign with 200 observations in total (\Tref{tab:obs}).}
\label{fig:TDC1-rung4}
\end{center}
\end{figure*}

\begin{figure*}[!htbp]
\begin{center}
\includegraphics[width=\linewidth]{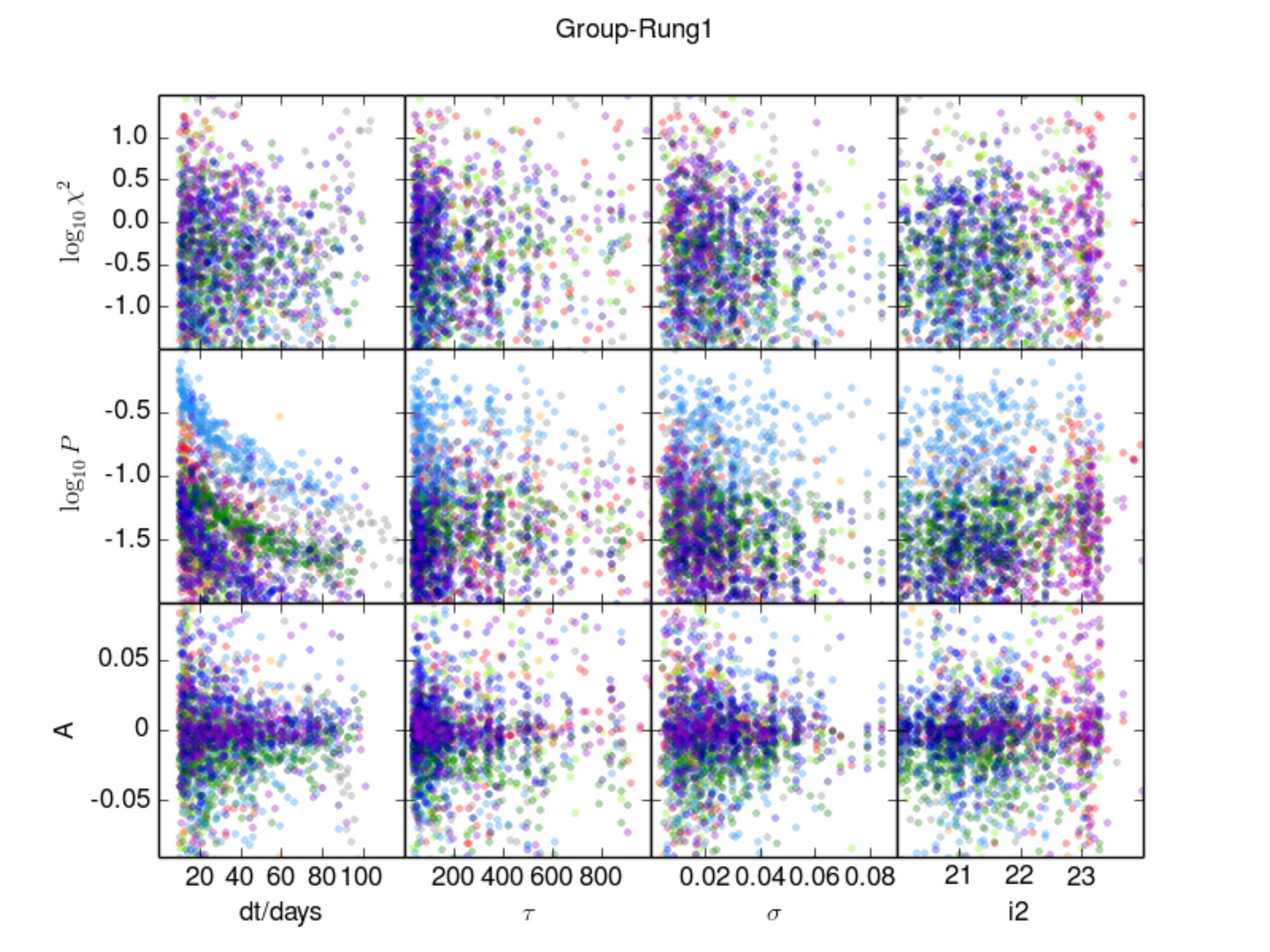}
\caption{Unfiltered results of Rung~1: individual metrics of each
``representative'' submission ($A_i$, $P_i$, $\chi_i^2$) as a function
of true time delay $dt$, the variability parameters of the intrinsic
quasar light curves ($\tau$, $\sigma$), and the magnitude of the fainter
image of each pair ($i_2$).  The color scheme is the same as that
described in the legend of \Fref{fig:TDC1-rung0}.}
\label{fig:swarm1}
\end{center}
\end{figure*}

\begin{figure*}[!htbp]
\begin{center}
\includegraphics[width=\linewidth]{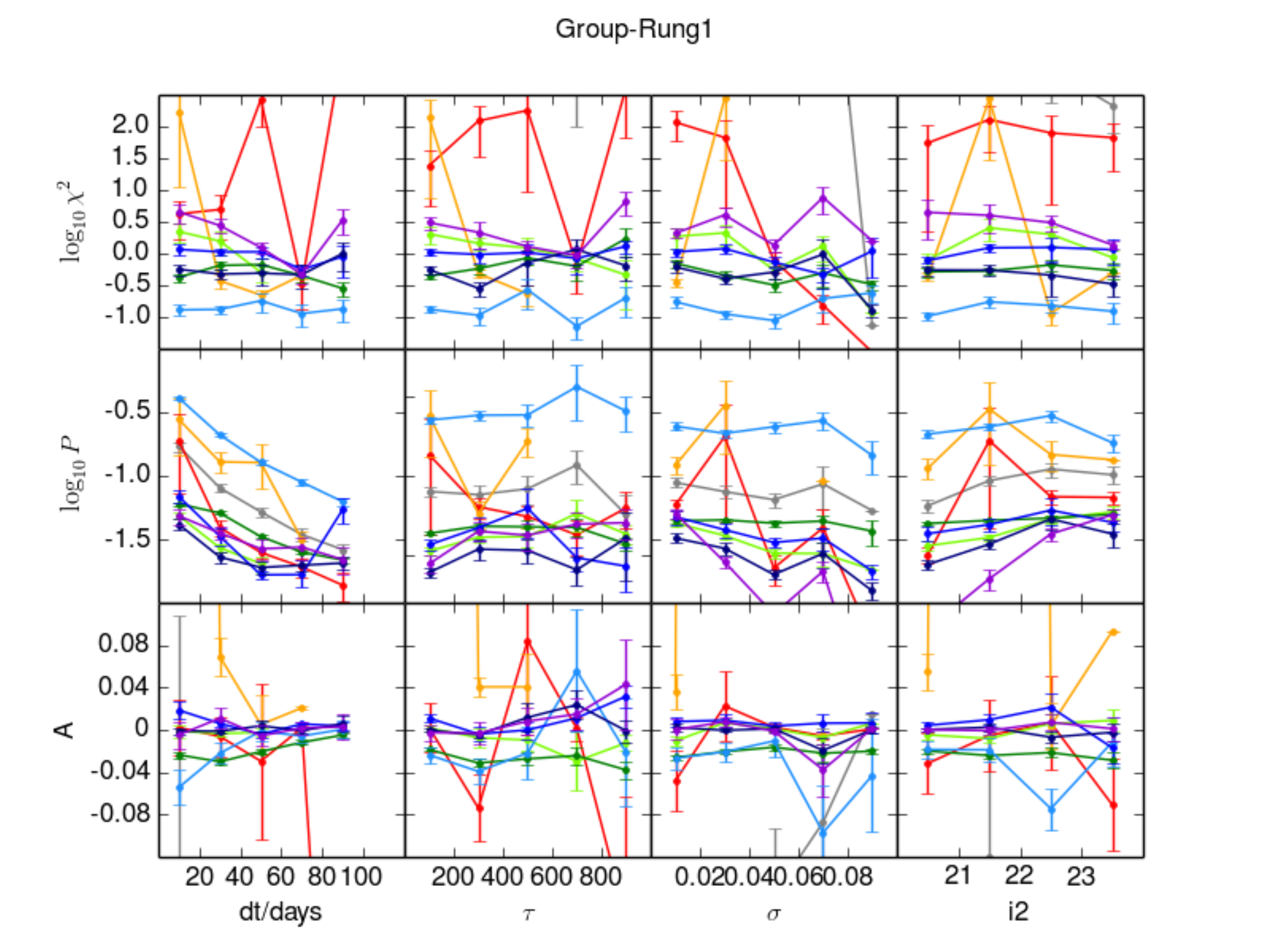}
\caption{Summary unfiltered statistics of the same data in
\Fref{fig:swarm1}, represented by the average statistics in bins of the
variable on the abscissa. The color scheme is the same as described in
the legend to \Fref{fig:TDC1-rung0}.}
\label{fig:swarm2}
\end{center}
\end{figure*}

\begin{figure}[!htbp]
\begin{center}
\includegraphics[width=\linewidth]{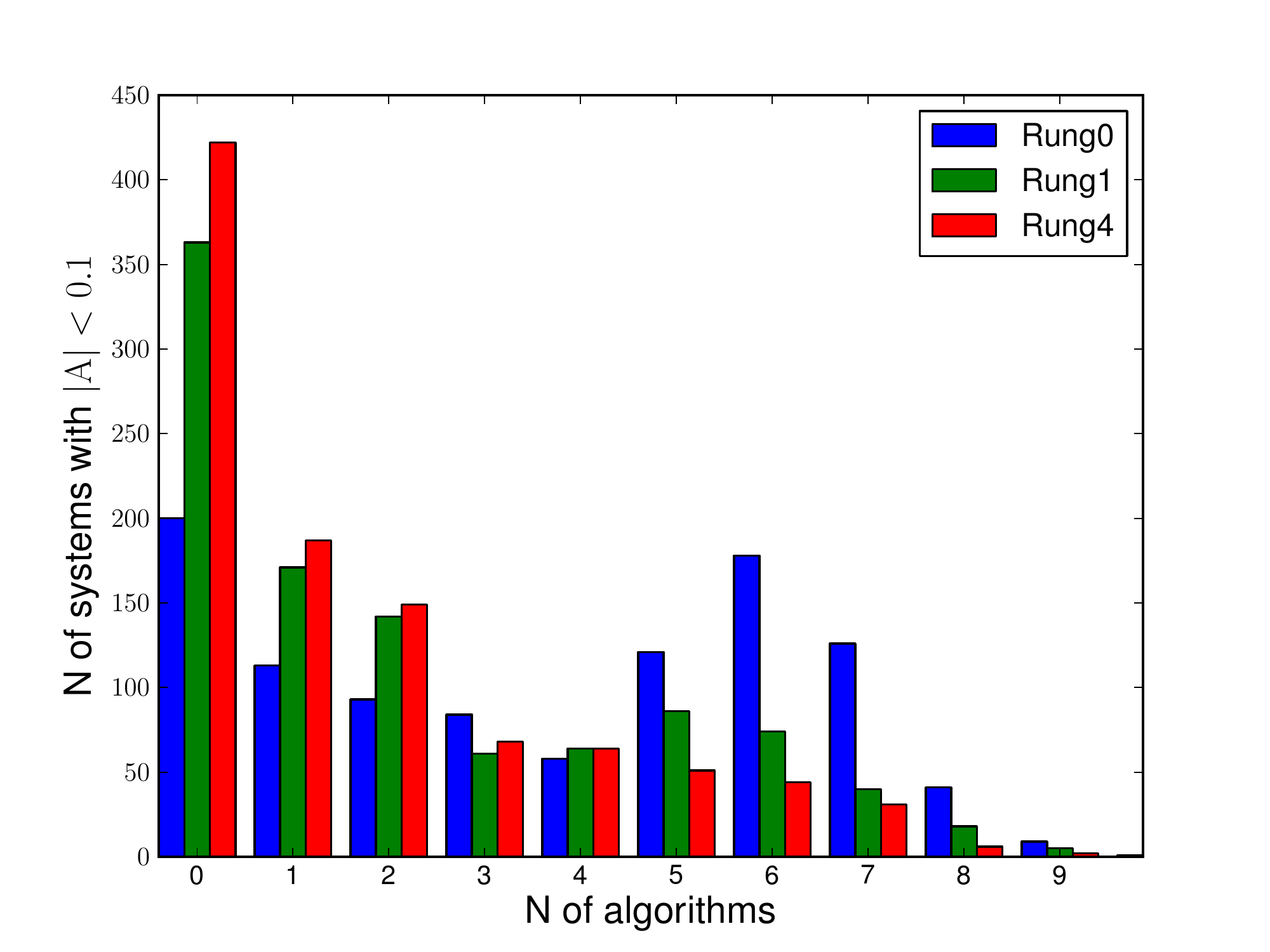}
\caption{Distribution of the number of systems for which
the time delay was successfully measured to a level of
$|A_i|<0.1$, plotted as a function of the number of algorithms (out
of 10) that measure the time delays to this level.  The plot shows
Rungs~0, 1, and~4, which represent COSMOGRAIL-like, ``optimistic''
LSST, and ``realistic'' LSST programs, respectively.  For Rung 0,
there were more than $\sim$200 systems for which none of the algorithms
achieved the desired $A$, but also a large number of systems for
which five, six, or seven of the algorithms successfully recovered
this level.  For Rungs~1 and~4, fewer of the systems were successfully
recovered at the $|A_i|<0.1$ level.}
\label{fig:NN014}
\end{center}
\end{figure}


\subsection{Lessons from TDC0 applied to TDC1}
\label{sec:analysis:lessons}

During the analysis of the TDC0 submissions, the ``Evil'' Team noticed that
several teams were affected by outliers: most of their submitted time
delay estimates were good, but a few differed from the truth by more
than would be expected, given their uncertainties.  To characterize
this, an additional metric $X$ was introduced: $X$~is the fraction of
pairs with $\chi^2_i<10$, i.e., the fraction without outliers.  $X=1$
means that none of the submitted delays is an outlier.  Outliers in
this category could stem from underestimated error bars, or for
example by convergence on the wrong solution in the presence of light
curve features (due to, e.g., microlensing) that are not taken into
account by the method's model.

We will return to the issue of outliers, and how they can be identified
based on lensing geometry or cosmological analysis, after we present the
main results of TDC1.  In this section, we give the unfiltered
statistics as well as the metrics calculated after points with
$\chi^2_i>10$ have been removed, in order to give an idea of how well a
method {\it could} do if outliers could be identified and rejected.

We also consider an additional cut, based only on the accuracy parameter
$|A_i|<0.1$, and the related quantity $X_A$, which counts the fraction
of systems satisfying this alternative criterion, \textbf{i.e., we take $|A_i|>0.1$
as outliers rather than $\chi^2_i>10$ in this case.}  This cut was chosen
to quantify the number of systems for which the time-delay would be much
more uncertain than the 3-5\% modeling error that can be obtained in the
reconstruction of the difference in gravitational potential between two
images in the best cases \citep{SuyuEtal2013,Suy++14}. In some sense
this cut filters out the systems that are not cosmologically consistent
and thus could be rejected by a joint cosmological analysis.

Finally, as a third way to illustrate the potential of each method once
outliers have been removed, we also consider the median, 16 and 84
percentile of the statistics $A_i,P_i$ and $\chi_i^2$ for each method,
as opposed to the means defined in \Sref{sec:intro}.


\subsection{Blind and non-blind submissions}

One of the main goals of this time delay challenge is to achieve a
true blind testing of the algorithms. To achieve this, TDC0 truth
files were not revealed until after the deadline of TDC1, lest they
give too much away about the data generation process. In addition,
upon requests from each ``Good'' Team we provided only minimal
feedback after each submission, in the form of the metrics listed
above rounded to two significant digits. This was deemed to be a
reasonable compromise between preserving the blindness of the
challenge, and helping teams to identify coding errors that had
nothing to do with their actual chosen algorithms. Only submissions
made prior to any feedback were considered truly blind, even though
resubmissions by the teams who decided to take advantage of this
opportunity were accepted.  Resubmissions were considered not fully
blind for the purpose of this analysis. Note that all of the
``representative'' submissions referred in later sections were made
fully blind.


\subsection{Basic statistics}

The metrics for each submission are shown in
Tables~\ref{tab:unfiltered_table} and~\ref{tab:filtered_table},
separated by challenge rung. In order to visually compare the different
algorithms in a relatively clear manner, we have chosen to show only one
submission for each team. This ``representative'' algorithm was chosen
by each team after the true time delays were unblinded, and therefore it
is somewhat indicative of the best performance of each method. Results
for all the other submissions are available at the TDC website.
Importantly, it should be kept in mind that this is a multi-dimensional
problem, and there is not necessarily a ``best'' submission, not even
within each method. Rather, each submission is a tradeoff between
competing needs of achieving low $P$ and $A$, while keeping $\chi^2$
reasonable and $f$ and $X$ as high as possible. Some of the statistics
are mathematically inter-dependent. For example, $\chi^2$ and $P$ both
contain the submitted uncertainty estimates: teams could decide to
reduce their $\chi^2$ at the price of increasing their $P$, and vice
versa.

The metrics obtained by these submissions are plotted in
Figures~\ref{fig:TDC1-rung0}--\ref{fig:TDC1-rung4}.  The plots show the
metrics that have been computed directly from the submitted values,
together with the recomputed metrics after rejecting the outliers using
the $\chi_i^2<10$ cut.  The corner plots in
Figures~\ref{fig:TDC1-rung0}--\ref{fig:TDC1-rung4} also show a shaded
region that represents the TDC1 soft targets that were estimated in
\paperone as the metric values needed for methods to be competitive,
namely:
\begin{itemize}
\item $f > 0.5$
\item $\chi^2 < 1.5$
\item $|A| < 0.03$ [goal 0.002]
\item $P < 0.03$
\end{itemize}

As discussed in \paperone, in this exploratory challenge, these
targets were deemed sufficient given the current lensed quasar sample
of a few tens of systems. In the long run, for samples of thousands of
lenses, a desirable goal is to improve the accuracy or bias to
sub-percent level ($|A|<0.2$\%, see \citet{HLinprep} for the cosmological
requirement derivation), such that the contribution of time
delay measurement to the error budget of cosmological parameters would
be smaller than the projected statistical uncertainties. We emphasize
that these targets are approximate and only with a fully cosmological
challenge would they be translated into a single indicator of
performance, as we outline in the final section of this paper.

As is shown in the figures, most of the algorithms achieved the $|A|$
and $\chi^2$ criteria, especially after the rejection of outliers in the
submissions.  The ``Evil'' Team's baseline method had a large fraction of
outliers, but once those were rejected, it did not perform significantly
worse than many of the ``Good'' Teams submissions. The criterion that proved
more difficult to meet was the one on the success fraction $f$, where
teams were typically closer to the threshold for TDC0 (shown also in the
cornerplot as a lighter shaded region) than for TDC1. As we discuss
below, this is due to the strategy that most teams followed, i.e. to
have high standards of acceptance in order to reduce outliers. Notably,
for many of the methods $|A|$ is at the sub-percent level -- well below the
target of 0.03 -- which is very promising in view of future cosmological
studies.

Interestingly, the ``evil'' light curves did not yield significantly
poorer statistics than the regular ones. From this comparison we infer
that the methods used are generally robust to small and realistic
unknown light curve systematics like the ones introduced by the ``Evil''
Team. This is encouraging and bodes well for the application of the
methods to real data.


\subsection{Trends with intrinsic properties of the lightcurves and implications for future work}

We now investigate how the quality of the inferred time delays depends on
the intrinsic properties of the light curves. We wish to discover
general trends that are not inherent to the peculiarities of each
method. In order to carry out this investigation, in \Fref{fig:swarm1}
we plot the individual accuracy, precision and goodness of fit of each
submission ($A_i$, $P_i$ and $\chi^2_i$) as a function of true time
delay, the variability parameters of the intrinsic quasar light curves
($\tau$ and $\sigma$), and the magnitude of the fainter image of each
pair ($i_2$). In this illustration we show the results for Rung~1; the
other rungs give similar results. \Fref{fig:swarm2} shows summary
statistics of the same data, represented by the average statistics in
bins of the variable on the abscissa -- the color scheme is the same as
described in the legend to \Fref{fig:TDC1-rung0}.

We can see in these figures a few global trends. The most prominent
appears to be between $P$ and the true time delay. $P$ decreases with
time delay consistent with the time delay uncertainty being
approximately constant in days, as expected if the absolute precision is
driven by the sampling of the light curves. Qualitatively, $P_i$ and
$A_i$ also appear to decrease (i.e. improve) as $\sigma$ increases, also
as expected: the light curves with the highest variability amplitudes
should be easier to interpret and therefore should yield higher
precision and fewer outliers.

Remarkably, we see very little dependency on $i_2$, as if the signal to
noise ratio of the fainter image is not as important, once it is passes
some minimum threshold. This suggests that the simulated data are of
sufficient quality and that the photometric uncertainty is subdominant
with respect to the uncertainties introduced by microlensing and
sampling. The weak dependency on the magnitude of the fainter image
$i_2$ implies that the statistics we derive from the TDC1 dataset are
very similar to what we would have derived from a random subset of OM10.
In fact, by recomputing weighted averages of the statistics to match the
OM10 $i_2$ magnitude distribution, we verified that the changes of the
statistics would have been comparable to their uncertainty.

Finally, we investigated the level of agreement between the algorithms
to see whether success was due solely to the properties of the light
curves or whether it depended on the specifics of each algorithm. The
results are shown in \Fref{fig:NN014} for three representative
rungs. Clearly, some light curves do not contain enough information
for any method to be successful (hence the peak at zero). In Rung~0,
there is a bump around 6 indicating that for very good light curve a
majority of the methods are successful. However, as the quality of
data degrades in the next rungs it appears that there is a continuum
distribution. Therefore we conclude that different methods pick up
different features of the light curves and accuracy varies widely
between methods.


\section{Implications for Observing Strategy}
\label{sec:strategy}

By comparing the results from the different TDC1 rungs, we can now
answer the following question: How does time delay measurement accuracy
depend on observing cadence, season length and campaign length?

\begin{figure*}[!ht]
  \begin{minipage}[b]{\linewidth}
    \begin{minipage}[b]{0.48\linewidth}
      \centering\includegraphics[width=\linewidth]{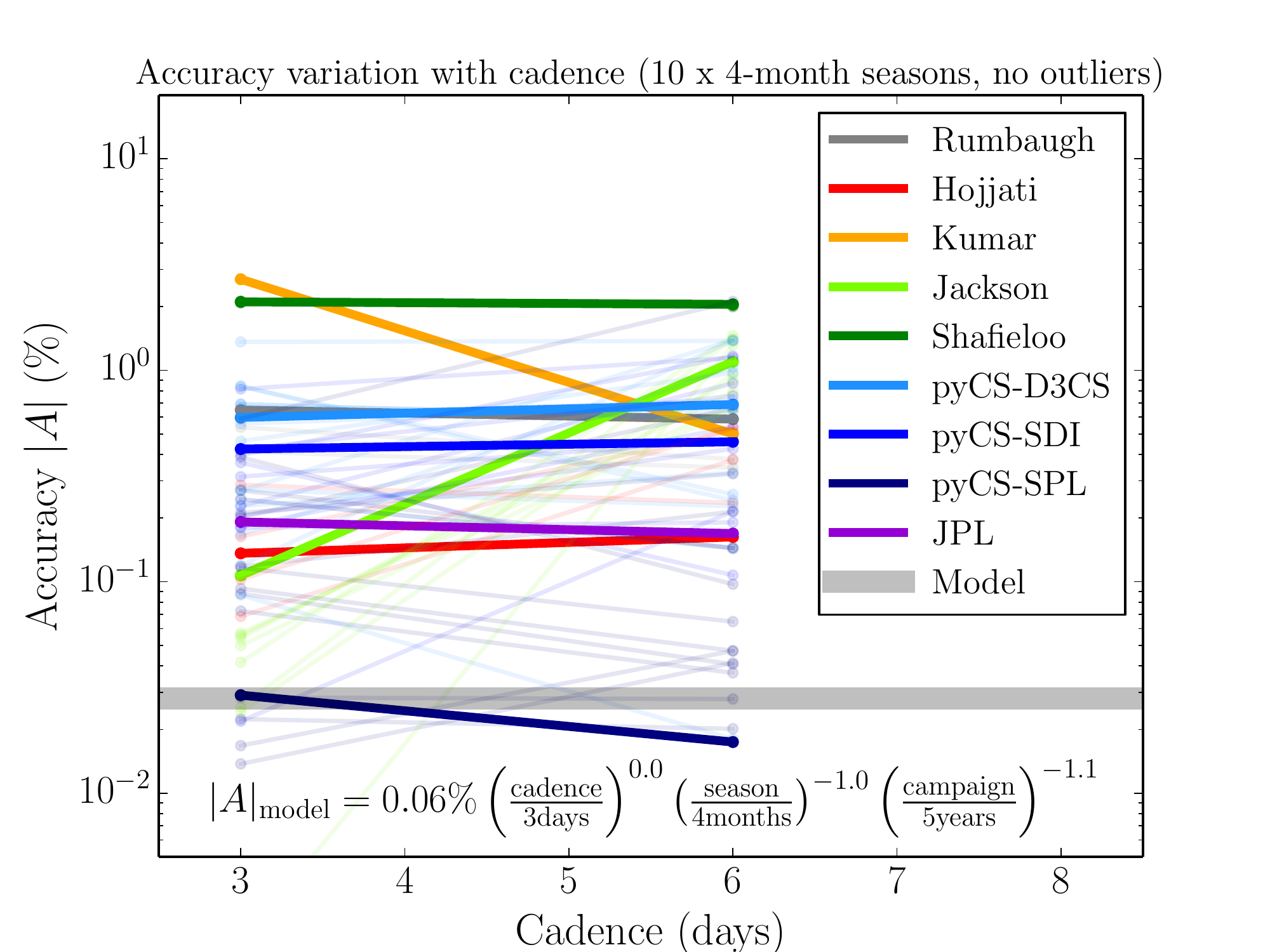}
    \end{minipage} \hfill
    \begin{minipage}[b]{0.48\linewidth}
      \centering\includegraphics[width=\linewidth]{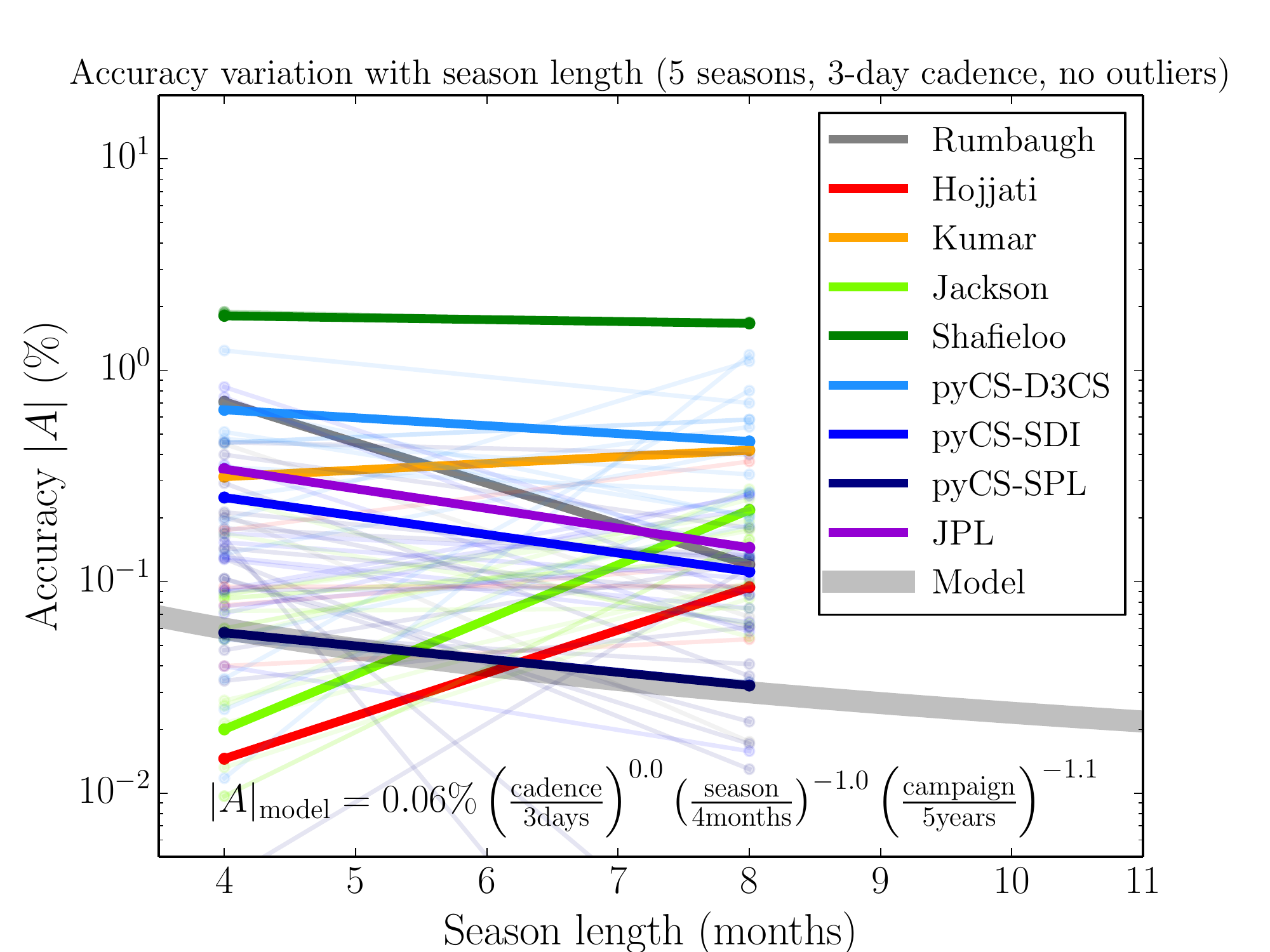}
    \end{minipage}
  \end{minipage}
  \begin{minipage}[b]{\linewidth}
    \begin{minipage}[b]{0.48\linewidth}
      \centering\includegraphics[width=\linewidth]{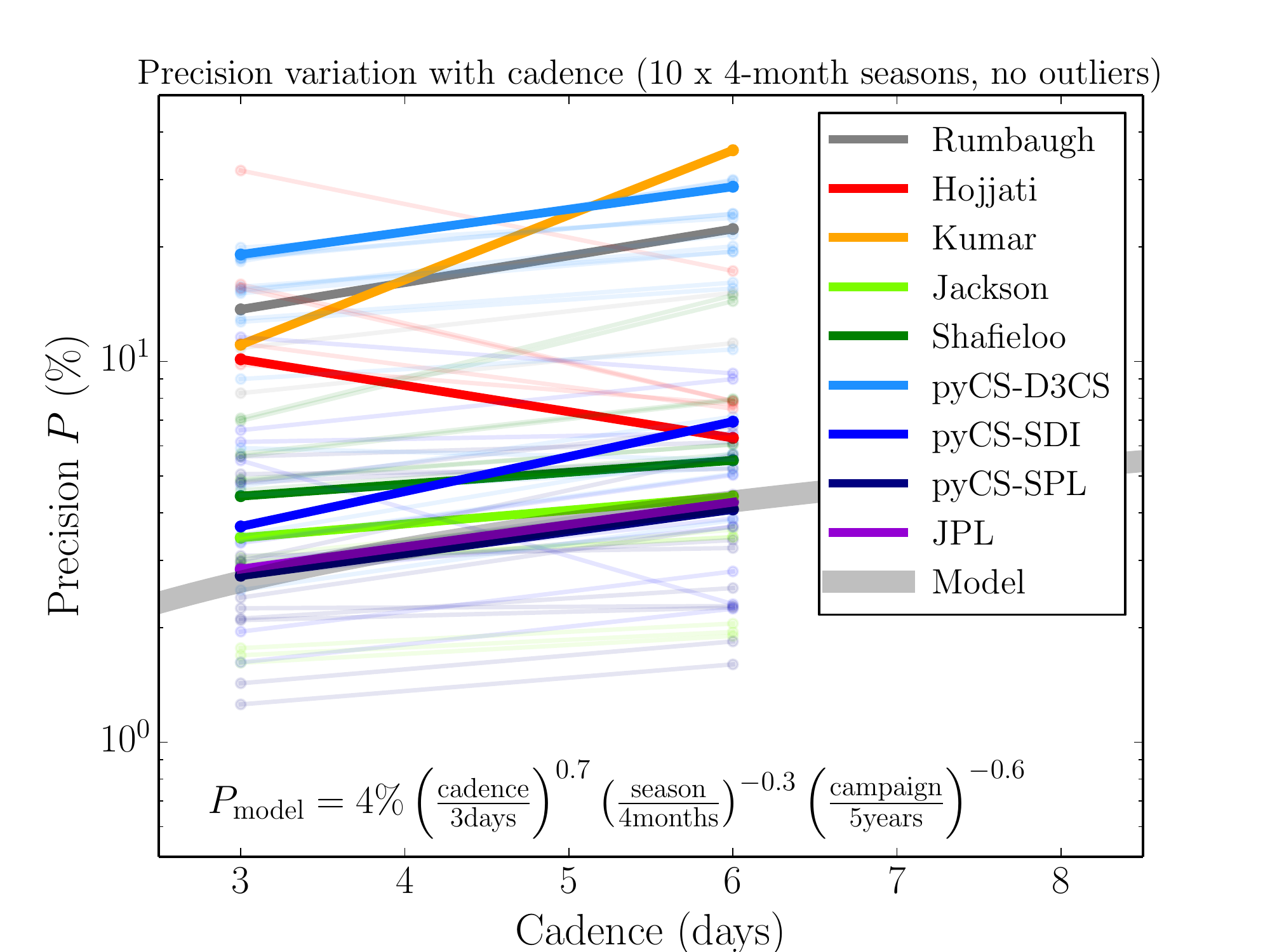}
    \end{minipage} \hfill
    \begin{minipage}[b]{0.48\linewidth}
      \centering\includegraphics[width=\linewidth]{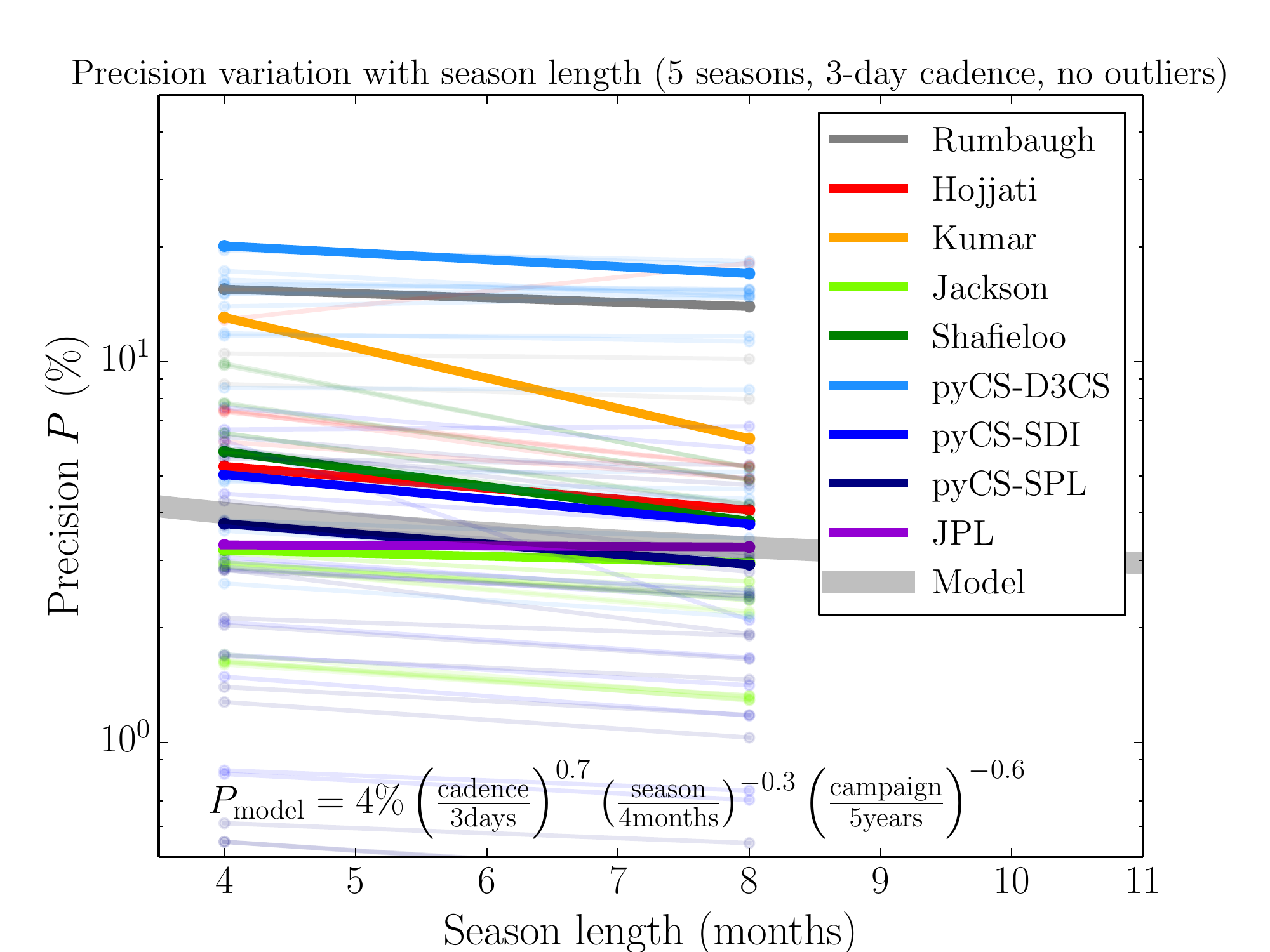}
    \end{minipage}
  \end{minipage}
  \begin{minipage}[b]{\linewidth}
    \begin{minipage}[b]{0.48\linewidth}
      \centering\includegraphics[width=\linewidth]{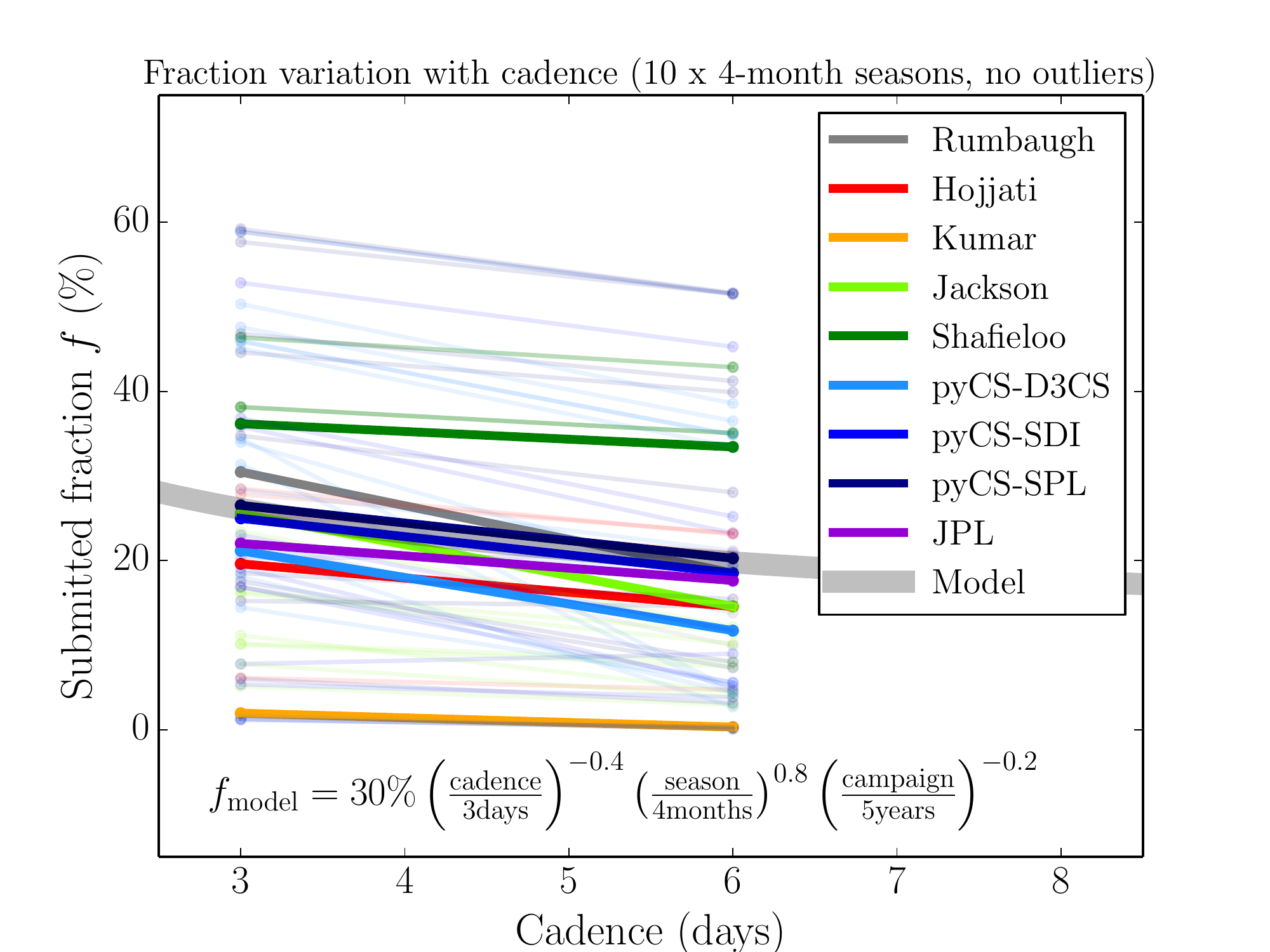}
    \end{minipage} \hfill
    \begin{minipage}[b]{0.48\linewidth}
      \centering\includegraphics[width=\linewidth]{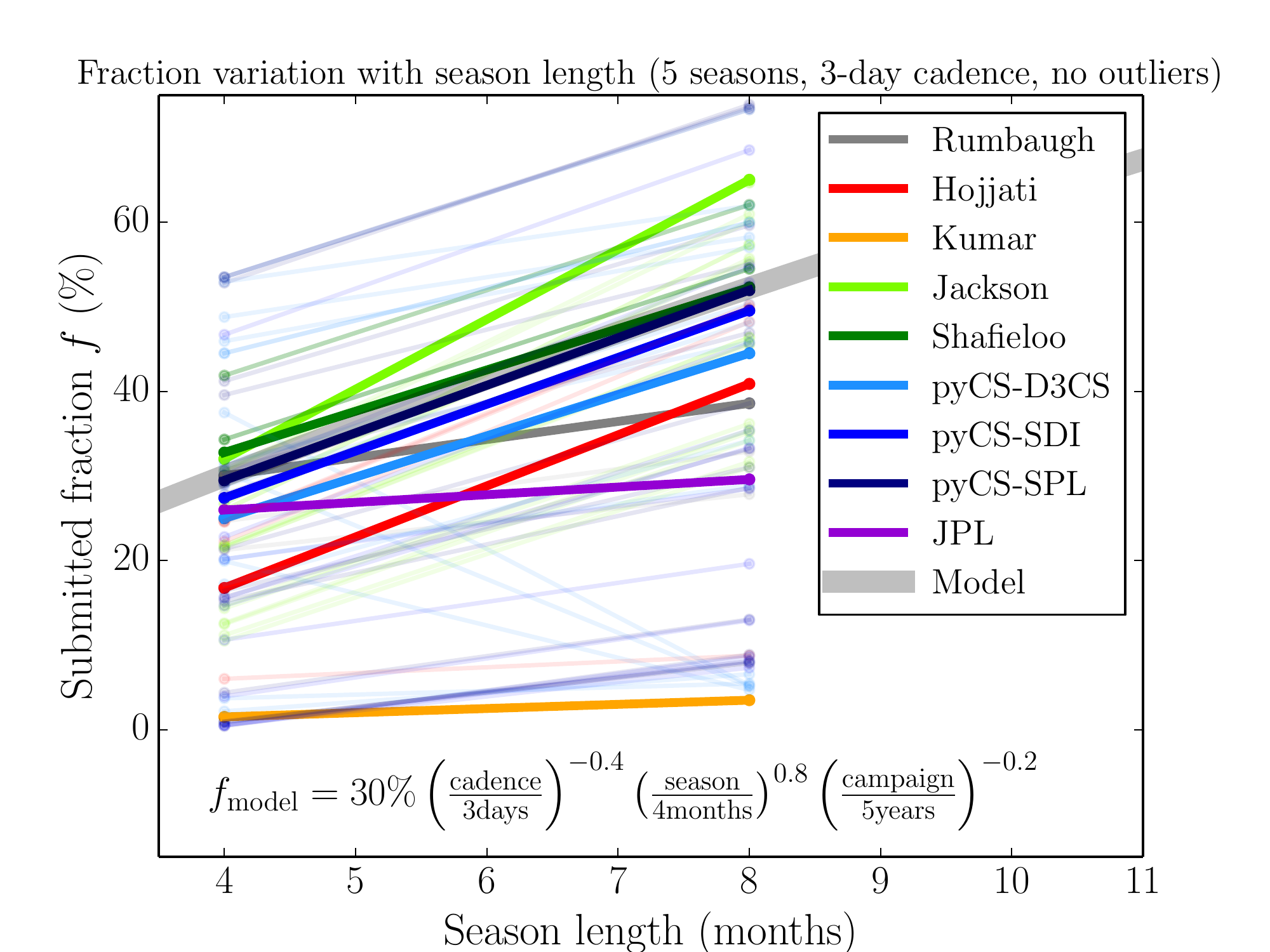}
    \end{minipage}
  \end{minipage}
\caption{Changes in accuracy $A$ (top row), precision $P$ (middle row)
and success fraction $f$ (bottom row)   with cadence (left) and season
length (right), seen in the different TDC1 submissions. The gray
approximate power law model was derived by visual inspection of the
pyCS-SPL results; the signs of the indices were pre-determined according
to our expectations.}
\label{fig:obsdep}
\end{figure*}

\Fref{fig:obsdep} shows the variation in the $|A|$, $P$ and $f$ metrics
with cadence and season length, assuming outliers to have been rejected
by $|A_i| > 0.1$.  Each pair of connected points plotted in the
panels of this figure represents a simple test where the control
variable (cadence or season length) is varied, while keeping the others
constant. Campaign length and cadence regularity were also investigated
in a similar manner, but the results -- which are less striking --- are
not shown here. The 6 tests we carried out in total are summarized in
\Tref{tab:ABtests}. The top two rows in the table correspond to the
plots shown in the left and right columns of the figure, respectively.

\begin{table*}
\begin{center}
\begin{tabular}{cll} \hline\hline
 Rungs &  Variable parameter            &  Fixed parameters \\ \hline
  1,4  &  Cadence (3,6 days)            &  4-month seasons, 10-year campaign   \\
  0,3  &  Season (4,8 months)           &  3-day cadence, 5-year campaign \\ \hline
  3,4  &  Cadence (3,6 days)            &  4-month seasons, 200 epochs length   \\
  0,1  &  Season (4,8 months)           &  3-day cadence, 400 epochs length   \\
  1,3  &  Campaign (5,10 years)         &  3-day cadence, 4-month seasons \\
  2,3  &  Cadence dispersion (0,1 days) &  3-day cadence, 4-month season, 5-year campaign  \\
\hline\hline
\end{tabular}
\caption{Exploring time delay estimation performance against observing
strategy. The tests defined in the top two rows (above the line) are
illustrated in \Fref{fig:obsdep}.}
\label{tab:ABtests}
\end{center}
\end{table*}

\Fref{fig:obsdep} shows some interesting diversity between methods.
Despite this, some approximate general trends can be seen. Greater
accuracy and success fractions seem to be associated primarily with
longer seasons, but there is considerable scatter between submissions,
perhaps due to residual outliers in some cases.  In most methods, little
dependence of accuracy on cadence, campaign lengths beyond 5 years, or
the regularity of the sampling was seen. The success fraction seems to
be somewhat dependent on cadence but less so on campaign length. In
general, the trends in precision with cadence and season length seem to
be less marked, and show less scatter, than those in accuracy and
success fraction. In general, cadence seems to be the most important
factor for precision.

While the variation of time delay measurement with observing strategy
seems to be somewhat algorithm-dependent, we can nevertheless hope to
capture the general trends just described. Focusing on the the PyCS-SPL
results, we derived a very approximate power-law model for how the $A$,
$P$ and $f$ metrics varied with the main three quantities that describe
the observing strategies in the rungs, mean cadence (cad), mean season
length (sea), and campaign length (camp). We find:
\begin{align}
|A|_{\rm model} &\approx 0.06\% \left(\frac{\rm cad} {\rm 3 days}  \right)^{0.0}
                          \left(\frac{\rm sea}  {\rm 4 months}\right)^{-1.0}
                          \left(\frac{\rm camp}{\rm 5 years} \right)^{-1.1} \notag \\
  P_{\rm model} &\approx 4.0\% \left(\frac{\rm cad} {\rm 3 days}  \right)^{ 0.7}
                         \left(\frac{\rm sea}  {\rm 4 months}\right)^{-0.3}
                         \left(\frac{\rm camp}{\rm 5 years} \right)^{-0.6} \notag \\
  f_{\rm model} &\approx 30\% \left(\frac{\rm cad} {\rm 3 days}  \right)^{-0.4}
                        \left(\frac{\rm sea}  {\rm 4 months}\right)^{ 0.8}
                        \left(\frac{\rm camp}{\rm 5 years} \right)^{-0.2} \notag
\end{align}
We can see that in this model,  the accuracy metric $A$ is the most
sensitive to the observing strategy. It is also the case that it is the
metric most sensitive to how the outliers are rejected. Rejecting
outliers that have $\chi_i^2>10$ gives similar conclusions to those drawn
here, but slightly different model parameters, in the sense that there
is even stronger dependence of $A$ on the observing strategy. In both
cases the dependence of $A$ on cadence is relatively weak. The season
length and campaign length seem to be more important parameters:
doubling either of these results in approximately a factor of two
improvement in $A$. We note that constraining the total number of
observations weakens these dependencies somewhat: for example, at fixed
cadence, lengthening the season means shortening the campaign, and in
our model, $|A|$ then decreases only as the ratio of the season length to
the campaign length to the power of 0.1. The results of the fixed epoch
number tests in \Tref{tab:ABtests} bore this out.

The precision and success fraction metrics' dependence on observing
strategy is weaker, but it is interesting to note that the precision
depends more strongly on cadence than the season length, while the
opposite is true for the success fraction.  This can be understood
qualitatively as the presence of large gaps reducing the overlap between
light curves, making it more difficult to reliably and uniquely identify
common features between them. Conversely, if the signal is properly
identified, then the precision is driven by the total number of
observation points, i.e. a combination of cadence and campaign duration.
As a rough rule of thumb, we might have in mind that season length
largely determines bias, while cadence controls precision. The precision
of an ensemble average parameter, such as the cosmological parameters,
may yet depend primarily on season length, however, through the success
fraction.

These simple model conclusions represent small extrapolations -- we did
not, for example, test doubling the season length and cadence
simultaneously -- but they represent a first approximation to the
response of the more accurate time delay estimation routines to
variations in observing strategy.

Finally, we note briefly the implications of this model for the sample
of lensed quasars that was forecast for LSST by \citet{OM10}. Rung~4
represents something like the ``universal cadence'' planned for LSST
\citep{LSSTpaper}, albeit with slightly shorter seasons. A cadence of 6
days would be well within the reach of such a strategy, but would
require using observations from most of the filters in the set. While in
this work we have  only simulated and analyzed single filter
lightcurves,  AGN variability has been observed to be significantly
correlated across the optical and near infra-red bands \citep[see e.g.,][]{SchmidtEtal2012}, and microlensing variability is expected, and
observed, to vary smoothly with wavelength due to source size effects
\citep[e.g.,][]{PoindexterEtal2008}. With sufficiently sophisticated
algorithms we might expect to be able to measure time delays from
multi-filter light curves with fidelity not dissimilar to that shown
by the TDC1 methods tested here.  The 3-day cadence of Rung~1 could be
achieved by LSST without changing the total number of visits; the impact
of such a strategy on the various different LSST science cases would
need to be investigated. We take Rungs~1 and~4 to span the range of
possibilities for the LSST time sampling.

Our model suggests that, if outliers with $|A_i| > 0.1$ can be rejected
(perhaps during a joint analysis of the ensemble), the cadence is effectively
unimportant for time delay measurement bias, and with LSST we might expect to
achieve an accuracy metric of $|A| = 0.03-0.06\%$. Such a small time delay
measurement bias is well below the systematic errors expected from lens
modeling. Meanwhile, the expected  precision achievable per lens in the Rung 1
and 4 cadences would be 2.6--4.3\%, and the  success fractions would be
20--26\%. The mock lenses used in this data challenge were not quite randomly
drawn from the OM10 catalog, but instead had approximately uniformly
distributed $i_3$ image magnitudes within four broad magnitude bins
(\Sref{sec:tdc1:sample}). Correcting for this, we find that we might, with the
present-day algorithms (tested here and represented by our simple model),
expect to be able to make time delay measurements with the above accuracy in
at least 20\% of an LSST sample of 1990 lenses selected to have $i_3 < 23.3$
and $10 < \Delta t < 120$ days. This would correspond to a well-measured
sample of around 400 lensed quasars. We must expect these numbers to be
refined as the LSST observing strategy is defined, and further time delay
measurement tests are carried out.


\section{Discussion}
\label{sec:discussion}

In this section, we give a brief analysis of each method's performance,
discussing how they performed and what can be improved in the future. We
note that the performance of each method must be evaluated in
multi-dimensional metric space. Each ``Good'' Team had to make choices with
respect to which metric to optimize. Some teams decided to favor
inclusiveness (high $f$) at the cost of a higher fraction of outliers
(lower $X$) or lower precision $P$, and vice-versa. In fact, some of the
teams submitted multiple entries spanning the range of parameter space,
and illustrating these competitive needs. Therefore, at this stage it is
not possible, nor useful, to identify a ``best'' submission, not even
within each method. It is more fruitful to understand the tradeoffs and
explore the range covered by each method, and then identify areas for
improvement.


\subsection{Gaussian Processes, by Hojjati \& Linder}

The GP method attained its twin goals of an automated fitting pipeline and
very good fit accuracy. The main issue to address is one of outliers,
which can be handled in two ways: global clipping and image
information.  This team found that the outliers were not due to
multi-modal fit distributions -- indeed the fits often have better
likelihood for the data than the truth. However, the cosmology derived
from the outliers would be discrepant from the cosmology from the global
fit ensemble, and in this way, outliers could be recognized and
clipped.  Another approach would be to use information such as image
separation (not provided in TDC1) to recognize and discard discrepant
fits. While these considerations would lower the accepted fraction of
fits, the correction of the mean function discussed in
\Sref{sec:response:HL} raises the fraction over those given here. This,
and a set of new but simple selection criteria (no limits on precision
were imposed by this team for TDC1 submissions), discussed in a
follow-up paper by \cite{HLinprep}, give considerable improvement in the
precision and fraction, and further improvement in accuracy.


\subsection{FOT, by Romero-Wolf \& Moustakas}

The unblinding of the TDC1 simulated data provided valuable information
on the behavior of this team's Bayesian inference algorithm. For the
most part, the technique identified catastrophic outliers. However, some
light curve pairs still resulted in large contributions to the $\chi^2$
estimator. Identifying this subset of outliers that pass the quality
cuts has provided valuable insight into the behavior of this technique,
and will allow for future refinement and development to reduce the
probability of mis-reconstructions.


\subsection{Smoothing and Cross-Correlation, by Aghamousa \& Shafieloo}

Throughout the challenge this team's main concern was to achieve
a high $f$ value without having any outliers. This was achieved with
$f>0.3$ for all five rungs. This conservative approach yielded average
$\chi^2$ values of around~$0.5-0.9$ for different Rungs with $P$ of
about $0.03$ to~$0.06$. As noted before, since $\chi^2$ and $P$ are
correlated, by simply dividing all estimated errors by a factor of
$\sqrt{2}$, $\chi^2$ of $\sim 1$ and $P$ of $\sim 0.02-0.04$ could be
achieved trivially. After the true time delays were revealed, a
calibration bias of 0.5 days for all the submissions was discovered,
resulting in $A\sim 1.8-2.5\%$ (the method had been calibrated only on
TDC0 Rung~0, owing to lack of time). By adding a calibration
correction of $0.5$ days to all this team's submissions' delay
estimates, the bias was removed, improving $A$ to $0.1-0.6\%$. To
summarize, this method seems promising in both reliability and
precision, and is automated in all steps. There is also the potential
to improve the error estimation by doing appropriate simulations for
each set of light curves separately.


\subsection{Supervised Pelt, by Jackson}

After the release of the true time delays, this submission was
re-examined to try to understand the reasons for the most severe
errors, especially those in which the true time delay differed from
the inference by $>3\sigma$ (between 9 and 18 cases in each rung out
of a few hundred submitted). In four of the worst cases, the problem
appeared to be unrealistically low errors fitted during the resampling
process, possibly due to a small number of anomalous points, and not
corrected by eye. This suggests that for a given set of light-curves,
a minimum error based on the fits to the ensemble should be adopted. A
significant fraction of the remaining severe errors were characterised
by a Pelt statistic vs.\ time delay plot with a relatively bumpy and
irregular minimum, even when the eye detected a good fit in terms of
the number of coincident points of inflection.  This is more difficult
to quantify, but suggests that an addition to the resampling-derived
error based on the shape of the Pelt statistic may be useful.


\subsection{PyCS {\tt d3cs}, {\tt spl} and {\tt sdi}}

The {\tt d3cs} classification of the light-curve pairs into different
confidence categories proved valuable. All the resulting ``doubtless''
({\tt dou}) submissions ($f=0.31$, averaging accross all rungs) are free
from any catastrophic outliers. As an example, none of the point
estimates from the vanilla {\tt spl} method is farther than $3.7
\sigma_i$ or $12.0$ days from the truth. For this same method, the less
pure {\tt doupla} submission ($f=0.65$) is contaminated by $1.0\%$ of
delays that are off by more than 20 days, or, alternatively, $5
\sigma_i$. Interestingly, the {\tt d3cs} estimates for time delays
shorter than 50 days are systematically biased low, leading to a
significant $A$ of approximately $-0.03$ for {\tt d3cs}. We speculate
that this bias is perceptual and due to users involuntarily trying to
maximize the overlap in the light curves. The {\tt sdi} and {\tt spl}
techniques were not influenced by this bias in their initial conditions,
and both reached a high accuracy, consistent with being unbiased. For
these two numerical techniques, the $\chi^2$ metric values are close to
unity, suggesting adequate to slightly over-estimated delay
uncertainties. The implemented simplifications to the original
techniques from \citet{TewesEtal2013a} seem therefore  acceptable for the
level of complexity present in the TDC1 data.


\subsection{Difference-smoothing, by Rathna Kumar, Stalin, \& Prabhu}

From the TDC1 feedback, it was realized that this procedure
overestimates the uncertainties in the measured time delays, and hence
was more prone to reporting catastrophic failures. This problem can be
solved by using a Gaussian filter of width equal to the median rather
than the mean temporal sampling of the light curves in the process of
simulating light curves having known time delays. With this choice, the
intrinsic variability in the simulated light curves does not get
smoothed out on short timescales. Also, there were a few cases in the
submissions where the measured and true time delays were discrepant at
the level of $\chi_i^2 > 10$. This points to a need to increase the
plausible range of time delays around the measured delay over which the
simulated light curves are generated to at least the 3$\sigma$
confidence interval implied by the inferred uncertainty, rather than the
2$\sigma$ confidence interval used in the TDC1 submissions. The time
delay measurements can be improved further by exploring a range of
reasonable values of free parameters, and selecting those which result
in the smallest uncertainty in the measured time delay. These changes
are now being rigorously tested on the TDC1 light curves and will be
described in the paper by \citet{2014arXiv1404.2920R} during the
revision process.


\subsection{DeltaTBayes, by Tak, Meng, van Dyk, Siemiginowska, Kashyap,
\& Mandel}

This team considered TDC1 to be a great opportunity to develop and
improve their Bayesian approach. Considering the team's late entry into
the challenge, the pragmatic Bayesian perspective was taken
\citep{Lee+11}, developing the approximate Gibbs sampling scheme
(algorithm1) and applying it only to the most realistic rung (Rung~4).
The main advantage of this pragmatic approach was the fast convergence
of its Markov chains, saving some computational time, a desirable
characteristic for analyzing large number of data sets. The method
performs well in terms of precision and accuracy. However it produces
error bars that are smaller than those from a fully-Bayesian approach,
though larger than an empirical Bayesian approach, leading to a
relatively high~$\chi^2$. To be balanced, several Gibbs sampling schemes
are being tested for the future.


\section{Summary and Conclusions}
\label{sec:summary}

In the next decade, dedicated efforts and the LSST survey will deliver
thousands of light curves for lensed quasars, ushering in a revolution
in time-delay cosmology \citep{Tre++13}. In order to prepare for and
make the most of this wealth of data, it is essential to ascertain
whether current algorithms are sufficiently accurate, fast, and precise.
It is also important to investigate the optimal observing strategies for
time delay determination, both in dedicated monitoring campaigns and for
LSST.

In order to investigate these two issues, we carried out the first
strong lens time delay challenge (TDC). After the preliminary time delay
challenge TDC0 \citep{PaperI}, the challenge ``Evil'' Team simulated
several thousand time delay light curves and made them available to the
community on the challenge website. Seven ``Good'' Teams responded to the
challenge, and blindly measured the time delays for TDC1 using 9
independent algorithms.  A simple method implemented by the
``Evil'' Team as a baseline was also included. Our main findings from
analyzing the the blind TDC1 submissions can
be summarized as follows.

\begin{itemize}

\item The measurement of time delays
from thousands of realistic light curves in manageable amounts of CPU
and investigator time has been demonstrated.
This is a considerable achievement given that
traditionally this process has been carried out only for very small
numbers of light curves (allowing investigators to spend significant
amounts of time on each system). Several independent approaches were
successful, ranging from cross-correlation, to scatter minimization, to
data modeling with Gaussian Processes and other suitable sets of basis
functions. Some methods relied heavily on visual inspection, while
others were almost completely automated.

\item In Rung~0 -- which simulates the typical observing parameters of a
dedicated monitoring campaign like COSMOGRAIL -- the best current
algorithms can recover time delays with negligible bias (often
sub-percent) and 3\% precision for over 50\% of the light curves. The
error bars are generally reasonable, resulting in $\chi^2$ of order
unity, while the fraction of outliers is also just a few percent. These
were the requirements for a method to be competitive, as described in
\paperone.  When enough information was present in the light curves,
typically 6 independent algorithms were able to recover time delays
within 10\% of the truth.

\item As the data quality was degraded in the subsquent Rungs~1-4
(emulating some observing strategies possible with LSST), the fraction
of usable light curves also decreased, hovering between 20 and 30\%.
Outliers became more common, although they can be contained by suitably
conservative algorithms, or by visual inspection. Once outliers are
excluded, the algorithms perform as well as in Rung~0, albeit with a
smaller fraction of the light curves (10-30\%) yielding robust results
with competitive precision and accuracy. A success fraction of 20\%
translates to an expected sample size of around 400 lensed quasars
detected and measured by LSST to very high accuracy -- well within the
systematic error requirements of time delay cosmography.

\item We have derived approximate scalings for the time delay metrics as
a function of observing parameters. Season and campaign length appear to
be the dominant terms controlling accuracy (or bias) and success rate,
while the precision of the time delay is most strongly related to the
cadence and campaign duration.

\end{itemize}

Much has been learned from this first blind time delay challenge, and
the results provide useful guidance and reference for designing future
experiments and improving the measurement algorithms. However, it should
be emphasized that this challenge was designed to be somewhat
simplistic. In particular, TDC1 consisted of a pure time delay
estimation challenge from light curves alone: teams were not given the
image positions, nor the deflector and source redshifts. It is likely
therefore that the results of this challenge might be overly
pessimistic. In real life, investigators will have access to the full
lensing configurations, and will be able to use this information as a
prior for their time delay inference (for example using the lensing
geometry for quads).

Furthermore, a fully cosmological challenge should enable outlier
rejection based on cosmological self-consistency in a joint analysis
of the ensemble of lenses. It should be possible to identify and
reject outliers that lead to cosmological parameters (chiefly H$_0$)
that are inconsistent with those inferred from the majority of
sample. Another limitation of the simplicity of TDC1 is that the
metrics measure how well an algorithm performs on time-delay
estimation, not directly on cosmological parameter inference.

Given the encouraging results of TDC1, we plan to overcome these two
limitations in the future. In the short term, we plan to translate the
simple metrics adopted here into a full cosmological estimation tool by
introducing the available additional information, and justifiable
assumptions about the underlying lens models. In the longer term, we
plan to organize a second time delay challenge, to further test our
ability to handle outliers, and to investigate the measurement of time
delays from multi-band data,
{\bf and in which more information will be provided for each system with the
ultimate goal for the ``Good'' Teams of inferring cosmological parameters,
rather than just time delays.}

The TDC0 and TDC1 data will remain available at
\url{http://timedelaychallenge.org} for any team who might be interested
in using them for developing algorithms for strong lens time delay
measurement.


\acknowledgments

We acknowledge the LSST Dark Energy Science Collaboration for hosting
several meetings of the ``Evil'' Team, and the private code repository
used in this work. We thank the referee for constructive criticism
which helped improved this paper. TT, CDF, KL acknowledge support from
the National Science Foundation collaborative grant ``Collaborative
Research: Accurate cosmology with strong gravitational lens time
delays'' (AST-1312329 and AST-1450141).  TT gratefully acknowledges
support by the Packard Foundation through a Packard Research
Fellowship. KL is supported by China Scholarship Council. The work of
PJM was supported by the U.S. Department of Energy under contract
number DE-AC02-76SF00515. VB and FC are supported by the Swiss
National Science Foundation (SNSF). MT acknowledges support by the DFG
grant Hi 1495/2-1. A. A and A. S. wishes to acknowledge support from
the Korea Ministry of Education, Science and Technology,
Gyeongsangbuk-Do and Pohang City for Independent Junior Research
Groups at the Asia Pacific Center for Theoretical Physics. A.S. would
like to acknowledge the support of the National Research Foundation of
Korea (NRF-2013R1A1A2013795).  EL is supported by DOE grant
DE-SC-0007867 and Contract No. DE-AC02-05CH11231. AH is supported by
an NSERC grant and thanks the Institute for the Early Universe, Korea
for computational resources.  AA, AS, AH, EL thank IBS Korea for
hospitality. The work of LAM and ARW was carried out at the Jet
Propulsion Laboratory, California Institute of Technology, under a
contract with the National Aeronautics and Space Administration. KM is
supported at Harvard by NSF grant AST-1211196.




\appendix

\section{Factors affecting the RMS microlensing magnification}

\begin{figure*}[!ht]
\begin{minipage}{0.48\linewidth}
\centering\includegraphics[width=\linewidth]{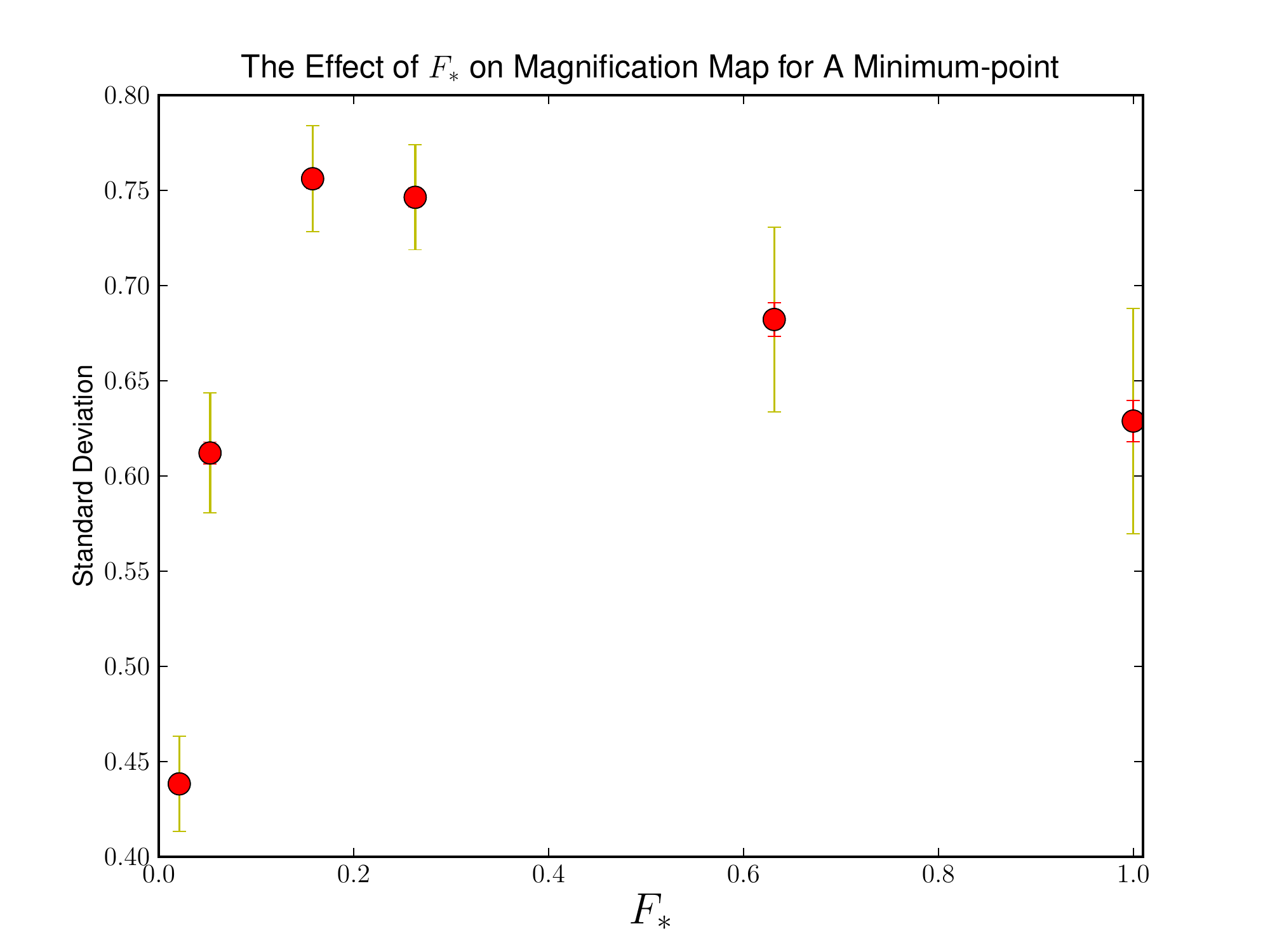}
\end{minipage}\hfill
\begin{minipage}{0.48\linewidth}
\centering\includegraphics[width=\linewidth]{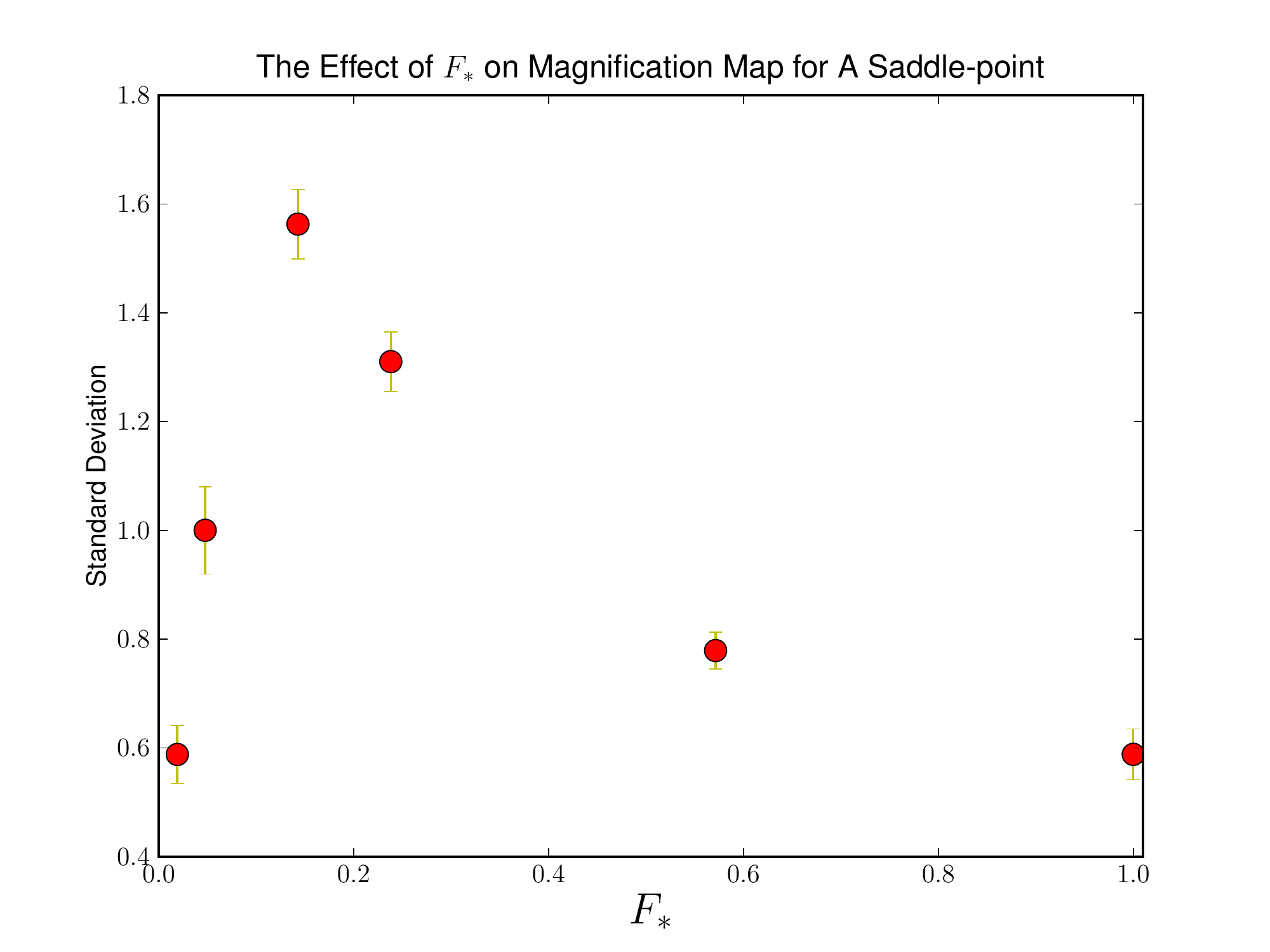}
\end{minipage}
\caption{Mean Standard Deviation of the magnification
map as a function of $F_*$. Each point is the result from 30
realisations with different position seeds. We show two errors:
Standard Deviation (yellow) and Standard Deviation of the mean value
(red). Top figure is for a minimum-image with $\kappa=0.475,
\gamma=0.425$. Bottom figure is for a saddle-image with $\kappa=0.525,
\gamma=0.575$. Both have the same macro magnification $\mu=10$.}
\label{fig:fstar}
\end{figure*}

How sensitive is the distribution of mock lightcurves to the random
realizations of the positions of the stars in the lens?  We generated 30
star field realizations, over fields 30 Einstein Radii ($R_E$) by 30
$R_E$ in area, with different random seeds for each fixed $F_*$ or
$\kappa$, and calculated the mean of their standard deviations as a
characteristic measure of the rms fluctuation in the microlensing
magnification. \Fref{fig:fstar} shows how this rms fluctuation varies as
a function of $F_*$. The top panel shows the case where the image arises
at the minimum of the time delay surface (where the eigenvalues of the
Hessian matrix are both positive and the image has positive parity); the
bottom panel shows the case where the image arises at a saddle point of
the time delay surface (where the eigenvalues have opposite signs and
the parity is flipped compared to the original source). For both
figures, significant trends, increasing when $F_*$ is small, and
decreasing at larger $F_*$, are apparent. These can be explained as
follows.

\begin{figure*}[!ht]
\begin{minipage}{0.48\linewidth}
\centering\includegraphics[width=\linewidth]{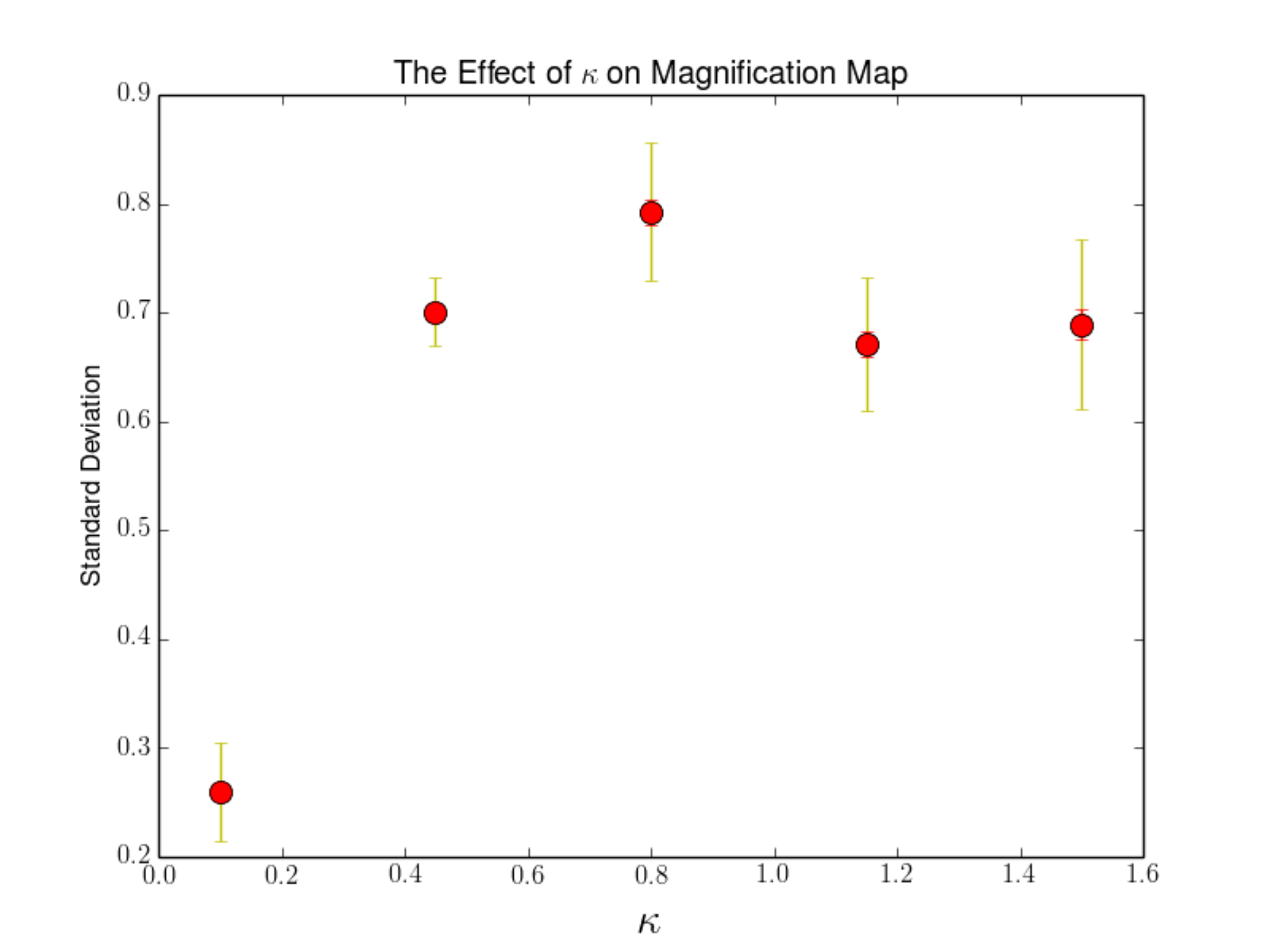}
\end{minipage}\hfill
\begin{minipage}{0.48\linewidth}
\centering\includegraphics[width=\linewidth]{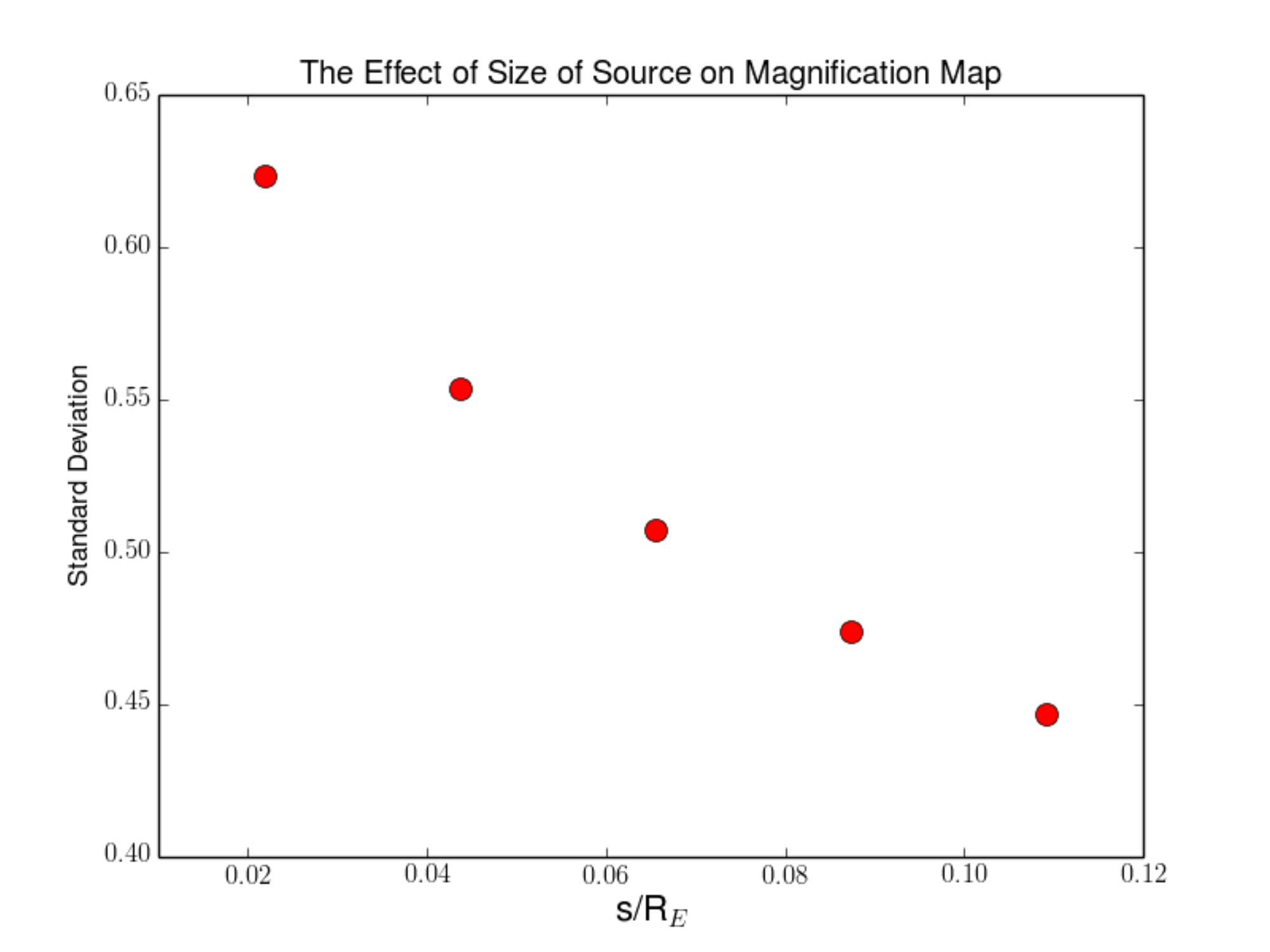}
\end{minipage}
\caption{Mean Standard Deviation of the magnification map as a function
of local convergence $\kappa$ (left) and source size $s$  (right). In
the left panel, $\kappa=\gamma$ and $F_*=0.1$ are fixed for each point,
while in the right panel $\kappa=\gamma=0.45$, $F_*=0.1$ are fixed for
each point.}
\label{fig:kappasourcesize}
\end{figure*}

At small $F_*$, when there are few stars, sparsely distributed in the
field, the magnification of each position is dominated by the nearest
individual star, and the variation of the map increases with more
stars that bring more caustics. However, when $F_*$ grows large, the
magnification at any position becomes less affected by the addition of
more stars, and the magnification and demagnification attributed to
different stars will average away.  The saddle-point images are more
vulnerable to demagnification and hence show larger variations in
their magnification maps \citep[see][for more on the differences of
microlensing between minima and saddle-point images]{MinSaddle}.

The lefthand panel of \Fref{fig:kappasourcesize} shows the effect of the
macrolens convergence $\kappa$ on the standard deviation of the source
plane magnification map. $\kappa$ affects the variation in two ways,
changing the stellar density fraction, and also the macro magnification.
These effects appear to approximately balance each other at
high~$\kappa$. At low convergence, the magnification and shear are also
low, and the microlensing effects weaker.

Meanwhile, the righthand panel of \Fref{fig:kappasourcesize} shows the
rms microlensing magnification fluctuation as a function of source size.
As expected, the fluctuations are  smoothed out at large source size,
reducing the amplitude of the microlensing fluctuations and ensuring
that the average microlensing magnification is unity.

\end{document}